\documentclass[twoside,12pt]{article}
\usepackage{amssymb,graphicx,macros2e}
\usepackage{times}
\skip\footins 39pt plus 4pt minus 2pt
\def\footnoterule{\kern-19pt\hrule width.5in\kern18.6pt}%
\newcommand{\text}[1]{\mbox{#1}}
\newcommand{\dotsb}{\ldots}
\renewcommand{\baselinestretch}{1}
\begin{document}

\newcommand{\qed}{\hfill\raisebox{-0.5mm}[0mm][0mm]{$\Box$}}
\newcommand{\st}{\!\!\!/}
\newcommand{\ksl}{k\!\!\!/}
\newcommand{\dsl}{\partial\!\!\!/}
\newcommand{\Asl}{A\!\!\!/}
\newcommand{\Dsl}{D\!\!\!\!/}
\newcommand{\half}{\frac{1}{2}}
\newcommand{\tbt}{{\bar \theta}\theta}
\newcommand{\A}{\alpha}
\newcommand{\B}{\beta}
\newcommand{\G}{\gamma}
\newcommand{\D}{\delta}
\newcommand{\E}{\varepsilon}
\newcommand{\T}{\theta}
\newcommand{\ts}{
   {\raisebox{-0.5ex}{\parbox{0.5ex}{
      \setlength{\unitlength}{0.5ex}
      \begin{picture}(1,6)
         \thinlines
         \put(0.5,-1){\line(0,1){7}}
      \end{picture}
   }}}
}
\newcommand{\lts}{
   {\raisebox{0ex}{\parbox{0.5ex}{
      \setlength{\unitlength}{0.5ex}
      \begin{picture}(1,12)
         \thinlines
         \put(0.5,-0.25){\line(0,1){12}}
      \end{picture}
   }}}
}
\newcommand{\N}{\mathbb{N}}
\newcommand{\R}{\mathbb{R}}
\newcommand{\Rscript}{\scriptstyle\mbox{\scriptsize \rm I}\!\mbox{\scriptsize\rm R}}
\newcommand{\rme}{{\mathrm{e}}}
\newcommand{\rmd}{{\mathrm{d}}} 
\newcommand{\tr}{\mbox{tr}}
\newcommand{\nn}{\nonumber}
\newcommand{\weiter}{\nonumber \\ & & }
\newcommand{\nicht}[1]{}
\newlength{\lang}
\newcommand {\eq}[1]{(\ref{#1})}
\newcommand {\Eq}[1]{Eq.\hspace{0.55ex}(\ref{#1})}
\newcommand {\Eqs}[1]{Eqs.\hspace{0.55ex}(\ref{#1})}
\newcommand {\Sec}[1]{section~\ref{#1}}
\newcommand {\ds}{\displaystyle}
\newcommand {\tx}{\textstyle}
\newcommand {\scr}{\scriptstyle}
\newcommand {\scrscr}{\scripscriptstyle}
\newcommand {\ind}[1]{\mathrm{#1}}
\newcommand {\p}{\partial}
\newcommand{\trint}{\int\!\!\!\int\!\!\!\int}
\newcommand{\bn}{:\hspace*{-0.5ex}}
\newcommand{\en}{\hspace*{-0.5ex}:}
\newcommand{\1}{\,{\bf 1}\,}
\newcommand{\llongrightarrow}{-\!\!\!-\!\!\!\!\rightarrow}
\newcommand{\lllongrightarrow}{-\!\!\!-\!\!\!-\!\!\!\!\rightarrow}
\newcommand{\llllongrightarrow}{-\!\!\!-\!\!\!-\!\!\!-\!\!\!\!\rightarrow}
\newcommand{\lllllongrightarrow}{-\!\!\!-\!\!\!-\!\!\!-\!\!\!-\!\!\!\!\rightarrow}
\newcommand{\llllllongrightarrow}{-\!\!\!-\!\!\!-\!\!\!-\!\!\!-\!\!\!-\!\!\!\!\rightarrow}
\newcommand{\lllllllongrightarrow}{-\!\!\!-\!\!\!-\!\!\!-\!\!\!-\!\!\!-\!\!\!-\!\!\!\!\rightarrow}
\newcommand{\llllllllongrightarrow}{-\!\!\!-\!\!\!-\!\!\!-\!\!\!-\!\!\!-\!\!\!-\!\!\!-\!\!\!\!\rightarrow}
\newcommand{\lllllllllongrightarrow}{-\!\!\!-\!\!\!-\!\!\!-\!\!\!-\!\!\!-\!\!\!-\!\!\!-\!\!\!-\!\!\!\!\rightarrow}
\newcommand{\llllllllllongrightarrow}{-\!\!\!-\!\!\!-\!\!\!-\!\!\!-\!\!\!-\!\!\!-\!\!\!-\!\!\!-\!\!\!-\!\!\!\!\rightarrow}
\newcommand{\lllllllllllongrightarrow}{-\!\!\!-\!\!\!-\!\!\!-\!\!\!-\!\!\!-\!\!\!-\!\!\!-\!\!\!-\!\!\!-\!\!\!-\!\!\!\!\rightarrow}
\newcommand{\llllllllllllongrightarrow}{-\!\!\!-\!\!\!-\!\!\!-\!\!\!-\!\!\!-\!\!\!-\!\!\!-\!\!\!-\!\!\!-\!\!\!-\!\!\!-\!\!\!\!\rightarrow}
\newcommand{\setcurrentlabel}[1]{\def\@currentlabel{#1}}
\newbox{\expbox}
\newlength{\explength}
\newcommand{\EXPhelp}[6]{%
\sbox{\expbox}{\ensuremath{#4#1}}%
\settowidth{\explength}{\rotatebox{90}{\ensuremath{#4\left#5\usebox{\expbox}\right#6}}}%
\ensuremath{#4\left#5\usebox{\expbox}\right#6%
_{{#4\hspace*{0.4ex}}\hspace*{-0.22\explength}#2}%
^{{#4\hspace*{0.4ex}}\hspace*{-0.22\explength}#3}}}
\newcommand{\EXPdu}[3]{%
\mathchoice{\EXPhelp{#1}{#2}{#3}{\displaystyle}{<}{>}}%
{\EXPhelp{#1}{#2}{#3}{\textstyle}{<}{>}}%
{\EXPhelp{#1}{#2}{#3}{\scriptstyle}{<}{>}}%
{\EXPhelp{#1}{#2}{#3}{\scriptscriptstyle}{<}{>}}}
\newcommand{\EXP}[2]{%
\mathchoice{\EXPhelp{#1}{#2}{}{\displaystyle}{<}{>}}%
{\EXPhelp{#1}{#2}{}{\textstyle}{<}{>}}%
{\EXPhelp{#1}{#2}{}{\scriptstyle}{<}{>}}%
{\EXPhelp{#1}{#2}{}{\scriptscriptstyle}{<}{>}}}
\newcommand{\BRAhelp}[6]{%
\sbox{\expbox}{\ensuremath{#4#1}}%
\settowidth{\explength}{\rotatebox{90}{\ensuremath{#4\left#5\usebox{\expbox}\right#6}}}%
\ensuremath{#4\left#5\usebox{\expbox}\right#6%
_{{#4\hspace*{0.4ex}}\hspace*{-0.2\explength}#2}%
^{{#4\hspace*{0.4ex}}\hspace*{-0.2\explength}#3}}}
\newcommand{\BRA}[2]{%
\mathchoice{\BRAhelp{#1}{}{#2}{\displaystyle}{(}{)}}%
{\BRAhelp{#1}{}{#2}{\textstyle}{(}{)}}%
{\BRAhelp{#1}{}{#2}{\scriptstyle}{(}{)}}%
{\BRAhelp{#1}{}{#2}{\scriptscriptstyle}{(}{)}}}
\newbox{\atbox}
\newlength{\atlengtha}
\newlength{\atlengthb}
\newcommand{\AThelp}[4]{%
\sbox{\atbox}{\ensuremath{#3#1}}%
\settoheight{\atlengtha}{\ensuremath{#3\usebox{\atbox}}}%
\settodepth{\atlengthb}{\ensuremath{#3\usebox{\atbox}}}%
#3\addtolength{\atlengtha}{0.1ex}
#3\addtolength{\atlengthb}{0.75ex}%
\addtolength{\atlengtha}{\atlengthb}%
#1\rule[-\atlengthb]{0.1ex}{\atlengtha}_{\raisebox{0.12ex}%
{\ensuremath{\,#4#2}}}}%
\newcommand{\AT}[2]{%
\mathchoice{\AThelp{#1}{#2}{\displaystyle}{\scriptstyle}}%
{\AThelp{#1}{#2}{\textstyle}{\scriptstyle}}%
{\AThelp{#1}{#2}{\scriptstyle}{\scriptscriptstyle}}%
{\AThelp{#1}{#2}{\scriptscriptstyle}{\scriptscriptstyle}}}
\newlength{\picheight}%
\newcommand{\AxesPicture}[4]{%
\settowidth{\picheight}{\rotatebox{90}{$\tx#4$}}%
%\centerline%
{\arraycolsep0.5ex
\renewcommand{\arraystretch}{1.3}
$\begin{array}{cc}
\mbox{\parbox{\picheight}{\rotatebox{90}{$~~\tx#4$}}}
 & \mbox{\epsfxsize=#2\textwidth\parbox{#2\textwidth}{\epsfbox{#1}}} \\
~ & \mbox{$~~\tx#3$}
\end{array}$}}
\newbox{\picbox}
\newlength{\picwidth}
\newcommand{\AxesPictureRight}[4]{%
\sbox{\picbox}{\epsfxsize=#2\textwidth\parbox{#2\textwidth}{\epsfbox{#1}}}
\settowidth{\picheight}{\rotatebox{90}{$\tx#4$}}%
\settowidth{\picwidth}{\rotatebox{90}{\usebox{\picbox}}}
\centerline{%
\arraycolsep0.5ex
\renewcommand{\arraystretch}{1.3}
$\begin{array}{cr}
\mbox{\parbox{\picheight}{\rotatebox{90}{\parbox{\picwidth}{\hfill{$\tx#4$}}}}}
 & \mbox{\usebox{\picbox}} \\
~ & \mbox{$\tx#3$}
\end{array}$}}
\newcommand{\AxesPictureSpecial}[6]{%
\centerline{$\raisebox{#5\textwidth}{\rotatebox{90}{$\tx#4$}}~
{{\epsfxsize=#2\textwidth\parbox{#2\textwidth}{\epsfbox{#1}}}}$%
$\hspace*{-2ex}\raisebox{#6\textwidth}{$\tx#3$}$}}
\newcommand{\NOhere}{\parbox{1cm}{\epsfxsize=1cm\epsfbox{./eps/isno-ani.eps}}}
\newcommand{\NO}{\marginpar{$\!\!\parbox{1.8cm}{\epsfxsize=1.8cm\epsfbox{./eps/isno-ani.eps}}$}}

\def\beginincorrect{

\noindent%
{\unitlength1mm
\begin{picture}(100,5)
\put(0,0){\line(0,1){5}}
\put(0,5){\line(1,0){179}}
\put(179,0){\line(0,1){5}}
\put(78,2){\mbox{\normalsize*****incorrect*****}}
\end{picture}}\newline
\scriptsize}
\def\endincorrect{

\noindent%
{\unitlength1mm
\begin{picture}(100,5)
\put(0,0){\line(0,1){5}}
\put(0,0){\line(1,0){179}}
\put(179,0){\line(0,1){5}}
\put(74,1){\mbox{\normalsize*****end incorrect*****}}
\end{picture}}

\normalsize}

\newcommand{\comment}[1]{}

%%%%%%%%%%%FIGURES%%%%%%%%%%%%%%%%%%
%
%\newcommand{\fig}[2]{\epsfxsize=#1\epsfbox{./figures/#2.eps}}
%\newcommand{\Fig}[1]{\epsfxsize=\columnwidth\epsfbox{./figures/#1.eps}}
%
\newcommand{\bilderscale}{0.35}
\newcommand{\textbilderscale}{0.25}

\newcommand{\fig}[2]{\includegraphics[width=#1]{./figures/#2}}
\newcommand{\pfig}[2]{\parbox{#1}{\includegraphics[width=#1]{./figures/#2}}}
\newcommand{\Fig}[1]{\includegraphics[width=\columnwidth]{./figures/#1}}
\newlength{\bilderlength} 
\newcommand{\usebilderscale}{\bilderscale}
\newcommand{\bilderskip}{\hspace*{0.8ex}}
\newcommand{\textdiagram}[1]{%
\renewcommand{\usebilderscale}{\textbilderscale}%
\diagram{#1}\renewcommand{\usebilderscale}{\bilderscale}}
\newcommand{\diagram}[1]{%
\settowidth{\bilderlength}{\bilderskip%
\includegraphics[scale=\usebilderscale]{./figures/#1}\bilderskip}%
\parbox{\bilderlength}{\bilderskip%
\includegraphics[scale=\usebilderscale]{./figures/#1}\bilderskip}}
\newcommand{\Diagram}[1]{%
\settowidth{\bilderlength}{%
\includegraphics[scale=\usebilderscale]{./figures/#1}}%
\parbox{\bilderlength}{%
\includegraphics[scale=\usebilderscale]{./figures/#1}}}

%%%%%%%DEBUG-PREPRINT%%%%%%%%%%%%%
%
%\pagestyle{myheadings}
%\protect\markboth{June}{DRAFT,  \today}

\title{\vspace{-1.5cm}\sffamily\Large\bfseries 
Derivation of the Functional Renormalization Group
$\beta$-function at order $1/N$ for manifolds pinned by disorder}

\author{{\sffamily\bfseries Pierre Le Doussal$^{1}$ and Kay J\"org
Wiese$^{1,2}$} \medskip \\
\small $^{1}$ CNRS-Laboratoire de Physique Th\'eorique de l'Ecole 
Normale Sup\'erieure, 24 rue Lhomond, 75005 Paris, France\\
%\small $^{2}$ KITP, University of California at Santa Barbara, Santa Barbara,
%CA 93106-4030
\small $^{2}$ Institut f\"ur Theoretische Physik, Universit\"at zu K\"oln,
Z\"ulpicher Str. 77, 50937 K\"oln, Germany
}

\date{\small\today}
\maketitle
\begin{abstract}
\noindent In an earlier publication, we have introduced a method to
obtain, at large $N$, the effective action for $d$-dimensional
manifolds in a $N$-dimensional disordered environment. This allowed to
obtain the Functional Renormalization Group (FRG) equation for
$N=\infty$ and was shown to reproduce, with no need for ultrametric
replica symmetry breaking, the predictions of the M\'ezard-Parisi
solution. Here we compute the corrections at order $1/N$. We introduce
two novel complementary methods, a diagrammatic and an algebraic one,
to perform the complicated resummation of an infinite number of loops,
and derive the $\beta$-function of the theory to order $1/N$.  We
present both the effective action and the corresponding functional
renormalization group equations. The aim is to explain the conceptual
basis and give a detailed account of the novel aspects of such
calculations. The analysis of the FRG flow, comparison with other
studies, and applications, e.g.\ to the strong-coupling phase of the
Kardar-Parisi-Zhang equation are examined in a subsequent publication.
\end{abstract}

%%%%%%%%%%%%%%%%%%%%%%%%%%%%%%%%%%%%%%%%%%%%%%%%%%%%%%%%%%%%%%%%%%%%%%
%%%%%%%%%%%%%%%%%%%%%%%%%%%%%%%%%%%%%%%%%%%%%%%%%%%%%%%%%%%%%%%%%%%%%%
%%%%%%%%%%%%%%%%%%%%%%%%%%%%%%%%%%%%%%%%%%%%%%%%%%%%%%%%%%%%%%%%%%%%%%
%%%%%%%%%%%%%%%%%%%%%%%%%%%%%%%%%%%%%%%%%%%%%%%%%%%%%%%%%%%%%%%%%%%%%%
%%%%%%%%%%%%%%%%%%%%%%%%%%%%%%%%%%%%%%%%%%%%%%%%%%%%%%%%%%%%%%%%%%%%%%
%%%%%%%%%%%%%%%%%%%%%%%%%%%%%%%%%%%%%%%%%%%%%%%%%%%%%%%%%%%%%%%%%%%%%%
%%%%%%%%%%%%%%%%%%%%%%%%%%%%%%%%%%%%%%%%%%%%%%%%%%%%%%%%%%%%%%%%%%%%%%
%%%%%%%%%%%%%%%%%%%%%%%%%%%%%%%%%%%%%%%%%%%%%%%%%%%%%%%%%%%%%%%%%%%%%%
%%%%%%%%%%%%%%%%%%%%%%%%%%%%%%%%%%%%%%%%%%%%%%%%%%%%%%%%%%%%%%%%%%%%%%
%%%%%%%%%%%%%%%%%%%%%%%%%%%%%%%%%%%%%%%%%%%%%%%%%%%%%%%%%%%%%%%%%%%%%%
%                          Introduction                              %
%%%%%%%%%%%%%%%%%%%%%%%%%%%%%%%%%%%%%%%%%%%%%%%%%%%%%%%%%%%%%%%%%%%%%%
%%%%%%%%%%%%%%%%%%%%%%%%%%%%%%%%%%%%%%%%%%%%%%%%%%%%%%%%%%%%%%%%%%%%%%
%%%%%%%%%%%%%%%%%%%%%%%%%%%%%%%%%%%%%%%%%%%%%%%%%%%%%%%%%%%%%%%%%%%%%%
%%%%%%%%%%%%%%%%%%%%%%%%%%%%%%%%%%%%%%%%%%%%%%%%%%%%%%%%%%%%%%%%%%%%%%
%%%%%%%%%%%%%%%%%%%%%%%%%%%%%%%%%%%%%%%%%%%%%%%%%%%%%%%%%%%%%%%%%%%%%%
%%%%%%%%%%%%%%%%%%%%%%%%%%%%%%%%%%%%%%%%%%%%%%%%%%%%%%%%%%%%%%%%%%%%%%
%%%%%%%%%%%%%%%%%%%%%%%%%%%%%%%%%%%%%%%%%%%%%%%%%%%%%%%%%%%%%%%%%%%%%%
%%%%%%%%%%%%%%%%%%%%%%%%%%%%%%%%%%%%%%%%%%%%%%%%%%%%%%%%%%%%%%%%%%%%%%
%%%%%%%%%%%%%%%%%%%%%%%%%%%%%%%%%%%%%%%%%%%%%%%%%%%%%%%%%%%%%%%%%%%%%%
%%%%%%%%%%%%%%%%%%%%%%%%%%%%%%%%%%%%%%%%%%%%%%%%%%%%%%%%%%%%%%%%%%%%%%

\section{Introduction}\label{intro} In a series of recent articles we
have constructed the Functional Renormalization Group (FRG) method for
disordered systems, applied to specific situations and beyond one loop
\cite{ChauveLeDoussalWiese2000a,LeDoussalWiese2001v1,LeDoussalWiese2001,%
LeDoussalWieseChauve2002,LeDoussalWiese2002a,LeDoussalWiese2003a,%
LeDoussalWieseChauve2003,LeDoussalWiese2003b,%
RossoKrauthLeDoussalVannimenusWiese2003}.  This method is, apart from
mean field theory \cite{MezardParisi1991,MezardParisi1992} using
Replica Symmetry Breaking (RSB) and some rare exactly solvable cases
\cite{BrunetDerrida2000a,BrunetDerrida2000}, the only known analytical
method which promises to handle the strong coupling glass phase of
disordered elastic systems
\cite{ChauveLeDoussalWiese2000a,LeDoussalWiese2001v1,LeDoussalWiese2001,%
LeDoussalWieseChauve2002,LeDoussalWiese2002a,LeDoussalWiese2003a,%
LeDoussalWieseChauve2003,LeDoussalWiese2003b,%
RossoKrauthLeDoussalVannimenusWiese2003,DSFisher1985,Fisher1985b,DSFisher1986,Nattermann1987,NarayanDSFisher1990,NattermanStepanowTangLeschhorn1992,NarayanDSFisher1992a,NarayanDSFisher1992b,NarayanDSFisher1993a,NarayanDSFisher1993b,BlatterFeigelmanGeshkenbeinLarkinVinokur1994,ErtasKardar1994,ErtasKardar1996,BalentsBouchaudMezard1996,Kardar1997,LeschhornNattermannStepanowTang1997,BucheliWagnerGeshkenbeinLarkinBlatter1998,DSFisher1998,DincerDiplom,Scheidl2loopPrivate,ScheidlDincer2000,NattermannScheidl2000,GorokhovFisherBlatter2002,SchwarzFisher2002}.
Such systems, modeled by an elastic manifold (of internal dimension
$d$) with a $N$-component displacement field $u(x)$ (i.e.\ $x \in
{\mathbb{R}}^d$ and $u (x) \in \mathbb{R}^{N}$), are of high interest
for numerous experiments
\cite{Gruner1988,LemerleFerreChappertMathetGiamarchiLeDoussal1998,PrevostRolleyGuthmann1999,PrevostRolleyGuthmann2002,MoulinetGuthmannRolley2002,BlatterFeigelmanGeshkenbeinLarkinVinokur1994,NattermannScheidl2000,GiamarchiLeDoussalBookYoung}. This
so-called random manifold model still offers great theoretical
challenges and a strong motivation is the hope to gain insight into
glassy physics. In addition, the $d=1$ case maps onto the much studied
Kardar-Parisi-Zhang growth equation \cite{KPZ}.  It exhibits a strong
coupling phase for which the upper critical dimension is still under
debate
\cite{LassigKinzelbach1997,MarinariPagnaniParisi2000,Laessig1995,Wiese1998a,Wiese2003a}.

Higher loop studies of the statics of disordered elastic systems
allow, in principle, a systematic dimensional expansion, in the
simplest case around $d=4$. They are however of a rather different
nature than in standard field theory for pure critical systems
\cite{ChauveLeDoussalWiese2000a,LeDoussalWieseChauve2003,ChauveLeDoussal2001,BucheliWagnerGeshkenbeinLarkinBlatter1998,DincerDiplom,Scheidl2loopPrivate,ScheidlDincer2000,BalentsLeDoussal2002,BalentsLeDoussal2003,LeDoussalWiesePREPf,LeDoussalWiesePREPb}.
Thermal fluctuations are found to be formally irrelevant in these
glass phases, suggesting that the physics is controlled by a zero
temperature fixed point. However before this fixed point is reached,
the zero temperature effective action is found to become non-analytic
\cite{DSFisher1986}.  Although this allows to evade the so-called
dimensional reduction \cite{EfetovLarkin1977} which makes naive
perturbation theory useless and yields unphysical results, it also
generates amazing new subtleties in the field theory.  These were
analyzed in a number of papers
\cite{ChauveLeDoussalWiese2000a,LeDoussalWieseChauve2003,ChauveLeDoussal2001,BucheliWagnerGeshkenbeinLarkinBlatter1998,DincerDiplom,Scheidl2loopPrivate,ScheidlDincer2000,BalentsLeDoussal2002,BalentsLeDoussal2003,LeDoussalWiesePREPf,LeDoussalWiesePREPb},
and although some solutions to the puzzles were proposed the physics
still remains to be elucidated.

An interesting limit where one can hope to gain insight into these
formidable problems is the large-$N$ limit.  Since $N=\infty$ is
formally the mean-field limit, it allows a direct confrontation
between the FRG method and mean field methods. A solution of the
$O(N)$ random manifold model for $N=\infty$ was proposed by M\'ezard
and Parisi, using a saddle point with spontaneous replica symmetry
breaking \cite{MezardParisi1991,MezardParisi1992}. As in other models
of glasses, spontaneous RSB can be related to ergodicity breaking of
the Gibbs measure into several ground states
\cite{MezardParisiVirasoro}.  Although it offers a rather elegant way
out of dimensional reduction, it is by no means clear that systems
with (large but) finite $N$ should exhibit such a tremendous
degeneracy of low energy states; and there are in fact indications to
the contrary \cite{Middleton2001}.

It is thus crucial to develop another line of attack, even in that
limit.  This is what we have achieved in a previous publication, where
we have computed the effective action of the theory at large
$N$. There we have derived the $\beta$-function of the field theory to
dominant order, i.e.\ for $N=\infty$
\cite{LeDoussalWiese2001v1,LeDoussalWiese2001,LeDoussalWiese2003b}. We
have discovered that beyond the Larkin length the FRG flow freezes (at
least for specific initial conditions) and that most of the features
of the M\'ezard Parisi solution can be recovered. Interestingly
however, in this formulation there is no need for a spontaneous RSB
ansatz with ultrametric structure.  Thus one may hope that it could be
more adapted to real world situations than the RSB calculations. Such
RSB calculations of fluctuations around the mean-field solution have
been attempted for the random-manifold problem only in the case of
1-step non-marginal RSB (with disappointing result
\cite{Goldschmidt1993}) and offer, in full generality, extreme
complications, as is illustrated by several studies for spin glasses
\cite{DeDominicisKondorTemesvari1994,CarlucciDeDominicisTemesvari1996,DeDominicisEtAlBookYoung,BrezinDeDominicis1998,BrezinDeDominicis2001,BrezinDeDominicis2002a,BrezinDeDominicis2002b,DeDominicisBrezin2004}.

The next challenge is thus to extend the FRG in a large-$N$ expansion
beyond the dominant order ($N=\infty$). This is the aim of the present
paper. Since this is a complicated calculation, and involves
developing new methods which are of interest by themselves, this paper
is restricted to the calculation of the effective action and
derivation of the $\beta$-function to order $1/N$. This is performed
at $T=0$ and at finite temperature. The analysis of the resulting FRG
flow, comparison with other studies, and applications, e.g.\ to the
strong-coupling phase of the KPZ equation is involved and is the
subject of a forthcoming publication.

The outline of the paper is as follows. In section \ref{genform} we
give the general formulation of the $1/N$ expansion for the effective
action of the random manifold.  Details and generalizations are given
in appendices \ref{largeNgeneral}, \ref{toy}, and \ref{app:Details
random manifold}. In section \ref{s:review} we summarize the main
results for $N=\infty$.  Section \ref{s:graphmethod} explains the
 derivation of the $1/N$ correction by a graphical method,
which introduces a new type of diagrammatics. Section
\ref{s:algebramethod} explains the principle of a second and
complementary method based on the algebra of 4-replica
tensors. Section \ref{results} contains the full result for the
effective action to order $1/N$, first expressed in bare parameters,
then as a function of the renormalized dimensionless disorder. This
allows, in section \ref{s:beta1overN}, for a derivation of the
$\beta$-function at $T=0$. The structure of the finite-$T$
$\beta$-function is indicated, and details given in appendix
\ref{app:detailsbeta}. A fool-proof diagrammatic version for finite
temperature is given in appendix \ref{excludedrepformalism}. More
details on the two main methods are given respectively in appendix
\ref{s:moregraphics} (for the diagrammatic method, including an
alternative derivation of the $T=0$ $\beta$-function) and in appendix
\ref{app:Algebra} (for the algebraic method). Appendix
\ref{app:integrals} contains a list of all integrals. A table
summarizing the notation is found in appendix \ref{sec:Notation}.

%%%%%%%%%%%%%%%%%%%%%%%%%%%%%%%%%%%%%%%%%%%%%%%%%%%%%%%%%%%%%%%%%%%%%%
%%%%%%%%%%%%%%%%%%%%%%%%%%%%%%%%%%%%%%%%%%%%%%%%%%%%%%%%%%%%%%%%%%%%%%
%%%%%%%%%%%%%%%%%%%%%%%%%%%%%%%%%%%%%%%%%%%%%%%%%%%%%%%%%%%%%%%%%%%%%%
%%%%%%%%%%%%%%%%%%%%%%%%%%%%%%%%%%%%%%%%%%%%%%%%%%%%%%%%%%%%%%%%%%%%%%
%%%%%%%%%%%%%%%%%%%%%%%%%%%%%%%%%%%%%%%%%%%%%%%%%%%%%%%%%%%%%%%%%%%%%%
%%%%%%%%%%%%%%%%%%%%%%%%%%%%%%%%%%%%%%%%%%%%%%%%%%%%%%%%%%%%%%%%%%%%%%
%%%%%%%%%%%%%%%%%%%%%%%%%%%%%%%%%%%%%%%%%%%%%%%%%%%%%%%%%%%%%%%%%%%%%%
%%%%%%%%%%%%%%%%%%%%%%%%%%%%%%%%%%%%%%%%%%%%%%%%%%%%%%%%%%%%%%%%%%%%%%
%%%%%%%%%%%%%%%%%%%%%%%%%%%%%%%%%%%%%%%%%%%%%%%%%%%%%%%%%%%%%%%%%%%%%%
%%%%%%%%%%%%%%%%%%%%%%%%%%%%%%%%%%%%%%%%%%%%%%%%%%%%%%%%%%%%%%%%%%%%%%
%                      General formula                               %
%%%%%%%%%%%%%%%%%%%%%%%%%%%%%%%%%%%%%%%%%%%%%%%%%%%%%%%%%%%%%%%%%%%%%%
%%%%%%%%%%%%%%%%%%%%%%%%%%%%%%%%%%%%%%%%%%%%%%%%%%%%%%%%%%%%%%%%%%%%%%
%%%%%%%%%%%%%%%%%%%%%%%%%%%%%%%%%%%%%%%%%%%%%%%%%%%%%%%%%%%%%%%%%%%%%%
%%%%%%%%%%%%%%%%%%%%%%%%%%%%%%%%%%%%%%%%%%%%%%%%%%%%%%%%%%%%%%%%%%%%%%
%%%%%%%%%%%%%%%%%%%%%%%%%%%%%%%%%%%%%%%%%%%%%%%%%%%%%%%%%%%%%%%%%%%%%%
%%%%%%%%%%%%%%%%%%%%%%%%%%%%%%%%%%%%%%%%%%%%%%%%%%%%%%%%%%%%%%%%%%%%%%
%%%%%%%%%%%%%%%%%%%%%%%%%%%%%%%%%%%%%%%%%%%%%%%%%%%%%%%%%%%%%%%%%%%%%%
%%%%%%%%%%%%%%%%%%%%%%%%%%%%%%%%%%%%%%%%%%%%%%%%%%%%%%%%%%%%%%%%%%%%%%
%%%%%%%%%%%%%%%%%%%%%%%%%%%%%%%%%%%%%%%%%%%%%%%%%%%%%%%%%%%%%%%%%%%%%%
%%%%%%%%%%%%%%%%%%%%%%%%%%%%%%%%%%%%%%%%%%%%%%%%%%%%%%%%%%%%%%%%%%%%%%

\section{$1/N$ expansion of the effective action: General
formula}\label{genform}

\comment{**** FINAL POUR PIERRE+KAY  **********} 

We start from the partition function of an interface ${\cal Z}_V=\int {\cal
D}[u]\, \rme^{- {\cal H}_V[u]/T}$ in a given
% realization of a random potential
sample, with energy
\begin{equation}\label{lf18}
{\cal H}_V[u] = \int_q \frac{1}{2} (q^2 + m^2) u(-q) \cdot u(q) +
\int_x V(x,u(x)) \ ,
\end{equation} 
where $\int_q \equiv \int \frac{\rmd^d
q}{(2 \pi)^d}$, $\int_x \equiv \int \rmd^d x$ and $u \cdot v =
\sum_{i=1}^N u^i v^i$. The $O(N)$ indices will be specified only when
strictly necessary, and below additional replica indices for the
replicated field $u_a^i$ will be introduced, $a=1,\dots ,n$. 
The small confining mass $m$ provides a
scale. To obtain a non-trivial large-$N$ limit, one defines the rescaled
field $v=u/\sqrt{N}$ and chooses the distribution of the random
potential to be rotationally invariant, e.g.\ its second cumulant as
\begin{equation}\label{correlator}
\overline{V(x,u) V(x',u')} = R(u-u') \delta_{x x'} = N
B((v-v')^2) \delta_{x x'} 
\end{equation}
in terms of a function $B(z)$. Higher connected cumulants are scaled
as
\begin{equation}\label{hcc}
\overline{V(x_1,u_1) \dots V(x_p,u_p)}^{\mathrm{conn}} = N
\delta_{x_1,\dots ,x_p} S^{(p)}(v_1,\dots ,v_p)\ ,
\end{equation} 
with $\delta_{x_{1},\dots ,x_{p}} :=
\prod_{i=2}^{p}\delta^{d} (x_{1}-x_{i})$.

Physical observables can be obtained for any $N$ from the replicated
action at $n=0$ with a source $J=\sqrt{N} j$ as
\begin{eqnarray}\label{lf19}
{\cal Z}[J]&=& \int {\cal D} [u] {\cal D}[ \chi] {\cal D}[\lambda]
\,\rme^{ - N {\cal S}[u,\chi,\lambda,j] }
 \\ 
 {\cal S}[u,\chi,\lambda,j] &=&  \frac{1}{2 T} \int_q  (q^2 + m^2)
v_a(-q) \cdot v_a(q) \nonumber \\ 
&&+ \int_x  U(\chi(x)) 
- \frac{1}{2} i \lambda_{ab}(x) [\chi_{ab}(x) -
v_a(x) \cdot v_b(x)] - j_a(x) \cdot v_{a}(x) \ ,
\end{eqnarray}
where the replica matrix field $\chi(x) \equiv \chi_{ab}(x)$ has been
introduced through a Lagrange multiplier matrix field
$\lambda_{ab}(x)$. Here and below summations over repeated replica (and $O(N)$)
indices $a,b=1,\dots ,n$ is implicit. The bare interaction matrix
potential
\begin{equation}\label{bpot}
U(\chi) = - \frac{ 1}{2 T^2}
\sum_{ab} B(\tilde{\chi}_{ab}) - \frac{1}{3! T^3} \sum_{abc}
S(\tilde{\chi}_{ab}, \tilde{\chi}_{bc}, \tilde{\chi}_{ca}) +
\dots
\end{equation}
depends only on the matrix
\begin{equation}\label{chicomb}
\tilde{\chi}_{ab}
:=\chi_{aa} + \chi_{bb} - \chi_{ab} - \chi_{ba}
\end{equation}
and has a cumulant
expansion in terms of sums with higher numbers of replicas.

The effective action functional $\Gamma[u]$ is defined as the Legendre
transform of ${\cal W} [J] = \ln {\cal Z}[J]$ and satisfies
\begin{equation}\label{Legendre}
\Gamma[u] + {\cal W}[J] = \int J(x) \cdot u(x) \ .
\end{equation}
Since $\Gamma[u]$ defines the renormalized vertices, its zero-momentum
limit defines the {\it renormalized disorder}, the quantity on which
we focus here. Thus we only need the result (per unit volume) for a
{\it uniform} configuration of the replica field $u_a(x)=u_a =\sqrt{N}
v_a$, which takes the form:
\begin{equation}\label{lf20}
\tilde{\Gamma}(v) := \frac{1}{L^d N} \Gamma(u) = \frac{1}{2 T} m^2
v_a^2 + \tilde{U}(v v)\ ,
\end{equation}
where $v v$ stands for the matrix $v_a\cdot v_b$. This defines the
renormalized disorder potential $\tilde{U}(v v)$ and, whenever it can
be expanded, up to a constant,
\begin{equation}\label{cumulants1}
\tilde{U}(v v)  = -\frac{1}{2 T^2} \sum_{ab} \tilde{B}(v_{ab}^2)
- \frac{1}{3! T^3} \sum_{abc}
\tilde{S}(v_{ab}^2,v_{bc}^2,v_{ca}^2) + \dots 
\ .
\end{equation}
It defines the {\it renormalized cumulants} $\tilde{B}(z)$, $\tilde S
(\dotsb)$ etc.. Here and in the following we denote 
\begin{equation}\label{vabdef}
v_{ab} := v_a -
v_b\ .
\end{equation}
We aim at calculating the effective action up to terms of order
$O(1/N^2)$, i.e.\ the first two terms in the expansion:
\begin{equation}\label{cumulants2}
\tilde{U}(v v)  = \tilde{U}^0(v v) + \frac{1}{N} \tilde{U}^1(v v) +
O(1/N^2) \ .
\end{equation}
Details of the calculation, as well as expressions for non-uniform
fields are given in Appendix \ref{app:Details random manifold}. For
the leading term we find, from a saddle-point evaluation
\cite{LeDoussalWiese2003b}:
\begin{eqnarray}
 \tilde{U}^0(v v) &=& U(\chi_v) 
+ \frac{1}{2} \sum_{n=1}^{\infty} \frac{n}{n+1} I_{n+1} \tr 
 \left[ -2 T \partial_\chi U(\chi_v)\right]^{n+1}  \\
I_{n} &:=& \int_k \frac{1}{(k^2 + m^2)^n}  \label{defIn}\ .
\end{eqnarray}
The trace acts on replica matrices and $\chi_v$ satisfies the
self-consistent equation
\begin{eqnarray}
 \chi_v^{ab} &=& v_{a} v_b + T \int_k G_v^{ab}(k) 
= v_{a} v_b + T I_1 \delta_{ab} + T \sum_{n=1}^{\infty} I_{n+1}
(\left[-2 T \partial_\chi U(\chi_v) \right]^{n})_{ab}  \\ 
 G_v(k) &=& \left[ (k^2 + m^2) \delta + 2 T \partial_\chi U(\chi_v)
\right]^{-1}\ . \label{chiv}
\end{eqnarray}
Note that for $d<2$ no UV cutoff is necessary (apart for a constant
term in the free energy), while for $2<d<4$ an UV cutoff is necessary
(and implicit in the following) only for $I_1$ \footnote{To obtain a
correct continuum limit, $T$ therefore should be scaled as $T =
\tilde{T}/\Lambda^{d-2}$, when $\Lambda$ is taken to infinity.}.

One also finds a compact and very useful self-consistent equation for
the derivative of the zero-th order potential:
\begin{eqnarray}\label{lf15}
 \partial_{ab} \tilde{U}^0(v v) & = & \p_{ab} U(\chi_v) \nonumber \\
& = & \p_{ab} U \left( v v + T I_1 \delta + T \sum_{n=1}^{\infty}
I_{n+1}  
(-2 T \partial \tilde{U}^0(v v))^{n}  \right)\ .
\end{eqnarray}
Everywhere we denote by $\partial_{ab} U(\phi) := \partial_{\phi_{ab}}
U(\phi)$ the simple derivative of the function $U(\phi)$ with respect
to its matrix argument $\phi_{ab}$. (Note that $\partial_{ab} \tilde
U(v v)$ is a {\it first derivative} of $\tilde U(v v)$ with respect to
the matrix element $v_a \cdot v_b$.)

Next, from calculations of the fluctuations around the saddle point,
one obtains the $1/N$ correction, which can be expressed in terms of
the zero-th order quantities as:
\begin{eqnarray}\label{lf25}
 \tilde{U}^1(v v) &=& \frac{1}{2} \int_q \tr \left(\ln\left[ \delta_{ac}
\delta_{bd} + 2 T \partial_{\chi_{ab}} \partial_{\chi_{ef}} U(\chi_v)
(T \Pi_v^{ef,cd}(q) + v_e G_v^{fc}(q) v_d + v_f G_v^{e d}(q) v_c )
\right] \right) \qquad \\ \label{lf26} \Pi_v^{ef,cd}(q) &=& \int_k G_v^{ed}(k)
G_v^{f c}(q-k) \\ 
G_v(k) &=& [ (k^2 + m^2) \delta + 2 T \partial_{v v}
\tilde{U}^0(v v) ]^{-1}\ ,
\end{eqnarray}
where here the trace acts in the space of replica pairs, i.e.\ $\tr
( M)=\sum_{ab} M_{ab,ab}$. Note that $\tilde{U}^0(v v)$ can also be
replaced by the full $\tilde{U}(v v)$ in the expression of
$\tilde{U}^1(v v)$ with the same accuracy (i.e.\ at leading order in
$1/N$).

The saddle-point equation (\ref{lf15}) for the zero-th order and the
result for the $1/N$-correction (\ref{lf25}) are still formal as they
encode the full renormalized disorder distribution. To yield the
renormalized disorder cumulants via
(\ref{cumulants1}),(\ref{cumulants2}) they must be expanded in the
number of replica sums, i.e.\ in cumulants.  In the following Section
\ref{s:review}, we recall the results for $N=\infty$, and proceed with
the non-trivial evaluation of (\ref{lf25}) via a graphical method in
Section \ref{s:graphmethod}, and via an algebraic method in Section
\ref{s:algebramethod}.

%%%%%%%%%%%%%%%%%%%%%%%%%%%%%%%%%%%%%%%%%%%%%%%%%%%%%%%%%%%%%%%%%%%%%%
%%%%%%%%%%%%%%%%%%%%%%%%%%%%%%%%%%%%%%%%%%%%%%%%%%%%%%%%%%%%%%%%%%%%%%
%%%%%%%%%%%%%%%%%%%%%%%%%%%%%%%%%%%%%%%%%%%%%%%%%%%%%%%%%%%%%%%%%%%%%%
%%%%%%%%%%%%%%%%%%%%%%%%%%%%%%%%%%%%%%%%%%%%%%%%%%%%%%%%%%%%%%%%%%%%%%
%%%%%%%%%%%%%%%%%%%%%%%%%%%%%%%%%%%%%%%%%%%%%%%%%%%%%%%%%%%%%%%%%%%%%%
%%%%%%%%%%%%%%%%%%%%%%%%%%%%%%%%%%%%%%%%%%%%%%%%%%%%%%%%%%%%%%%%%%%%%%
%%%%%%%%%%%%%%%%%%%%%%%%%%%%%%%%%%%%%%%%%%%%%%%%%%%%%%%%%%%%%%%%%%%%%%
%%%%%%%%%%%%%%%%%%%%%%%%%%%%%%%%%%%%%%%%%%%%%%%%%%%%%%%%%%%%%%%%%%%%%%
%%%%%%%%%%%%%%%%%%%%%%%%%%%%%%%%%%%%%%%%%%%%%%%%%%%%%%%%%%%%%%%%%%%%%%
%%%%%%%%%%%%%%%%%%%%%%%%%%%%%%%%%%%%%%%%%%%%%%%%%%%%%%%%%%%%%%%%%%%%%%
%                          Review of N=oo                            %
%%%%%%%%%%%%%%%%%%%%%%%%%%%%%%%%%%%%%%%%%%%%%%%%%%%%%%%%%%%%%%%%%%%%%%
%%%%%%%%%%%%%%%%%%%%%%%%%%%%%%%%%%%%%%%%%%%%%%%%%%%%%%%%%%%%%%%%%%%%%%
%%%%%%%%%%%%%%%%%%%%%%%%%%%%%%%%%%%%%%%%%%%%%%%%%%%%%%%%%%%%%%%%%%%%%%
%%%%%%%%%%%%%%%%%%%%%%%%%%%%%%%%%%%%%%%%%%%%%%%%%%%%%%%%%%%%%%%%%%%%%%
%%%%%%%%%%%%%%%%%%%%%%%%%%%%%%%%%%%%%%%%%%%%%%%%%%%%%%%%%%%%%%%%%%%%%%
%%%%%%%%%%%%%%%%%%%%%%%%%%%%%%%%%%%%%%%%%%%%%%%%%%%%%%%%%%%%%%%%%%%%%%
%%%%%%%%%%%%%%%%%%%%%%%%%%%%%%%%%%%%%%%%%%%%%%%%%%%%%%%%%%%%%%%%%%%%%%
%%%%%%%%%%%%%%%%%%%%%%%%%%%%%%%%%%%%%%%%%%%%%%%%%%%%%%%%%%%%%%%%%%%%%%
%%%%%%%%%%%%%%%%%%%%%%%%%%%%%%%%%%%%%%%%%%%%%%%%%%%%%%%%%%%%%%%%%%%%%%
%%%%%%%%%%%%%%%%%%%%%%%%%%%%%%%%%%%%%%%%%%%%%%%%%%%%%%%%%%%%%%%%%%%%%%

\section{Review of the results for $N=\infty$}
\label{s:review}
\comment{**** FINAL POUR PIERRE + KAY **********} 

In this section we review the main results at $N=\infty$. Details can
be found in \cite{LeDoussalWiese2003a}.

\subsection{Self-consistent equation at $N=\infty$}\label{spNoo}
\begin{figure}[t] \centerline{ \fig{12cm}{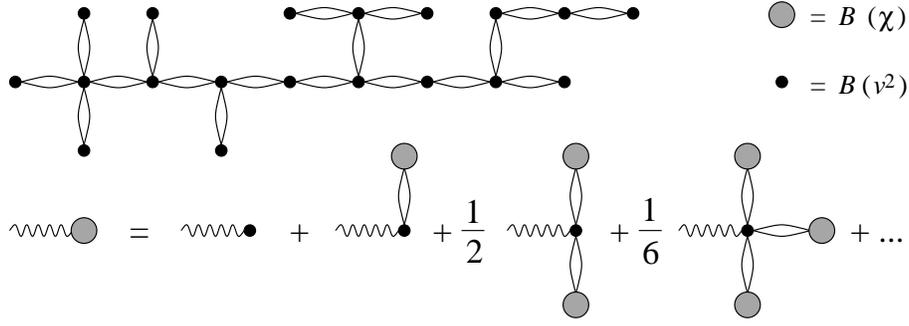} }
\caption{Top: typical $T=0$ contribution to $\tilde B
(v_{ab})$. Bottom: self-consistent equation at leading order for
$\tilde B' (v_{ab}^{2})=B' (\chi_{ab} )$. The wiggly line denotes a
derivative, and is combinatorially equivalent to choosing one $B$. At
finite $T$ one can attach an additional arbitrary number of tadpoles
to any $B$.  Also note that no loop made out of 3 propagators appears:
this would be a contribution to the third cumulant (3-replica term),
not calculated here; it is given in
\protect\cite{LeDoussalWiese2003b}.} \label{fig1}
\end{figure}

We start by recalling the cumulant expansion and only derive the
result for the second cumulant. Higher cumulants are given in
\cite{LeDoussalWiese2003a}. One studies a bare model, where only the
second cumulant is non-zero:
\begin{equation}\label{lf17b}
U(\chi) = - \frac{1}{2 T^2} \sum_{ab}
B(\tilde{\chi}_{ab}) 
\end{equation}
and calculates the renormalized disorder (\ref{cumulants1}).
%\begin{equation}\label{lf17}
% \tilde{U}^0(\chi) = - \frac{1}{2 T^2} \sum_{ab}
%\tilde{B}(\tilde{\chi}_{ab}) - \frac{1}{6 T^3} \sum_{abc}
%\tilde{S}(\tilde{\chi}_{ab},\tilde{\chi}_{bc},\tilde{\chi}_{ac}) +
%\dotsb
%\end{equation}
(We will drop the index zero on the cumulant functions, indicating the
leading order). The self-consistent equation (\ref{lf15}) can be
expanded in sums with increasing numbers of replicas. We only need:
\begin{eqnarray}\label{lf27}
\left[ - 2 T \partial \tilde{U}^0(vv)\right]_{ab} &=&
\frac{2}{T} (\delta_{ab} \sum_c \tilde{B}'_{ac} - \tilde{B}'_{ab})
+\dotsb \\ 
 \left[(- 2 T \partial \tilde{U}^0(vv))^2\right]_{ab} &=&
\frac{4}{T^2} \left( \delta_{ab} \sum_{ef} \tilde{B}'_{ae} \tilde{B}'_{af}
- \tilde{B}'_{ab} \sum_f (\tilde{B}'_{af} + \tilde{B}'_{bf}) + \sum_c
\tilde{B}'_{ac} \tilde{B}'_{cb} \right) +\dotsb \ , \label{lf28} 
\end{eqnarray}
where $\tilde{B}'_{ab}=\tilde B'(v_{ab}^2)$ (recall
$v_{ab}^2:=(v_a-v_b)^2$). Here and below the dropped terms contain
sums with too many replicas to contribute to the final result for the
self-consistent equation of the second cumulant (2-replica term). One
thus has:
\begin{eqnarray} \label{lf29}
 \tilde{\chi}_v^{ab} &=& v_{ab}^2 + 2 T I_1 (1-\delta_{ab}) + 4 I_2
\Big[ \frac{1}{2} \sum_c (\tilde{B}'_{ac} + \tilde{B}'_{bc} ) -
\frac{1}{2} (\tilde{B}'_{aa} + \tilde{B}'_{bb}) + \tilde{B}'_{ab} -
\delta_{ab} \sum_c \tilde{B}'_{ac} \Big] +\dotsb\ . \qquad 
\end{eqnarray}
The self-consistent equation becomes:
\begin{equation}\label{lf30}
 \tilde{B}'(v_{ab}^2) - \delta_{ab} \sum_c \tilde{B}'(v_{ac}^2) +
\mbox{3-replica terms} = B'(\tilde\chi_v^{ab}) - \delta_{ab} \sum_c
B'(\tilde\chi_v^{ac})
\end{equation}
and can be solved by appropriate Taylor expansion of the r.h.s.. It is
solved for $a \neq b$:
\begin{equation}\label{lf31}
 \tilde{B}'(v_{ab}^2) + \frac{1}{T} \sum_g \tilde{S}'_{abg} +
\sum_{gh}\dotsb  + \dotsb  = B'(\tilde\chi_v^{ab})\ .
\end{equation}
It is then easy to see that the second cumulant satisfies a closed
equation at any $T$, 
\begin{equation}\label{saddlepointforB}
  \tilde{B}'(v_{ab}^2) = B'\left(v_{ab}^2 + 2 T I_1 + 4 I_2
(\tilde{B}'(v_{ab}^2) - \tilde{B}'(0))\right)\ ,
\end{equation}
with no other contributions from higher cumulants at any $T$. A more
detailed derivation is given in \cite{LeDoussalWiese2003a}.

\subsection{Derivation of the FRG equation at $N=\infty $} \noindent
From the previous section the renormalized second cumulant of the
disorder $\tilde{B}'(x)$ satisfies the self-consistent equation
\begin{equation}\label{lf72}
\tilde{B}'(x) = B'\left(x + 2 T I_1 + 4 I_2 (\tilde{B}'(x) -
\tilde{B}'(0))\right)  \ .
\end{equation}
It implies
\begin{equation}\label{lf73}
 \tilde{B}'(0) = B'(2 T I_1) \ ,
\end{equation}
as well as
\begin{equation}\label{lf74}
 \tilde{B}''(x) = B''(x + 2 T I_1 + 4 I_2 (\tilde{B}'(x) -
\tilde{B}'(0))) [ 1 + 4 I_2 \tilde{B}''(x) ]\ .
\end{equation}
We now derive the corresponding exact FRG equation. 
Taking the derivative $m \partial_m$ gives:
\begin{eqnarray}
 m \partial_m \tilde{B}'(x) &=& B''(x + 2 T I_1 + 4 I_2 (\tilde{B}'(x)
- \tilde{B}'(0))) \nonumber \\ 
&&\times \left[2 T m \partial_m I_1 + 4 (m \partial_m I_2)
(\tilde{B}'(x) - \tilde{B}'(0)) + 4 I_2 m \partial_m \tilde{B}'(x) - 4
I_2 m \partial_m \tilde{B}'(0)
\right] \nonumber \\ 
&=& \frac{\tilde{B}''(x)}{1 + 4 I_2 \tilde{B}''(x)} \nonumber \\
&&\times [2 m \partial_m T I_1 + 4 (m \partial_m I_2) (\tilde{B}'(x) -
\tilde{B}'(0)) + 4 I_2 m \partial_m \tilde{B}'(x) - 4 I_2 m \partial_m
\tilde{B}'(0) ]\ .\qquad 
\end{eqnarray}
This yields:
\begin{equation}
 m \partial_m \tilde{B}'(x) = \tilde{B}''(x) [ 2 m \partial_m T I_1 +
4 (m \partial_m I_2) (\tilde{B}'(x) - \tilde{B}'(0)) - 4 I_2 m
\partial_m \tilde{B}'(0) ]\ .
\end{equation}
Thus one has also:
\begin{equation}
 m \partial_m \tilde{B}'(0) =  \frac{\tilde{B}''(0)}{1 + 4 I_2 \tilde{B}''(0)}
2 m \partial_m (T I_1) \ .
\end{equation}
Hence one gets finally
\begin{equation} \label{beta finite T N=oo}
\mbox{$
\ds m\frac{\p }{\p m} \tilde B'(x) = \tilde B''(x)
\left[ 2 (m\frac{\p }{\p m} T I_1) \frac1{1+ 4 I_2 \tilde B''(0)}
+ 4 (m\frac{\p }{\p m} I_2)(\tilde B'(x)-\tilde B'(0))
\right]$}\ ,
\end{equation}
which can also be integrated once over $x$. As emphasized in
Ref. \cite{LeDoussalWiese2003a} it is exact at $N=\infty$ for any $d$
and correctly matches the 1-loop FRG equation obtained by Balents and
Fisher for any $N$ but only to $O(\epsilon)$, $\epsilon=4-d$. It can
be solved directly, or equivalently the self-consistent equation
(\ref{lf72}) can be inverted.  The corresponding solutions for various
models are discussed in \cite{LeDoussalWiese2003a} and compared with
the M\'ezard-Parisi solution \cite{MezardParisi1991} obtained in a
rather different manner through a replica-symmetry-breaking saddle
point.

Before discussing specific models, we now turn to the evaluation of
the effective action of the FRG and the $\beta$-function to the next
order in $1/N$.

%%%%%%%%%%%%%%%%%%%%%%%%%%%%%%%%%%%%%%%%%%%%%%%%%%%%%%%%%%%%%%%%%%%%%%
%%%%%%%%%%%%%%%%%%%%%%%%%%%%%%%%%%%%%%%%%%%%%%%%%%%%%%%%%%%%%%%%%%%%%%
%%%%%%%%%%%%%%%%%%%%%%%%%%%%%%%%%%%%%%%%%%%%%%%%%%%%%%%%%%%%%%%%%%%%%%
%%%%%%%%%%%%%%%%%%%%%%%%%%%%%%%%%%%%%%%%%%%%%%%%%%%%%%%%%%%%%%%%%%%%%%
%%%%%%%%%%%%%%%%%%%%%%%%%%%%%%%%%%%%%%%%%%%%%%%%%%%%%%%%%%%%%%%%%%%%%%
%%%%%%%%%%%%%%%%%%%%%%%%%%%%%%%%%%%%%%%%%%%%%%%%%%%%%%%%%%%%%%%%%%%%%%
%%%%%%%%%%%%%%%%%%%%%%%%%%%%%%%%%%%%%%%%%%%%%%%%%%%%%%%%%%%%%%%%%%%%%%
%%%%%%%%%%%%%%%%%%%%%%%%%%%%%%%%%%%%%%%%%%%%%%%%%%%%%%%%%%%%%%%%%%%%%%
%%%%%%%%%%%%%%%%%%%%%%%%%%%%%%%%%%%%%%%%%%%%%%%%%%%%%%%%%%%%%%%%%%%%%%
%%%%%%%%%%%%%%%%%%%%%%%%%%%%%%%%%%%%%%%%%%%%%%%%%%%%%%%%%%%%%%%%%%%%%%
%           Kay's section about the graphical method                 %
%%%%%%%%%%%%%%%%%%%%%%%%%%%%%%%%%%%%%%%%%%%%%%%%%%%%%%%%%%%%%%%%%%%%%%
%%%%%%%%%%%%%%%%%%%%%%%%%%%%%%%%%%%%%%%%%%%%%%%%%%%%%%%%%%%%%%%%%%%%%%
%%%%%%%%%%%%%%%%%%%%%%%%%%%%%%%%%%%%%%%%%%%%%%%%%%%%%%%%%%%%%%%%%%%%%%
%%%%%%%%%%%%%%%%%%%%%%%%%%%%%%%%%%%%%%%%%%%%%%%%%%%%%%%%%%%%%%%%%%%%%%
%%%%%%%%%%%%%%%%%%%%%%%%%%%%%%%%%%%%%%%%%%%%%%%%%%%%%%%%%%%%%%%%%%%%%%
%%%%%%%%%%%%%%%%%%%%%%%%%%%%%%%%%%%%%%%%%%%%%%%%%%%%%%%%%%%%%%%%%%%%%%
%%%%%%%%%%%%%%%%%%%%%%%%%%%%%%%%%%%%%%%%%%%%%%%%%%%%%%%%%%%%%%%%%%%%%%
%%%%%%%%%%%%%%%%%%%%%%%%%%%%%%%%%%%%%%%%%%%%%%%%%%%%%%%%%%%%%%%%%%%%%%
%%%%%%%%%%%%%%%%%%%%%%%%%%%%%%%%%%%%%%%%%%%%%%%%%%%%%%%%%%%%%%%%%%%%%%
%%%%%%%%%%%%%%%%%%%%%%%%%%%%%%%%%%%%%%%%%%%%%%%%%%%%%%%%%%%%%%%%%%%%%%

\section{Corrections in $1/N$, via the graphical method}
\label{s:graphmethod} 
In this section, we present a graphical method to calculate the
corrections to $\tilde B$ at order $1/N$. An algebraic method is
presented in the next section \ref{s:algebramethod}. Both methods are
completely independent, since they use orthogonal ideas. They were
performed independently by the authors, each on a different
continent. The agreement on the final result gives some confidence
that it is free from calculational errors.

\subsection{General considerations for a scalar field
theory}\label{gcgm}
\begin{figure}
\centerline{\fig{9cm}{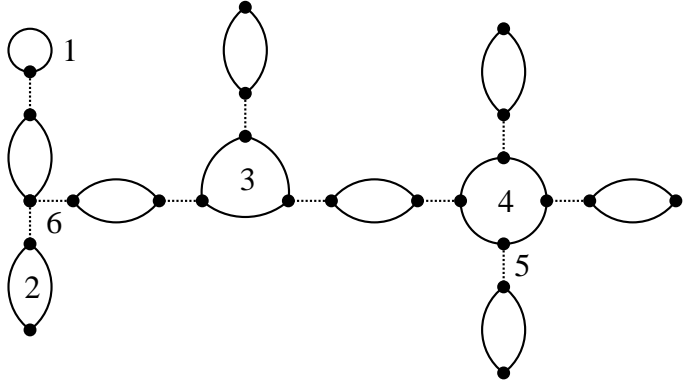}} \caption{A typical ``tree-like''
diagram contributing to the leading order in the $1/N$-expansion of
the renormalized potential $\tilde V$ (effective action). The numbers
1, 2 and 3 depict special features discussed in the main text.  There
is a factor $1/N$ per dotted line and a factor of $N$ per ``small
loop'' of propagators (solid lines), thus the overall factor is $N$ as
it should from (\ref{phiint}). This graph does not contain any ``big
loop'' (see text).} \label{largeNgen}
\end{figure}
Let us start with some general considerations about which graphs
contribute at a given order in $1/N$. For that purpose, we consider a
general (pure, no disorder) scalar field theory, with a $N$-component
field $u^i(x)$, $i=1\dotsb N$  and interaction
\begin{equation}\label{phiint}
{{\cal S}}_{\mathrm{int}} = N \int_{x} V \left({\textstyle\frac{ \vec
{u}^{2}}{N}} \right) \ ,
\end{equation}
where the ${u} {u} $-correlations of the free theory ($V=0$) are given
by
\begin{equation}\label{lf32}
\left< {u}^{i}(x){u}^{j} (y) \right> = \delta^{ij}\,C (x-y)\ ,
\end{equation}
with $i,j=1\dots N$ and $C (x-y)$ independent of $N$ is denoted by a
solid line in figure \ref{largeNgen}.  Graphically, we can denote this
by
\begin{equation}\label{lf33}
S_{\mathrm{int}} = b_{2} \diagram{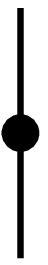} +
\frac{b_{4}}{N}\diagram{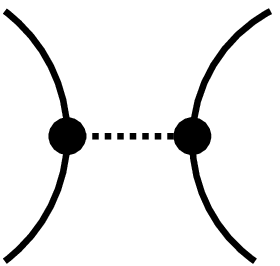} + \frac{b_{6}}{N^{2}} \diagram{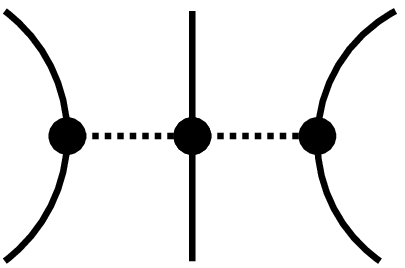} +
\dots \ ,
\end{equation}
where all coefficients $b_i$ are of order $1$; there is thus a factor
of $1/N$ per dotted line.

The renormalized  potential $\tilde V$ (effective action)  is given by
the sum of all 1-particle-irreducible diagrams. They must thus contain
``small loops''. To leading order  in  $1/N$ the following diagrams
are possible:
\begin{itemize}
\item Tadpoles, contracting any  $\vec {u}^{2}$ with itself
only. The factor of $N$ from the ``small loop'' (the $\sum_{i=1}^{N}$ over the
number of components) compensates the factor of $N$ from the argument
of $V (\vec {u}^{2}/N)$. (See 1 on figure \ref{largeNgen}.)
\item Closed ``small loops'' with 2, 3, 4 or more vertices, as denoted by the
same number on figure \ref{largeNgen}: Adding one more vertex gives a
factor of $N$, see Eq.~(\ref{phiint}); this is compensated by using an
additional $\vec {u}^{2}/N$.
\item Note that all vertices $\vec {u}^{2}$, $(\vec {u}^{2})^{2}$,
a.s.o.~contribute equivalently. (See 5, 6  on figure \ref{largeNgen}.)
\end{itemize}

Note that all these diagrams are ``tree-like'' diagrams, where the
branches (made of ``small loops'') are made out of the diagrams
through which no total momentum is running. We will call 
them ``tree-like'' in the following, to distinguish
them from normal trees. They are resummed by the saddle-point
equations (see appendix \ref{toy}) or from graphical inspection, as
done here: First of all, insertions of $V'$ into a line of propagators
act like a mass, leading to the replacement of $1/ ({k^{2}+m^{2}})$ by
\begin{equation}\label{lf34}
G (k):= \frac{1}{k^{2}+m^{2}+2\tilde V' {\left(\frac{\vec {u}^{2}}{N}
\right)}}\ .
\end{equation}
The effective potential $\tilde V$ is obtained from
\begin{eqnarray}\label{sps}
\tilde V' \left({\textstyle\frac{ \vec
{u}^{2}}{N}}\right) &=& V'\left(\chi  \right)\\
\chi &=& \frac{ \vec {u}^{2}}{N}+ \int_{k} G (k) =
\frac{{u}^{2}}{N}+ \int_{k} \frac{1}{k^{2}+m^{2}+2\tilde V'
{\textstyle \left(\frac{\vec {u}^{2}}{N} \right)}} \ .
\end{eqnarray}
Note that the derivatives are graphically understood as follows:
Choosing one vertex (derivative!) in the effective potential $\tilde
V$ is equivalent to having a bare vertex with the same derivative
taken (thus $V'$) and attaching to it loops made out of
correlation-functions. Attaching any number of such loops to $V$,
amounts to shifting its argument, as can be seen from
Taylor-expansion. In these loops, again derivatives (one needs
${u}^{2}$ to attach the loop) of the effective potential are
inserted. The latter can thus be written as $V'$ with shifted argument
to account for more things to be attached to this $V'$ or equivalently
using (\ref{sps}) to a $\tilde V'$.  This result coincides with
(\ref{B24}).\medskip

\begin{figure}
\centerline{\fig{13cm}{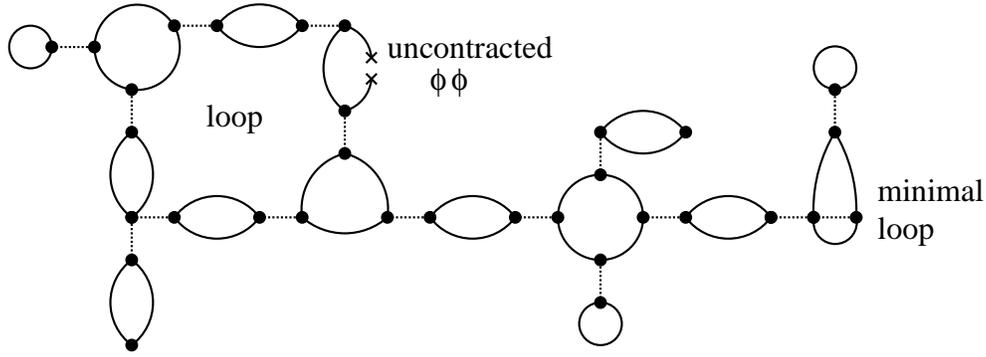}} \caption{Typical graphs which
contribute to the renormalized potential $\tilde V$ to subleading
order in $1/N$. The ``big loop'' (a loop made of loops, see main text)
accounts for a factor of $1/N$; the same is true for the ``minimal
loop'' on the right, which is a ``big loop'' in disguise (as explained
in the main text.) The given diagram is thus of order $1/N^{2}$}
\label{fig1oNphi}
\end{figure}

Diagrams at next order $1/N$ contain exactly one ``big loop'', see
figure \ref{fig1oNphi}.
Take a ``tree'' (as e.g.\ the object on Fig.\ \ref{largeNgen}) and
glue it together to form a ``big loop'' by identifying two vertices; this does
not change the factor of $N$ from the loops, but one looses one factor
of $N$ from the missing vertex. Note that also the ``minimal loop''
marked on figure \ref{fig1oNphi} belongs to the same class, even though
it looks different.  The ``big loop'' demands to sum a series of diagrams at
non-vanishing momentum, and then to carry through the integration over
momenta. However the simplification remains that any added
``tree-like'' branches are resummed by replacing the argument of $V$
from $\vec {u}^{2}/N$ to $\chi$. 

Another feature arises at subdominant order: In the ``big'' loop, one
may pick any given small loop (i.e.\ the loop made out of
correlation-functions) and replace one of the two
correlation-functions by $\vec {u}^{2}/N$: This means that the
corresponding fields ${u}$ did not get contracted. Since the remaining
correlation-functions force their indices to be equal, this gives a
factor of $\vec {u}^{2} = N \times (\vec {u}^{2}/N )$, thus
contributes the same factor of $N$. Note that these diagrams do not
contribute to the effective action at leading order, which is
treelike, since the resulting diagrams would be 1-particle
reducible. 

To resum the order $1/N$-diagrams one has to sum over loops of all
sizes. The result is 
\begin{eqnarray}\label{lf35}
\delta \tilde V \left({\textstyle \frac{\vec {u}^{2}}{N}}\right) &=&
-\frac{1}{2N} \sum_{n=1}^{\infty}\frac{1}{n} \int_{p} \left[- 2 V''
(\chi) \left( { I}_{2} (p) + 2\, {\textstyle \frac{\vec {u}^{2}}{N}\,
G (p) } \right)
\right]^{n}  \nonumber \\
&=& \frac{1}{2N} \int_{p}\ln \left[1+2 V'' (\chi)
\left( {I}_{2} (p) + 2\, {\textstyle \frac{\vec {u}^{2}}{N}\, G (p) }
\right)  \right]\\ 
{I}_{2} (p) &=& \int_{k} G (k+p) G (k) \ .
\end{eqnarray}
This can be compared to the results of appendix \ref{toy}, and more
specifically to formula (\ref{B.45}).

\subsection{Elastic manifolds in disorder: General considerations,
building blocks} \label{buildingblocks} Let us start the treatment of
the disordered model with some general considerations.  First to
organize the $1/N$-expansion, one may still use the diagrammatics of
the previous section, which shows the $O(N)$-index content. The same
diagrams still exist, but they now also have a complicated replica
content. The replica content can be explicated by using ``splitted
vertices'' instead of the unsplitted ones of the previous section.
The corresponding replica diagrammatics, which shows the replica
structure {\em only} was explained in details in
\cite{LeDoussalWieseChauve2003}.  This can be drawn as
\begin{equation}\label{drawing-split}
\sum_{ab} B \left(( v_{a}-v_{b})^{2}\right) =\ {}_{a}
\diagram{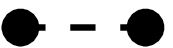}_{b}\ \ ,
\end{equation}
where a {\em dashed} line connects the two dots, standing for replicas $a$
and $b$. In order to avoid confusion, note that this dashed line is
different from the {\em dotted}  line used in figures \ref{largeNgen} and
\ref{fig1oNphi} as well as equation (\ref{lf33}) to show the $O
(N)$-structure.

Below we  introduce a third diagrammatics
which allows to track {\em both} the
replica and the $O(N)$ indices, not an easy task.
Before doing so let us explain a few points.

Since we are only interested in the corrections to the {\it 2-replica
part} $\tilde B_{ab}$ of the effective action, there are many $O(N)$
diagrams which do not contribute.

At dominant order (figure \ref{fig1}) small loops with three vertices
(see 3 in
figure \ref{largeNgen}), or more, do not contribute to $\tilde B_{ab}$, but to
the third cumulant, or higher, as can be seen from 
\begin{equation}\label{lf36}
\diagram{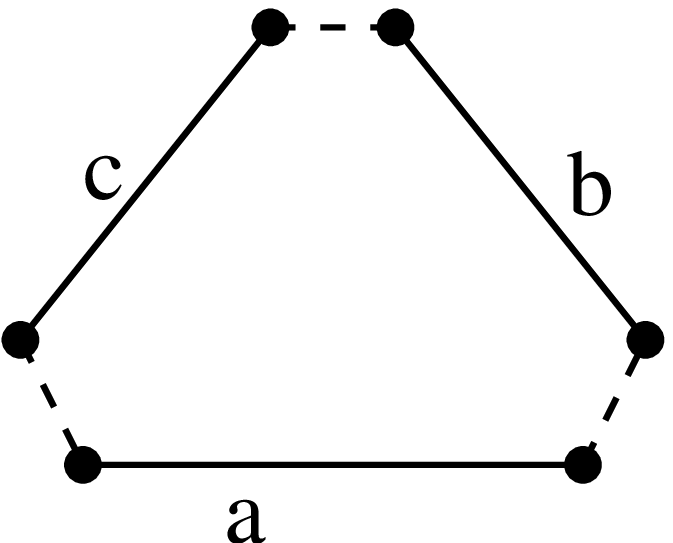}\ , 
\end{equation}
which is a diagram showing only replica indices,
$N$ indices being implicit. Note that solid lines identify 
replica indices. This is why in figure \ref{fig1}
only chain-diagrams and tadpoles (the latter are omitted for
simplicity of presentation) appear, rendering calculations appreciably
simpler.

\begin{figure}
\centerline{\pfig{.35\textwidth}{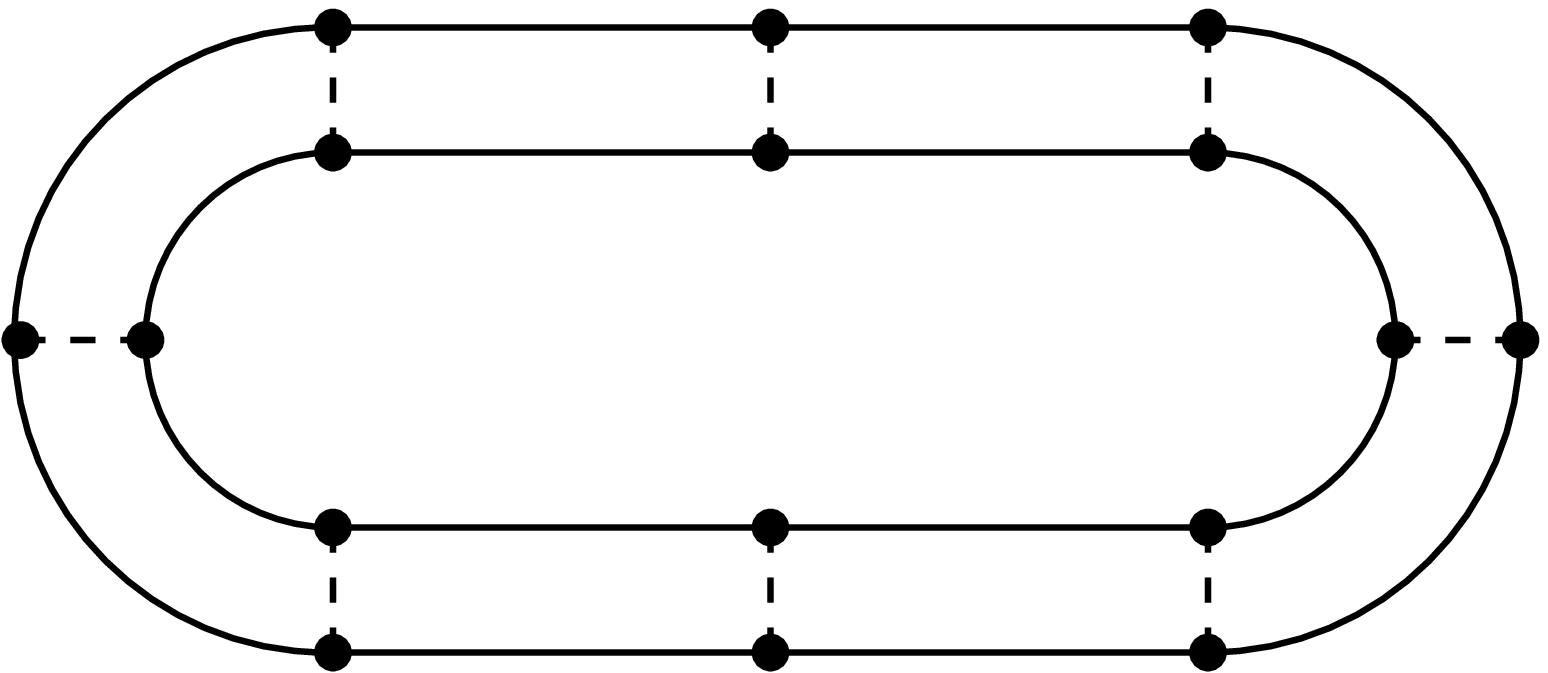} \qquad
$\longrightarrow$\qquad \pfig{.35\textwidth}{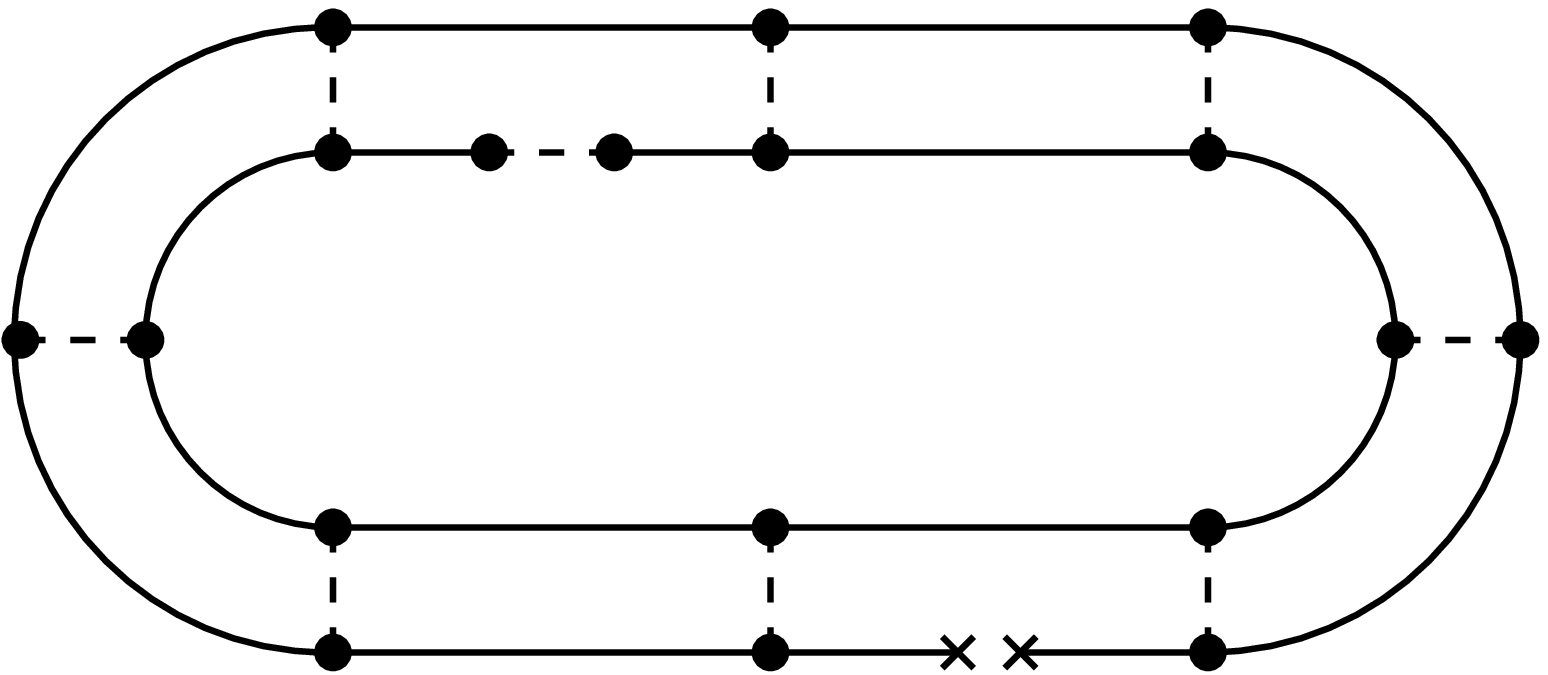}}
\caption{Example of a big loop made out of small loops, in splitted
replica notation (left). It contributes to the 2-replica part (i.e.\
the disorder) at order $1/N$. Note that there are two constraints more
than needed to have a 2-replica term. Thus two redundant constraints
can be ``wasted'', by cutting each of the solid lines exactly once, either by
inserting a $B'$ or leaving a $u u$ uncontracted, as is done on the
right.  } \label{f:bigloop}
\end{figure}
At subdominant order in $1/N$, only diagrams with one ``big'' loop
(general feature of the order $1/N$ discussed in the last section)
made of any number of ``small'' 2-loops and only exactly {\it two}
``small'' 3-loops can contribute to the renormalized second
cumulant. Each small 3-loop can also be replaced by an uncontracted $u
u $.  To understand this, consider the simplest ``big'' loop, i.e.\
the railroad diagram, which is drawn on the left of figure
\ref{f:bigloop} in splitted replica notation.  It contains exactly two
closed propagator lines, which over-constrain the replicas to be
equal. (These are the inner and outer solid lines on the left of
figure \ref{f:bigloop}). These over-constraints can be relaxed, by
cutting each line exactly once, in order not to get a higher replica
term, as is illustrated on the right of figure \ref{f:bigloop}. This
``cutting'' is possible by either inserting into a propagator a vertex
(which contains two replicas, thus is not ``replica-conserving'') or
by leaving one $u u$ uncontracted. This is the basic principle, whose
careful exploration leads to all of the diagrams at order $1/N$, as we
will discuss now.

In order to do so, we have to introduce a new powerful graphical
notation:
\begin{eqnarray}
\diagram{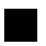} &=& B(\bar \chi)\\
\diagram{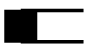}{_{a} \atop ^{b}} &=& 
 B'(\bar \chi_{ab}) \left(\vec{u}_{a}-\vec{u}_{b}   \right)^{2}\\
{_{a} \atop ^{b}}\diagram{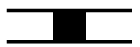}{_{a} \atop ^{b}}
 &=& B''(\bar \chi_{ab})\frac{1}{2}
\left[ \left(\vec{u}_{a}-\vec{u}_{b}   \right)^{2}\right]^2\ ,
\end{eqnarray}
where lines departing in the same direction belong to the same
vector-index, and a continuing line represents the same  replica. 

We also use the following short-hand notation
\begin{equation}\label{B''shorthand}
B'_{ab}:=B' (\bar \chi_{ab})
\end{equation}
and similar formulas for the higher derivatives $B''_{ab}:=B'' (\bar
\chi_{ab})$, etc.
Another frequently used shorthand is
\begin{equation}\label{B'0shorthand}
B'_{0}:=B'_{aa}\ , \qquad B''_{0}:=B''_{aa}\ , \qquad \mbox{etc.}
\ .
\end{equation}
In order to be able to resum the ``big'' loop, we will now introduce some
building blocks.  Since there can be any number of 2-loops in the ``big''
loop one needs to define the resummed chain
\begin{equation}\label{lf37}
{ _a \atop ^b}\diagram{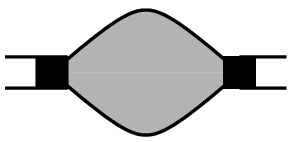}{ _a \atop ^b}\ : =
{ _a \atop ^b} \diagram{Bpp}{ _a \atop ^b} +
 { _a \atop ^b}\diagram{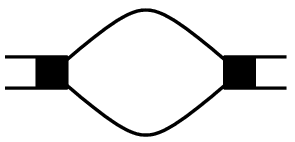}{ _a \atop ^b}+
 { _a \atop ^b}\diagram{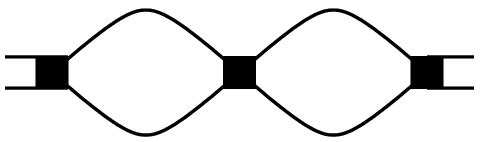}{ _a \atop ^b}+
 { _a \atop ^b}\diagram{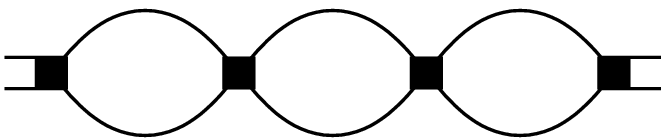}{ _a \atop ^b}+\dots 
\ .
\end{equation}
We have with momentum $p$ running through the diagram
\begin{equation}\label{lf38}
\stackrel{{p} }{\longrightarrow}
 { _a \atop ^b}\diagram{B1loop}{ _a \atop ^b} =
 4 I_{2}(p)\, B''(\bar \chi_{ab})\ 
{ _a \atop ^b} \diagram{Bpp}{ _a \atop ^b}
\end{equation}
with 
\begin{eqnarray}\label{I2p}
I_{2}(p) &=& \int_{k} \frac1{(k+p/2)^{2}+m^{2}} \frac1{(k-p/2)^{2}+m^{2}}\ .
\end{eqnarray}
Let us introduce a compact notation for the integrals, summarized in
appendix \ref{app:integrals}.  (All our notations and important
formulas are also summarized in a table in appendix
\ref{sec:Notation}.)

\begin{eqnarray}\label{Jik}
J_{\alpha \beta } (q) &\equiv& J^{q}_{\alpha \beta } :=
\int_{k}\frac{1}{((k+q/2)^{2}+m^{2})^{\alpha }}\frac{1}
{((k-q/2)^{2}+m^{2})^{\beta }}
\\
\label{I3}
I_{3}(p) &:=& J_{1,2} (p)= \int_{k} \frac1{\left(
(k+p/2)^{2}+m^{2}\right)^{2}} 
\frac1{(k-p/2)^{2}+m^{2}} \\\label{I4}
I_{4 } (p)&:=&  J_{2,2} (p)=\int_{k} \frac1{\left(
(k+p/2)^{2}+m^{2}\right)^{2}} \frac1{\left((k-p/2)^{2}+m^{2}
\right)^{2}}
\ .
\end{eqnarray}
Thus
\begin{eqnarray} \label{lf39}
\stackrel{{p} }{\longrightarrow} { _a \atop ^b}\diagram{B2loop}{ _a
\atop ^b} &=& \left[4 I_{2}(p)\, B''(\bar \chi_{ab}) \right]^{2} \ {
_a \atop ^b} \diagram{Bpp}{ _a \atop
^b} \\ 
 \stackrel{{p} }{\longrightarrow} { _a \atop ^b}\diagram{B3loop}{ _a
\atop ^b} &=& \left[4 I_{2}(p)\, B''(\bar \chi_{ab}) \right]^{3} \ {
_a \atop ^b} \diagram{Bpp}{ _a \atop ^b} \ ,
\end{eqnarray}
and so on. These chain-like diagrams form a geometric series, which is
resummed as
\begin{equation}\label{Bsummed}
{ _a \atop ^b}\diagram{Bsummed}{ _a \atop ^b} = \frac{1}{1-4
I_{2}(p)\, B''(\bar \chi_{ab})} \ { _a \atop ^b} \diagram{Bpp}{ _a
\atop ^b} =: \frac{1}{2} \left[(u_{a}-u_{b})^{2} \right] H_{ab}(p)
\left[(u_{a}-u_{b})^{2}\right]\ .
\end{equation}
We have introduced $H_{ab}(p)$, the ``effective'' $B''_{ab}$ after
resummation
\begin{equation} \label{Habp}
H_{ab}(p):=\frac{B''(\bar \chi_{ab})}{1-4 I_{2}(p)\, B''(\bar
\chi_{ab})} \ ,
\end{equation}
which we equivalently can express at leading order in $1/N$ through
$B''(\bar \chi_{ab}) = \tilde{B}_{ab}''/[1+4I_{2}\tilde{B}''_{ab} ]$
as
\begin{equation}\label{lf40}
H_{ab}(p):=  \frac{\tilde{B}''_{ab}}{1+4\left[I_{2}-I_{2}(p) \right]
\tilde{B}''_{ab}}\ .
\end{equation}
We also define $H_v (p)$ as
\begin{equation}\label{Hp}
H_{v}(p):=\frac{{B}'' (\bar \chi (v))}{1-4I_{2}(p) {B}'' (\bar \chi
(v))} =\frac{\tilde{B}'' (v^{2})}{1+4\left[I_{2}-I_{2}(p) \right]
\tilde {B}'' (v^{2})} \ ,
\end{equation}
valid again at leading order in $1/N$.  Note that the denominator in
(\ref{Hp}) reflects the renormalization of $\tilde B$: The divergent
integral $I_{2} (p)$ does not appear alone, but together with its
counter-term, the integral $I_{2} (p)$ subtracted at
$p=0$.\begin{figure}\renewcommand{\arraystretch}{2.2}
\begin{center}
\begin{tabular}{|c|c|}
\hline $\diagram{B}$ & $ B(\bar \chi_{ab})$\\\hline
$\displaystyle\diagram{Bp}{_{a} \atop ^{b}}$ & $ \displaystyle B'(\bar
\chi_{ab}) \left(\vec{u}_{a}-\vec{u}_{b} \right)^{2}$ \\\hline
$\displaystyle {_{a} \atop ^{b}}\diagram{Bpp}{_{a} \atop ^{b}} $ &
$\displaystyle B''(\bar \chi_{ab})\frac{1}{2} \left[
\left(\vec{u}_{a}-\vec{u}_{b} \right)^{2}\right]^2
\rule[-2.5ex]{0mm}{6ex} $ \\\hline
$_{a}\!-\!\!\!-\!\!\!-\!\!\!-\!\!\!-\,_{b} $ & $\displaystyle
\delta_{ab}C (p) =
\frac{\delta_{ab}}{p^{2}+m^{2}}\rule[-3ex]{0mm}{6ex}$ \\ \hline
$_{a}\!-\!\!\!-\!\!\!\!\times \times\!\!\!\!-\!\!\!-\,_{b} $ &
$v_{ab}^{2}$ \\ \hline $\displaystyle { _a \atop ^b}\diagram{Bsummed}{
_a \atop ^b}$ & $\displaystyle H_{ab} (p) =\frac{B''(\bar
\chi_{ab})}{1-4 I_{2}(p)\, B''(\bar \chi_{ab})}\rule[-3ex]{0mm}{6ex}$
\\\hline $_{a}\diagram{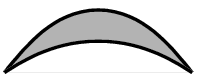}_{b}$ & $\delta_{ab} I_{2} (p) ( 1-4
I_{2}B''_{aa})$ \\\hline $_{a}\diagram{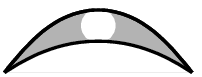}_{a}$ & $I_{2} (p)
(1+2 I_{2}B''_{aa}) ( 1-4 I_{2}B''_{aa})$ \\\hline
\end{tabular}
\end{center}
\caption{Building blocks of the perturbation theory. See main
text. The last two blocks only appear at finite temperature.}
\label{f:buildingblocks}
\end{figure}
It turns out that at zero temperature the above are the only building
blocks needed. However at $T>0$ one needs two more building blocks
which are quite non-trivial. As was shown in
\cite{LeDoussalWieseChauve2003} non-zero temperature diagrams contain
at least one replica ``sloop''; these are exactly the
over-constraining lines discussed above and on figure
\ref{f:bigloop}. There is one factor of $T$ for each such
``sloop''. The additional building blocks thus contain sloops: One
sloop at order $T$, and 2 sloops (an example is on the left of
figure \ref{f:bigloop}) at order $T^{2}$. Higher orders in $T$ are
only possible at order $1/N^{2}$, or higher.

To explain the construction of these additional building blocks, which
is subtle, one goes back to the diagrammatics showing only replica
indices.

The first building block is the ``moon-diagram''. This is the sum over
all diagrams, which at both ends have lines joining only one of the
both replicas, and which enforce the joined replicas to be equal:
\begin{eqnarray}\label{moondef}
_{a}\diagram{moon}_{b} &=& \diagram{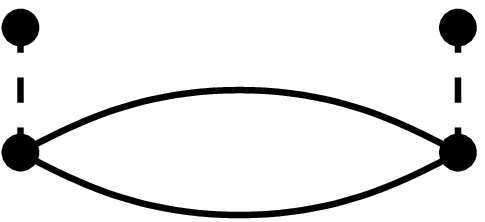}  + \diagram{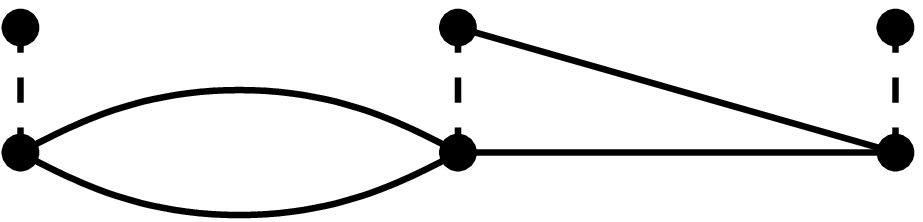} +
\diagram{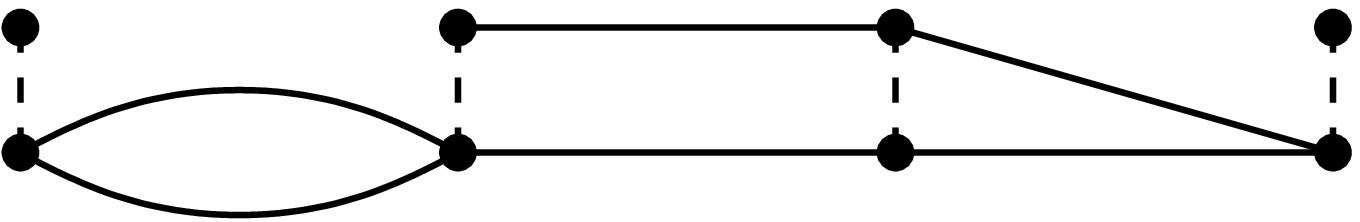}\nonumber \\
&&   + \diagram{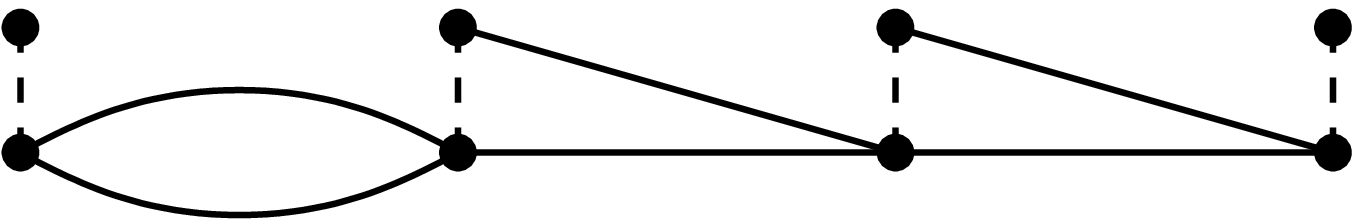} + \ \dotsb  
\ .
\end{eqnarray}
Note that we construct the chain from left to right. Otherwise the
graphical representation of the perturbation expansion is not
unambiguous, as can be seen from the following example
\begin{equation}
\diagram{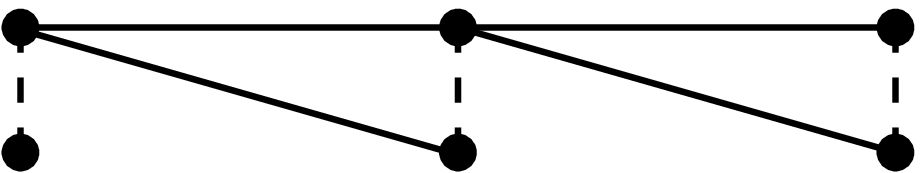}
\ .
\end{equation}
Drawing first the two left-most lines, the two right-most ones can no
longer be added, since in the middle, there is $R''(u_a-u_a)\equiv R''
(0)$, which does not depend on the field.  Conversely, {\em if we
decided} to first draw the two rightmost lines, then the two leftmost
could be drawn. Consequently the diagrams to be drawn in
(\ref{moondef}) would be different (actually they would be nothing but
the diagrams mirrored such that their left and right ends are
exchanged), even though the final result would be the same. This
phenomenon is detailed (for a different diagram) in appendix
\ref{2looptadpole}.  Note that there is {\em nor a contradiction, nor
an inconsistency of the approach.}  It merely means, that when using
these kind of rules, which have the advantage of simplifying
calculations importantly, one has to order the contractions. An
approach, which does not have this deficiency, but is very
complicated, is explained in appendix \ref{excludedrepformalism}. We
will use it there to recalculate diagram (\ref{moondef}).

We claim that
\begin{equation}\label{moontheorem}
_{a}\diagram{moon}_{b} = \diagram{moon1} + \diagram{moon2} =
\diagram{moon1} (1-4 I_{2} (p)B''_{aa}) \ . 
\end{equation}
The second identity is trivial perturbation theory. The non-trivial
statement is the first identity. To prove it, we remark that starting
with (recall we construct from left to right)
$\textdiagram{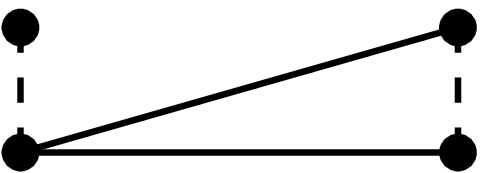} $
no further contraction can be made. 
Therefore, we have to start with
$\textdiagram{moon1}\ .$
At chain-length two, the only possibility is
\begin{equation}
\diagram{moon2}\ .
\end{equation}
At chain-length three, there are two and only two possible
prolongations, which have no additional free replica-indices:
\begin{equation}
\diagram{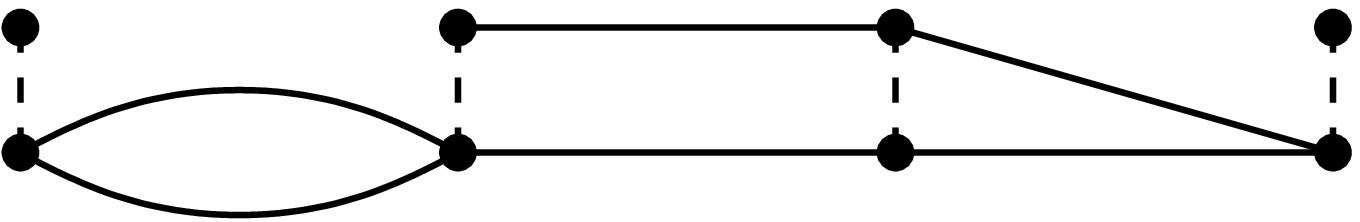} + \diagram{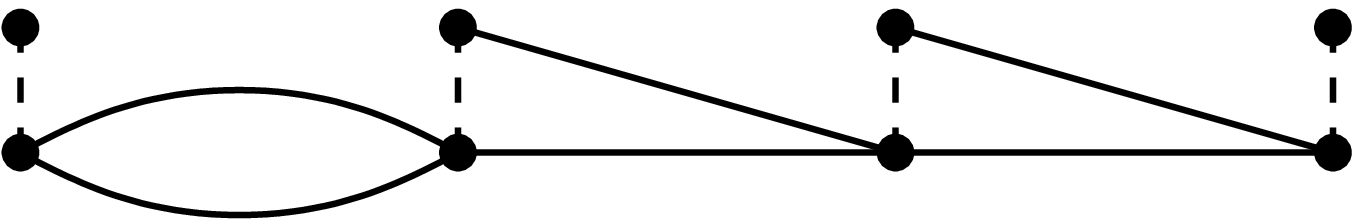} =0 \ .
\end{equation}
These diagrams cancel. The same is true for longer chains, since at
any intermediate position (i.e.\ not the first and not the last
lines), there is always the combination 
\begin{equation}
\diagram{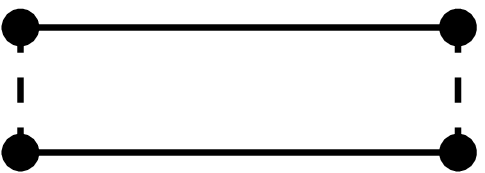}+ \diagram{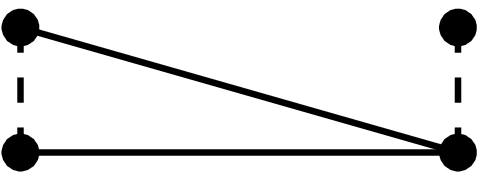}\ ,
\end{equation}
which when closed at the right end with 
$\textdiagram{h7}$
cancel. This completes the proof. \medskip 

The last diagram which we need is the ``half-moon-diagram''
$\diagram{halfmoon}$, which is similar to the moon-diagram, but does
not {\em enforce} the replicas at its ends to be equal. However it
will always be evaluated at coinciding replicas. (It thus contains
$\diagram{moon}$ as a subset.) We claim that
\begin{eqnarray}\label{halfmoontheorem}
{}_{a}\diagram{halfmoon}_{a} &=& \diagram{moon1} + \diagram{moon2} +
\diagram{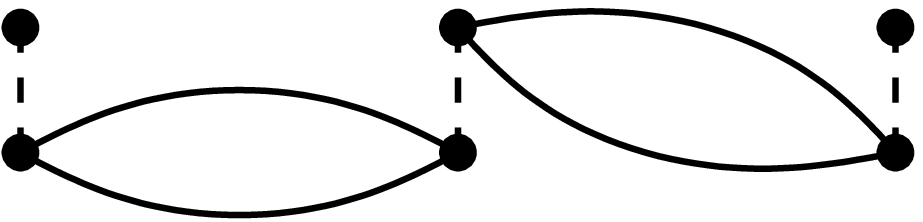} + \diagram{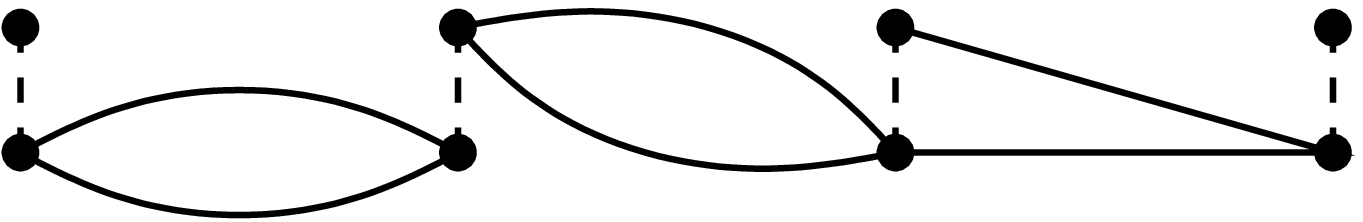}\nonumber \\
&=& \diagram{moon1} \left(1+2 I_{2} (p) B''_{aa} \right) (1-4 I_{2}
(p) B''_{aa})\ . 
\end{eqnarray}
This is proven by first remarking that all other diagrams can be
generated from those: 
Add left of a 
$\textdiagram{h7}$ or a $\textdiagram{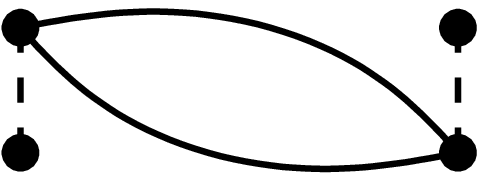}$
the combination
\begin{equation}
\diagram{h6}+ \diagram{h7}\ ;
\end{equation} they cancel pairwise.  The same is true for chains made
out of the combination $\textdiagram{h6}+ \textdiagram{h7}$.
Finally note that one cannot insert more $\textdiagram{monn8}$, since
they would lead to higher replica terms. 

All rules and building blocks are collected on figure \ref{f:buildingblocks}.

\subsection{Zero temperature ($T=0$)}\label{T=0} We start our
discussion with zero temperature, $T=0$. We have to construct all
diagrams with the topology of a loop in the large-$N$ limit. Note that
e.g.\ the diagram (\ref{lf38}) counts as a line in this
construction. The building blocks are given on figure
\ref{f:buildingblocks}.  At zero temperature, one needs all possible
constraints on the sum of replicas. This means that one can not use
$\diagram{moon}$ or $\diagram{halfmoon}$, which both contain one
non-replica conserving line $\textdiagram{moon1}$ at zero
temperature. At finite temperature, one can use one at order $T$ and
two at order $T^{2}$.

At $T=0$, we find the
following diagrams. Note that the notation is such that crossing lines
do not intersect.  All the diagrams correct the effective action $\tilde
B$ without any further combinatorial factor.
\begin{eqnarray}\label{1oN1}
\diagram{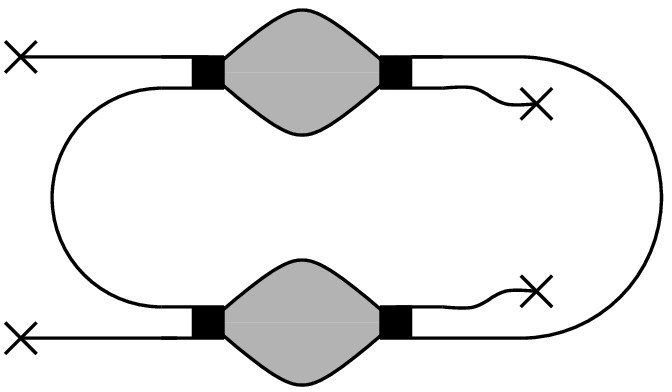}+\diagram{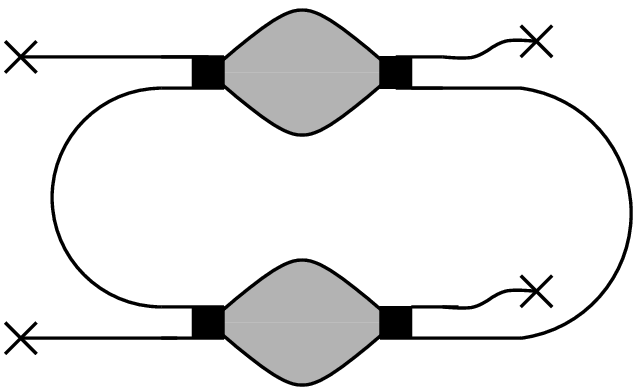}&=&8 
\int_{p}C (p)^{2} \left[H_{ab}(p)v_{ab}^{2} \right]^{2}
\\
\label{1oN2} \diagram{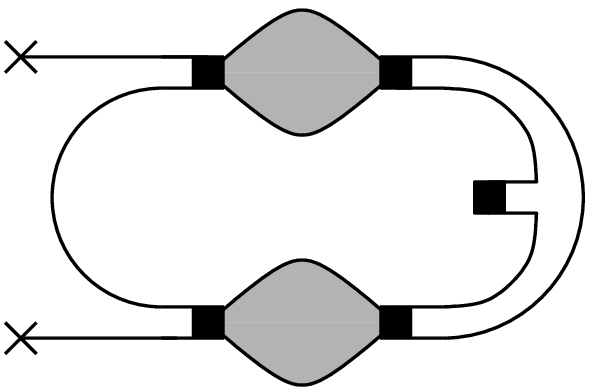}+\diagram{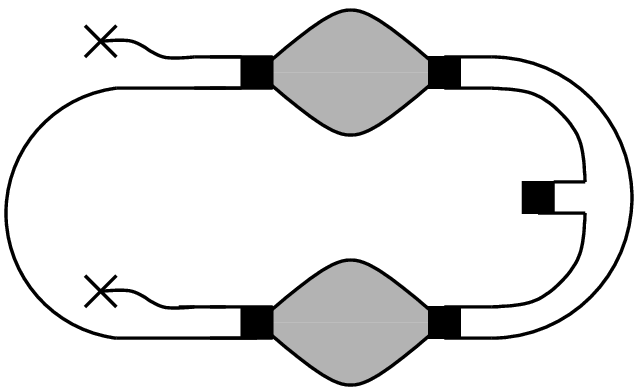}&=& 64 \int_{p}I_{3} (p)C
(p) H_{ab}(p)^{2}v_{ab}^{2} \left ( B' _{ab}-B'_{0} \right)
\\
\diagram{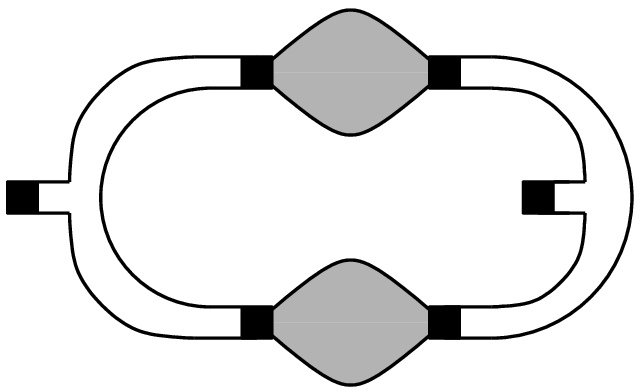}+\diagram{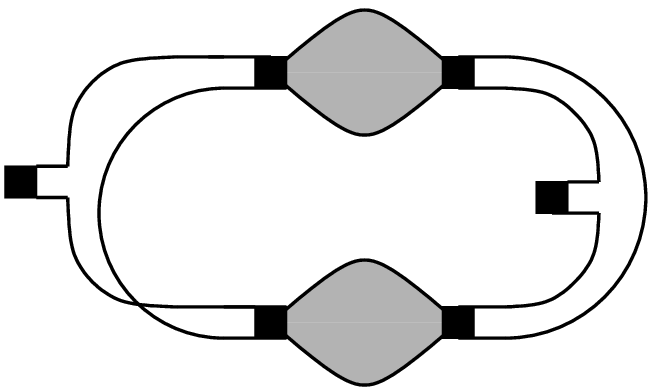}&=& 128 \int_{p} I_{3}(p)^2  H_{ab}(p)^{2}
     \left(B'_{ab}-B'_{0} \right)^{2}
\label{1oN3}\\ 
\label{1oN4}
\diagram{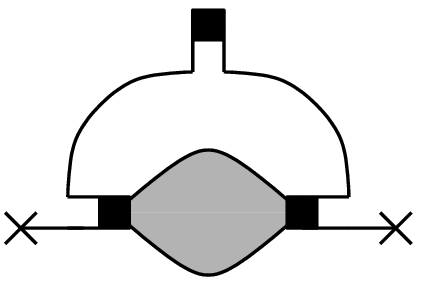}+\diagram{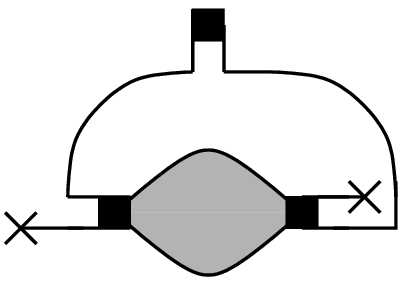}&=& 8 \int_{p}C (p)^{2}
 v_{ab}^{2}H_{ab}(p)\left(
B'_{ab}-B'_{0} \right)
\\
\diagram{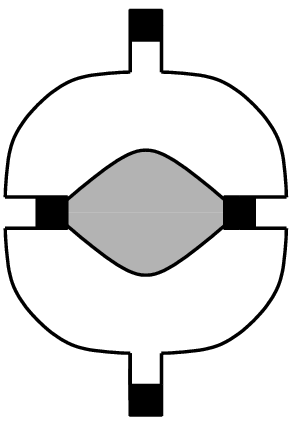}+\diagram{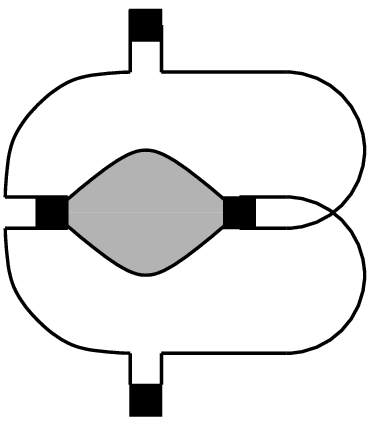} &=& 16 \int_{k}\int_{p}H_{ab}(p) 
\left(B'_{ab}-B'_{0} \right)^{2} I_{4} (p) \qquad  
\label{1oN5}
\ .
\end{eqnarray}
The combinatorial factors can, and have been checked by
straightforward calculating the diagrams with $H_{ab}$ replaced by
$B''_{ab}$, both by hand and computer-algebraically. 

The idea of how to construct these diagrams is straightforward: We
start by a closed chain of 2-loops, which disregarding the $O
(N)$-structure has been drawn on the left of figure
\ref{f:bigloop}. Then one has to cut each line exactly once. These
cuts can either be done at different positions in the ``big'' loop
(diagrams (\ref{1oN1}) to (\ref{1oN3})), or at the same position
(diagrams (\ref{1oN4}) and (\ref{1oN5})). Then there is the
possibility to either insert into a propagator a vertex (these are the
terms proportional to $B'_{ab}-B'_{0}$) or not to contract two fields
(the terms proportional to $v_{ab}^{2}$). The left and right diagrams
in each equation are distinguished by their twist: The left one is the
untwisted one (as the one drawn on the left of figure
\ref{f:bigloop}), the right one the twisted one, obtained by cutting
both lines between two neighboring vertices, and reglueing them
together with the two lines exchanged (this gives one single
propagator line running twice around.) Note that the twisted diagrams
do not appear in the final result, since we suppose analyticity for
$B$, such that e.g.\ $\lim_{a\to b} H_{ab} v_{ab}^{2}
=0$.\footnote{Note that this construction suggests how to construct
additional ``anomalous'' terms, known e.g.\ to be necessary at 2-loop
order \cite{LeDoussalWieseChauve2003,Wiese2003a}. We do not present
them here, but relegate their discussion to a subsequent publication.}

As discussed above other imaginable contributions are 3-replica terms,
where one is not using the maximal number of possible constraints on
the number of free replica-sums (one line cut twice instead of each
line cut once):
\begin{eqnarray}\label{lf42}
\diagram{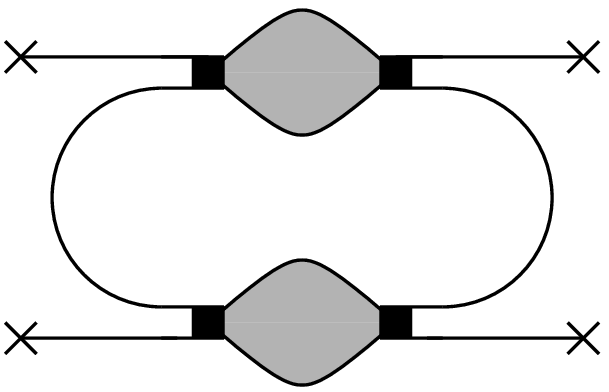} &=& \mbox{3-replica term}\\
\diagram{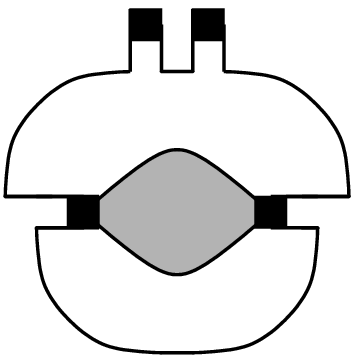} &=&\mbox{3-replica term}\label{lf43}\ .
\end{eqnarray}
We note contributions which vanish identically for completeness:
\begin{eqnarray}\label{lf44}
\diagram{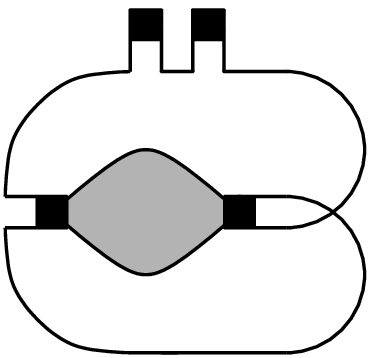} &=& 0\\
\diagram{1oN3p}&=& 0\label{lf45} \ .
\end{eqnarray}

\subsection{Corrections at order $T$}\label{order} We remark that
adding an additional line to any object already constructed at order
$1/N$ results into a diagram of higher topology in the large
$N$-limit. This means that the diagram does not contribute, a
statement which remains true to any order in perturbation theory. The
only remaining possibility is to proceed as before, but constraining
replica-indices to be the same by more than one propagator
(line). Examples are $\diagram{moon}$ and
$\diagram{halfmoon}$. Another example would be a ``circular railroad
diagram'', see left of figure \ref{f:bigloop} and equation
(\ref{lf51}) below.  Since we had two lines to ``waste'', it means
that there will be a term of order $T$ and $T^{2}$.  (As a
side-remark, we note that at order $1/N^{2}$, there will be terms of
up to order $T^{4}$, since for the leading term one has to cut up to 4
lines, a.s.o..)

We now give the order $T$-contributions: 
\begin{equation}\label{1oNT1a}
\diagram{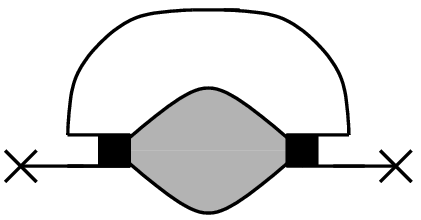}=4T \int_{p }\frac{v_{ab}^{2} H_{ab} (p)}{p^{2}+m^{2}} 
\end{equation} 
\begin{equation}\label{1oNT1b}
\diagram{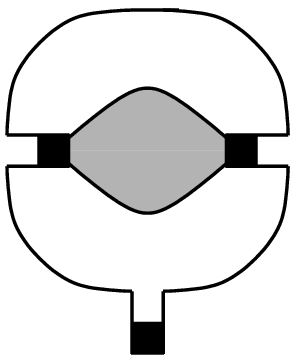}=16 T \int_{p } I_3 (p) H_{ab} (p)
[{B}'_{ab}-{B}'_{0}]
\ .
\end{equation}
These are the contributions, where one replica-line has been cut by
the insertion of either $v^{2}$ or $B'$, whereas the other one (on top
of the diagrams) is redundant. 

We can also use a double (redundant) line, using the moon diagram
\diagram{moon}. Starting from \Eq{1oNT1a}, and inserting it into the
bubble line, we find
\begin{equation}\label{1oNT1c}
\diagram{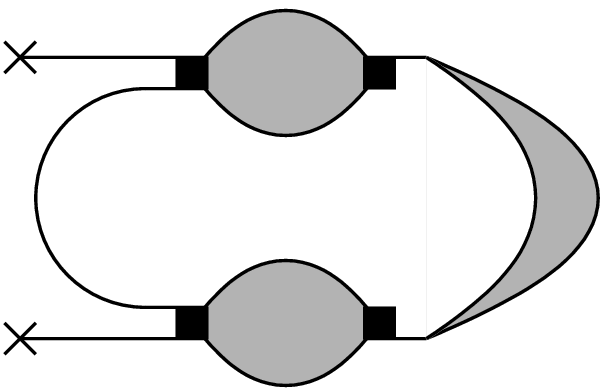}=8T \int_{p } C(p) v_{ab}^{2} H_{ab} (p)^{2} I_2 (p)
\left[1-4 I_{2} (p)B''_{aa}\right] \ .
\end{equation}
Starting the same procedure from (\ref{1oNT1b}) gives
\begin{equation}\label{1oNT1d}
\diagram{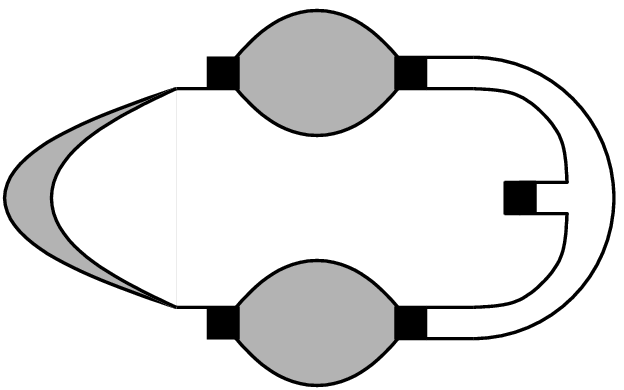}=32 T \int_{p } H_{ab} (p)^{2} ({B}'_{ab}-{B}'_{aa})
I_2 (p) I_{3} (p) \left[1-4 I_{2} (p) B''_{aa}\right]\ .
\end{equation}
Note that there is also a ``twisted'' version of (\ref{1oNT1b}) and
(\ref{1oNT1d}), which add up to 
\begin{equation}\label{1oNTanosimp}
\diagram{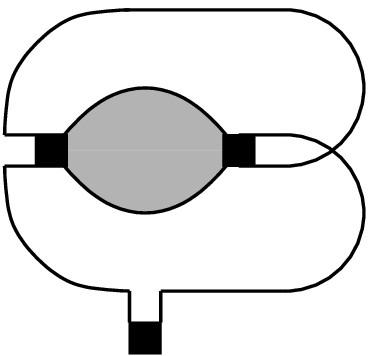}+\diagram{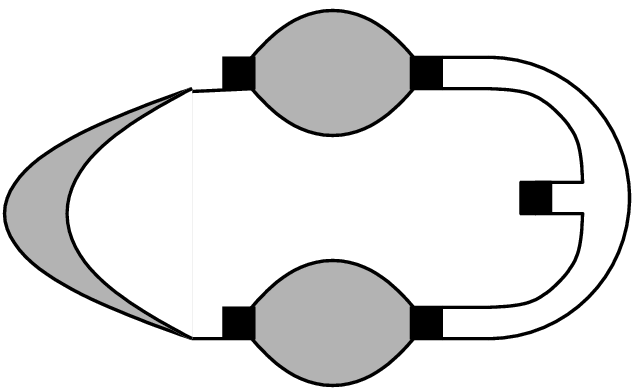}=- 16 T\int_{p} H_{aa} (p) B'_{ab}
I_{3} (p) \left[ 1+2I_{2} (p)B''_{aa} \right]\ .
\end{equation}
%{See program is in ~wiese/Mathematica/1overN/horrortest.nb}

\subsection{Corrections at order $T^{2}$}\label{falseT2} We continue
with diagrams at order $T^{2}$. There is one diagram, which does not
necessitate any cut in a replica-conserving line, thus has two
redundancies. It is the ``railroad-diagram'', i.e.\ a closed chain,
with at least one vertex. Note that contrary to what one might expect,
for one vertex this is {\em not} a product of two tadpoles summed at
leading order, even though it looks alike. However, the structure in
vector-indices makes it a loop in $N$, i.e.\ a subdominant term. Since
no vertex is marked, the sum is not a geometric series, but a
logarithm:
\begin{equation}\label{lf51}
\diagram{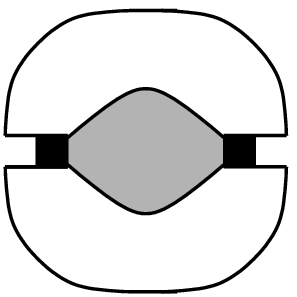}= \frac{T^{2}}{2} \int_{p} \sum _{n=1}^{\infty}
\frac{\left[ 4B'' (\bar \chi_{v} )I_{2} (p) \right]^{n}}{n} \ .
\end{equation}
This can be written either as a function of the bare disorder, or
using $B''(\bar \chi_{ab}) = \tilde{B}_{ab}''/ ( 1+4I_{2}\tilde{B}''_{ab}
)$ (valid at all $T$) as a function of the renormalized disorder:
\begin{eqnarray}\label{1oNT2a}
 \diagram{1oNT2a}&=& - \frac{T^{2}}{2}\int_{p} \ln \Big[1-4 I_{2} (p) B''
(\bar \chi_{v})  \Big] \nonumber\\ 
&=&-\frac{T^{2}}{2}\int_{p} \ln \left( \frac{ 1+ 4\tilde{B}''
(v)[I_{2}-I_{2} (p)]} {1+ 4\tilde{B}''(v)I_{2} } \right)
\ .
\end{eqnarray}
Note that this is the term where a double, completely
replica-conserving line goes around. 

More diagrams are possible, with one and two defects, i.e.\
replica-line cuttings, equivalent to inserting $\diagram{moon}$ or
$\diagram{halfmoon}$.  Using only one cut, both outer indices are
forced to be equal, and we have to insert $\diagram{halfmoon}$,
calculated in (\ref{halfmoontheorem}):
\begin{equation}\label{1oNT2b}
\diagram{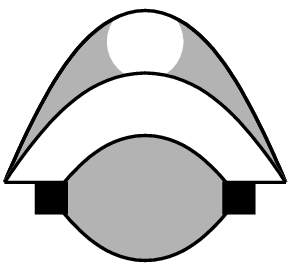}= 2 {T^{2}} \int_{p} I_{2} (p) H_{ab}(p) \left[1+2
I_{2} (p) B''_{aa} \right] \left[1-4 I_{2} (p) B''_{aa} \right] \ .
\end{equation}
The overall prefactor of \Eqs{1oNT2a} and \eq{1oNT2b} is such that the
term linear in $\tilde{B}''_{ab}$ comes with a factor of 2, and they both
add up to a factor of $4$, which can be checked with a simple 1-loop
calculation. 

The remaining term is obtained by cutting two-replica lines, using the
replica-conserving moon
$\diagram{moon}$.  This gives the contribution
\begin{eqnarray}\label{1oNT2c}
 \diagram{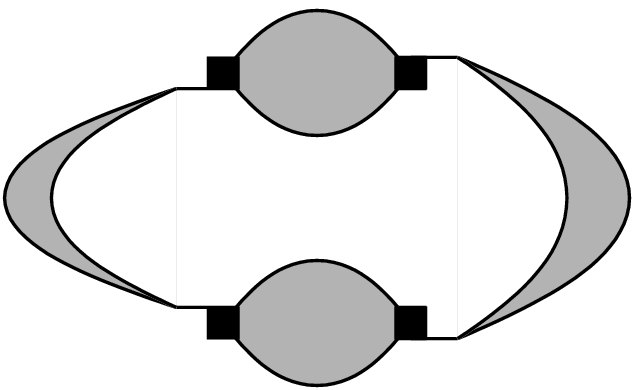}&=& 2 {T^{2}} \int_{p}\left[ I_{2} (p) H_{ab} (p)
(1- 4 I_{2} (p) B''_{aa}) \right]^{2}\ .
\end{eqnarray}

There is an additional anomalous term. It is nothing but a 1-loop
diagram, of the form $\textdiagram{h7}$, where the right-most vertex
is the sum of all diagrams at order $T^{2}$, as given by (\ref{lf51})
to (\ref{1oNT2c}), evaluated at coinciding replicas: 
\begin{eqnarray}\label{lf57}
{\cal A}^{T^{2}}&=&-16T^{2} I_{2} B'_{ab}\int_{p} \frac{ I_{2} (p)  [ 1+2
I_{2} (p)B''_{aa}]B'''_{aa}}{1-4 I_{2} (p)B''_{aa}} \nonumber \\
&=& -16T^{2} I_{2} B'_{ab}\int_{p}  I_{2} (p)  [ 1+6 I_{2} (p)H_{aa}
(p)]B'''_{aa}
\ .
\end{eqnarray}
This can be reexpressed as a function of $\tilde B$, see section
\ref{efakren}. One might suspect that longer chains can be constructed
to connect the $B'_{ab}$ with the derivative of the effective action
at order $T^{2}$, taken at coinciding indices. With the same arguments
as already made a couple of times above, the insertions of
$\textdiagram{h6}+\textdiagram{h7}$ pairwise cancel. Also note that
one could of course draw all these diagrams; there is however no clear
advantage of doing so. 

The alert reader will also wonder why we have not mentioned any such 
term at order $T^{0}$ or $T$. It turns out that the effective action
at order $T^{0}$, when derived once and taken at coinciding arguments,
actually vanishes (supposing analyticity!). At order $T$, the term in
question is nothing but (\ref{1oNTanosimp}).

One caveat is in order: Even though this procedure is simple and
elegant, there are many hidden traps. It is therefore good to check
this calculation by an explicit loop expansion, using the excluded
replica formalism. This has been done up to 8-loop order, and relies
heavily on computer algebraic support. The procedure can also be
formalized, leading to an additional more rigorous but somehow
elaborate approach, the ``Excluded Replica Formalism'', which is
presented in appendix \ref{excludedrepformalism}.

%%%%%%%%%%%%%%%%%%%%%%%%%%%%%%%%%%%%%%%%%%%%%%%%%%%%%%%%%%%%%%%%%%%%%%
%%%%%%%%%%%%%%%%%%%%%%%%%%%%%%%%%%%%%%%%%%%%%%%%%%%%%%%%%%%%%%%%%%%%%%
%%%%%%%%%%%%%%%%%%%%%%%%%%%%%%%%%%%%%%%%%%%%%%%%%%%%%%%%%%%%%%%%%%%%%%
%%%%%%%%%%%%%%%%%%%%%%%%%%%%%%%%%%%%%%%%%%%%%%%%%%%%%%%%%%%%%%%%%%%%%%
%%%%%%%%%%%%%%%%%%%%%%%%%%%%%%%%%%%%%%%%%%%%%%%%%%%%%%%%%%%%%%%%%%%%%%
%%%%%%%%%%%%%%%%%%%%%%%%%%%%%%%%%%%%%%%%%%%%%%%%%%%%%%%%%%%%%%%%%%%%%%
%%%%%%%%%%%%%%%%%%%%%%%%%%%%%%%%%%%%%%%%%%%%%%%%%%%%%%%%%%%%%%%%%%%%%%
%%%%%%%%%%%%%%%%%%%%%%%%%%%%%%%%%%%%%%%%%%%%%%%%%%%%%%%%%%%%%%%%%%%%%%
%%%%%%%%%%%%%%%%%%%%%%%%%%%%%%%%%%%%%%%%%%%%%%%%%%%%%%%%%%%%%%%%%%%%%%
%%%%%%%%%%%%%%%%%%%%%%%%%%%%%%%%%%%%%%%%%%%%%%%%%%%%%%%%%%%%%%%%%%%%%%
%       Pierre's section about the algebraic method                  %
%%%%%%%%%%%%%%%%%%%%%%%%%%%%%%%%%%%%%%%%%%%%%%%%%%%%%%%%%%%%%%%%%%%%%%
%%%%%%%%%%%%%%%%%%%%%%%%%%%%%%%%%%%%%%%%%%%%%%%%%%%%%%%%%%%%%%%%%%%%%%
%%%%%%%%%%%%%%%%%%%%%%%%%%%%%%%%%%%%%%%%%%%%%%%%%%%%%%%%%%%%%%%%%%%%%%
%%%%%%%%%%%%%%%%%%%%%%%%%%%%%%%%%%%%%%%%%%%%%%%%%%%%%%%%%%%%%%%%%%%%%%
%%%%%%%%%%%%%%%%%%%%%%%%%%%%%%%%%%%%%%%%%%%%%%%%%%%%%%%%%%%%%%%%%%%%%%
%%%%%%%%%%%%%%%%%%%%%%%%%%%%%%%%%%%%%%%%%%%%%%%%%%%%%%%%%%%%%%%%%%%%%%
%%%%%%%%%%%%%%%%%%%%%%%%%%%%%%%%%%%%%%%%%%%%%%%%%%%%%%%%%%%%%%%%%%%%%%
%%%%%%%%%%%%%%%%%%%%%%%%%%%%%%%%%%%%%%%%%%%%%%%%%%%%%%%%%%%%%%%%%%%%%%
%%%%%%%%%%%%%%%%%%%%%%%%%%%%%%%%%%%%%%%%%%%%%%%%%%%%%%%%%%%%%%%%%%%%%%
%%%%%%%%%%%%%%%%%%%%%%%%%%%%%%%%%%%%%%%%%%%%%%%%%%%%%%%%%%%%%%%%%%%%%%

\section{Corrections at order $1/N$, via the algebraic method}
\label{s:algebramethod}

\comment{*** VERSION FINALE POUR PIERRE, KAY est REPASS\'E *************} 

The calculation of the renormalized disorder to next order requires the
calculation of the following trace in the space of four replica matrices,
from (\ref{lf25}) and (\ref{lf26}):
\begin{eqnarray}
\tilde{U}^1(v v) &=&  \frac{1}{2} \int_q \tr [\ln ( \delta_{ac}
\delta_{bd} + {\cal M}_{ab,cd}) ] \label{start}\\ 
 {\cal M}_{ab,cd} &=& M_{ab,ef} N_{ef,cd}(q) \\
 M_{ab,cd} &=& (2 T \partial_{\chi_{ab}}  \partial_{\chi_{cd}}
U(\chi))|_{\chi=\chi_v} \\ 
 \overline{N}_{ef,cd}(q) &=& T \Pi_v^{ef,cd}(q) + G_v^{fc}(q) v_e \cdot v_d +
G_v^{e d}(q) v_f \cdot v_c \ .
\end{eqnarray}
$\Pi_{v}^{ef,cd} (q)$ has been defined in (\ref{lf26}).  Expression
(\ref{start}) can be computed by systematically expanding in sums over
increasing numbers of replicas, as was done at dominant order. However
the calculation is considerably more tedious in the present case. We
 give the main features here and relegate details to
Appendix \ref{app:Algebra}.

To obtain the correction to the second cumulant of the disorder, we
will only need the two-replica part of this function, which we denote
$P_2 \tilde{U}^1(v v)$ ($P_n X$ denotes the part of an expression $X$
which contains exactly $n$ free sums over replicas). Since the trace
already involves at least one replica sum, we can and will truncate
all expressions given below by discarding all terms with two or more
replica sums. This can be checked systematically and originates from
the fact that once a replica sum appears in an expansion, it can never
disappear later on.  As a result, we find that third and higher
cumulants do not appear in the correction to the second cumulant, as
they involve higher replica sums\footnote{This formal expansion in
replica sums assumes some analyticity property in a way which should
be analyzed later.}.

The expansion of the matrices $M$, $\overline{N}$ and ${\cal M}$ are
computed in Appendix \ref{app:Algebra}. It is crucial to write
explicitly all Kronecker-delta functions. The matrix ${\cal M}$ is
found to have the form:
\begin{equation}
 {\cal M}_{ab,cd} = m_{ab} \frac{1}{2} (\delta_{ad} \delta_{bc} +
\delta_{bd} \delta_{ac}) + (1 + m_{ab}) \overline{{\cal M}}_{ab,cd}
\ , \label{para1}
\end{equation}
where $m_{ab}$ is symmetric in $a,b$ and $\overline{{\cal M}}$ is
symmetric in $a,b$ and symmetric in $c,d$ (but not necessarily in
exchange of $(a,b)$ with $(c,d)$) and can thus be parameterized as
\begin{eqnarray}
 \overline{\cal M}_{ab,cd} &=& \delta_{a b c d} x_a + \delta_{a b c }
y_{ad} + \delta_{a b d} y_{ac} + \delta_{a c d} z_{ab} + \delta_{b c
d} z_{ba} + \delta_{a c} t_{a b d}
+ \delta_{a d} t_{a b c} +  \delta_{b c} t_{b a d} + \delta_{b d} t_{b
a c}\nonumber   \\ 
&& + \delta_{a b} \delta_{c d} u_{ac} + \delta_{a b} v_{acd} +
\delta_{c d} w_{abc} + g_{abcd}
\ ,
\end{eqnarray}
where all Kronecker-delta's have been written explicitly. Also note
that $v_{acd}$ is symmetric in $c$ and $d$; $w_{abc}$ is symmetric in
$a$ and $b$ and $g_{abcd}$ is symmetric in $a,b$ and symmetric in
$c,d$, whereas all others have no such symmetry. These matrices form a
closed algebra which is studied in Appendix
\ref{app:Algebra}. Unfortunately this algebra is rather large, even
though it is the smallest algebra sufficient for the present
calculation.  Note that we have explicitly separated the part
proportional to the (symmetrized) identity $1^{\ind{sym}}_{ab,cd}=1/2
(\delta_{ac} \delta_{bd} + \delta_{ad} \delta_{bc} )$.

The first preliminary step in the trace log calculation is to
prove\footnote{\underline{Proof:} Write
\newcommand{\one}{1\hspace*{-.6ex}\mathrm{l}} $ { \cal M}= \mathbb{A}
+ (\one +\mathbb{A}') \bar {\cal M} $, where $ \one_{ab,cd} =
\delta_{ac}\delta_{bd} $. Then $\tr [\ln (\one + {\cal M})]= \tr[\ln
(\one + \mathbb{A}+ (\one +\mathbb{A}')\bar {\cal M}) ]= \ln
\left(\det [\one + \mathbb{A}+(\one +\mathbb{A}')\bar {\cal M}]
\right) =\ln \left(\det [\one + \mathbb{A}] \right)+\ln \left(\det
[\one + (\one + \mathbb{A})^{-1}(\one + \mathbb{A}')\bar {\cal M}]
\right) $.  Further $\ln \left(\det [\one + \mathbb{A}] \right) =
\tr\left[\ln (1+\mathbb{A}) \right] = \half \sum_{ab}
\ln(1+m_{ab})+\half \sum_a \ln (1+m_{aa})$. Since $\bar {\cal
M}_{ab,cd}=\bar {\cal M}_{ba,cd} $, the following identity holds: $
\mathbb{A} \bar{\cal M} = \mathbb{A}' \bar {\cal M}$, such that $ \ln
\left(\det [\one + (\one + \mathbb{A})^{-1}(\one + \mathbb{A}')\bar
{\cal M}] \right)= \ln \left(\det [\one + \bar {\cal M}] \right) = \tr
[\ln (\one + \bar {\cal M})] = -\sum_{p>0} (-1)^{p} \tr \left[\bar {
\cal M}^{p} \right] $. q.e.d.} that 
\begin{equation}\label{5.7s}
 \tr \left[ \ln(\delta_{ac} \delta_{bd} + {\cal M}_{ab,cd})\right]=
\frac{1}{2} \sum_{ab} \ln(1 + m_{ab}) + \frac{1}{2} \sum_{a} \ln(1 +
m_{aa}) + \tr\left[ \ln \left( \delta_{ac} \delta_{bd} +
\overline{\cal M}_{ab,cd}]\right) \right]\ , 
\end{equation}
 where in the first two terms the $\ln$ simply acts
on numbers. This formula is valid \footnote{Note that the naive
identity $\tr [\ln (\delta_{ac} \delta_{bd} + {\cal M}_{ab,cd})] =
\tr[ \ln (\frac{1}{2} ( \delta_{ac} \delta_{bd} + \delta_{ad}
\delta_{bc} ) + {\cal M}_{ab,cd})]$ is incorrect.}  for a matrix
${\cal M}$ symmetric in $a,b$ and symmetric in $c,d$.  The last trace
log is equal to its usual series expression $\sum_{p \geq 1}
\frac{(-1)^{p+1}}{p}(\overline{\cal M}^q)^p_{ab,ab}$.

It does not seem possible to express the trace log in general but here it
is possible to expand it systematically in the number of replica sums. Let
us sketch the method that we  found most convenient. We write 
\begin{eqnarray}
 \tr [\ln ( 1 + \overline{\cal M})] &=& - \int_0^{1} \frac{\rmd
\lambda}{\lambda}\, \tr \, M^\lambda  \\ 
M^ \lambda &=& (1+ \lambda \overline{\cal M} )^{-1} - 1\ ,
\label{inversion}
\end{eqnarray}
and we want to compute all terms of $M^\lambda$, namely:
\begin{eqnarray}
M^\lambda_{ab,cd} &=& \delta_{a b c d} x^\lambda_a + \delta_{a b c }
y^\lambda_{ad} + \delta_{a b d} y^\lambda_{ac} + \delta_{a c d}
z^\lambda_{ab} + \delta_{b c d} z^\lambda_{ba} + \delta_{a c}
t^\lambda_{a b d}
+ \delta_{a d} t^\lambda_{a b c} +  \delta_{b c} t^\lambda_{b a d}
 + \delta_{b d} t^\lambda_{b a c}  \nn\\
&& + \delta_{a b} \delta_{c d} u^\lambda_{ac} + \delta_{a b}
v^\lambda_{acd} + \delta_{c d} w^\lambda_{abc} + g^\lambda_{abcd} 
\ . \label{para2}
\end{eqnarray}
We do this by using the algebra detailed in Appendix \ref{app:Algebra} 
to solve the
following equation, equivalent to (\ref{inversion}):
\begin{equation}\label{eqmat}
\lambda \overline{\cal M} + M^\lambda + \lambda \overline{\cal M}
M^\lambda = 0 \ .
\end{equation}
One projects onto each component $x,y,z,\dots $ and onto terms with an
increasing number of replica sums.  What we want are 
$x^\lambda,y^\lambda,z^\lambda,\dots $ as a function of the known
$x,y,z,\dots $, which parameterize $\overline{\cal M}$ (their expressions
are given below).

More specifically one writes:
\begin{eqnarray}
 x_a &=& P_0 x_a +  P_1 x_a + \dotsb  \\
 x^\lambda_a &=& P_0 x^\lambda_a + P_1 x^\lambda_a + \dotsb  \label{expsum}
\end{eqnarray}
and similarly for all other components $y,z,t,...$ of the matrices
$\overline{\cal M}$ and $M^\lambda$.  Using the algebraic rules for the
product of two matrices, it turns out that it is possible to solve
(\ref{eqmat}) for all components of $M^\lambda$ in an iterative
manner.  This is simplified since $P_0 x_a=0$ (see below).  First
we determine all zero-sum components $P_0 x^\lambda,P_0
y^\lambda,\dots $ from the corresponding $P_0 x,P_0 y,\dots $. It can be
done in the following order: First compute $P_0 y^\lambda_{ab}$
together with $P_0 y^\lambda_{aa}$, then similarly  $P_0
z^\lambda,P_0 x^\lambda,P_0 t^\lambda,P_0 v^\lambda,P_0 w^\lambda,P_0
g^\lambda$.  Second one can project (\ref{eqmat}) onto one-sum terms,
and determine $P_1 x^\lambda,P_1 y^\lambda$ etc\dots

At then end we need:
\begin{eqnarray}
 P_2\ \tr\  \ln [1 + \overline{\cal M}] &=& - \int_0^{1} \frac{\rmd
 \lambda} {\lambda}\,  P_2 (\tr\  M^\lambda)   \\ 
 P_2 (\tr\ M^\lambda) &=&\sum_a ( P_1 x^\lambda_a + 2 P_1
y^\lambda_{aa} + 2 P_1 z^\lambda_{aa} + 2 P_1 t^\lambda_{aaa} + P_1
u^\lambda_{aa}
+ P_1 v^\lambda_{aaa} + P_1 w^\lambda_{aaa} ) \nn\\
&& + \sum_{ab} ( 2 P_0 t^\lambda_{abb} + P_0 g^\lambda_{abab} ) 
\ .
\end{eqnarray}
The remaining integrations over $\lambda$ of each term are found to be of the form
$\int_0^1 \rmd \lambda \lambda^{p}/(1 + 2 y_{aa} \lambda)^q = {}_2F_1
( 1 + p, q, 2 + p, -2 y_{aa}) /(1+p)$ 
which for integer $p$ and $q$ can be expressed
as rational fractions.

The detailed calculation is however very tedious and has been
performed using Mathematica. The result is given in the following
 and is found to agree exactly with the graphical method of
the previous section. Here we give the zero-sum and one-sum components
of the matrix $\overline{\cal M}$ needed for the calculation of $P_2\
\tr\ \ln (1 + \overline{\cal M})$. They are calculated from
(\ref{start}), see Appendix \ref{app:Algebra}, and read:
\begin{eqnarray}
 P_0 x_a &=&  0 \quad , \quad  P_0 u_{ac} = - P_0 y_{ac} \quad , \quad
P_0 w_{acd} = 0 \\ 
 P_0 y_{ad}  &=& I_2(q) A^q_{aa} \overline{B}''_{ad} \quad , \quad
A^q_{aa} = \frac{2}{1 - 4 B''(2T I_1) I_2(q) }\\ 
 P_0 z_{ab} &=&  I_2(q) A^q_{ab} \overline{B}''_{ab} \quad , \quad
A^q_{ab} = \frac{2}{1 - 4 \overline{B}''_{ab} I_2(q) } \\ 
 P_0 t_{a b d} &=& - \frac{1}{T} A^q_{ab} \overline{B}''_{ab} 
\left[(v_b-v_a) v_d C(q)  - 2 I_3(q) (\tilde{B}'_{bd} -
\tilde{B}'_{ad}) \right] \\ 
 P_0 v_{acd} &=& - \frac{1}{T} A^q_{aa} \Big\{ \overline{B}''_{ad}
\left[ (v_d-v_a) v_c C(q)- 2 I_3(q) (\tilde{B}'_{dc} -
\tilde{B}'_{ac}) \right]
 \nonumber \\ 
&&\hphantom{- \frac{1}{T} A^q_{aa} \Big\{} + \overline{B}''_{ac}
\left[ (v_c -v_a) v_d C(q) - 2 I_3(q)
(\tilde{B}'_{cd} - \tilde{B}'_{ad}) \right] \Big\}  \\ 
 P_0 g_{abcd} &=& \frac{2}{T^2} C(q)^2 A^q_{ab} \overline{B}''_{ab}
(v_a - v_b)\left[  (\tilde{B}'_{bc} -  \tilde{B}'_{ac}) v_d +
(\tilde{B}'_{bd} -  \tilde{B}'_{ad}) v_c\right] \nonumber  \\ 
&& - \frac{4}{T^2} I_4(q) A^q_{ab} \overline{B}''_{ab} 
 \left(\tilde{B}'_{ad} \tilde{B}'_{bc} +
\tilde{B}'_{bd} \tilde{B}'_{ac} - \tilde{B}'_{ad} \tilde{B}'_{ac} -
\tilde{B}'_{bd} \tilde{B}'_{bc} \right)
\ ,
\end{eqnarray}
where we have defined $\tilde{B}'_{ab}=\tilde{B}'(v^2_{ab})$ and 
\begin{eqnarray}
&& \overline{B}''_{ab} = B'' (\overline{\chi}_{ab})\ , \qquad
\overline{B}'''_{ab} = B''' (\overline{\chi}_{ab}) \\ 
&& \overline{\chi}_v^{ab} = v_{ab}^2 + 2 T I_1 + 4 I_2
(\tilde{B}'_{ab} - \frac{1}{2} (\tilde{B}'_{aa} + \tilde{B}'_{bb}))
\end{eqnarray}
with $\overline{B}''_{aa}=B''(2 T I_1)$. All integrals $I_n$ and $I_n(q)$ are
defined in section \ref{buildingblocks} and Appendix
\ref{app:integrals} and $C(q)=1/(q^2+m^2)$. We recall that $B$ is the
bare second cumulant, $\tilde B$ the renormalized one and satisfies
$\tilde{B}'_{ab}=B'(\overline{\chi}_v^{ab})$ at dominant order, which
is sufficient for our purpose.  For convenience we use
$\overline{\chi}_v^{ab} := \tilde \chi_{ab}^{(0)}$ to denote the
zero sum part in the decomposition $\tilde \chi_v^{ab}= \tilde
\chi_{ab} + \delta_{ab} \chi_{a}$ in the notations of \cite{LeDoussalWiese2003a}.

For the 1-sum terms we need only the diagonal values:
\begin{eqnarray}
 P_1 x_a &=&   - I_2(q) A^q_{aa}  \sum_f \overline{B}''_{af} \\
 P_1 y_{aa} &=& \frac{8}{T} \frac{1}{\left[ 1 - 4 B''(2 T I_1)
I_2(q)\right]^2}
\Big[B''(2 T I_1) I_{3} (q) + T B'''(2 T I_1) I_2 I_2(q) \Big] \sum_f \tilde{B}'_{af} \nonumber \\
&& - \frac{2}{T} \frac{1}{1 - 4 B''(2 T I_1) I_2(q)} \sum_f
\overline{B}''_{af} \Big[ C(q)(v_a - v_f) v_a
- 2 I_3(q) (\tilde{B}'(0) - \tilde{B}'_{af}) \Big] \\
 P_1 z_{aa} &=& - P_1 u_{aa} = \frac{8}{T} \frac{1}{[1 - 4 B''(2 T
I_1) I_2(q)]^2}
\left[B''(2 T I_1) I_3(q) + T B'''(2 T I_1) I_2 I_2(q) \right] \sum_f \tilde{B}'_{af}\qquad \quad  \\
P_1 v_{aaa} &=& - \frac{8}{T^2} \frac{1}{1 - 4 B''(2 T I_1) I_2(q)}
\bigg[ I_4(q) \sum_f \overline{B}''_{af} (\tilde{B}'_{af} - \tilde{B}'(0))^2 \nonumber  \\
&& \hspace{4.7cm}+ C(q)^2 \sum_f \overline{B}''_{af} (\tilde{B}'_{af} -
\tilde{B}'(0)) (v_a - v_f) v_a \bigg]\ .
\end{eqnarray}
In addition since $t_{aaa}=0$ one has $P_1 t_{aaa}=0$; $P_1 w_{aaa}=0$
since $w_{abc}=0$ and it turns out that $P_1 g_{aaaa}$ is not needed.

Starting from these values one performs the algebra and obtains $P_2\
\tr\ \ln ( 1 + \overline{\cal M})$. To this one must add the
``simple'' part of $P_2 \ \tr\ \ln$ in (\ref{5.7s}) above which one
expands as follows:
\begin{eqnarray}
&& \!\!\!P_2 \left[ \frac{1}{2} \sum_{ab} \int_q \tr\  \ln(1 + m_{ab}) + \frac{1}{2} \sum_{a} \int_q \tr\  \ln(1 + m_{aa}) \right] \\
&& = \frac{1}{2} \sum_{ab} \int_q \ln(1 - 4 \overline{B}''_{ab}
I_2(q)) - \frac{8}{T} \int_q \frac{1}{1 - 4 B''(2 T I_1) I_2(q)} \Big[
I_3(q) B''(2 T I_1) + T I_2 B'''(2 T I_1) I_2(q) \Big] \tilde{B}'_{af}
\nn
\ ,
\end{eqnarray}
where $m_{ab}$ is computed in appendix \ref{app:Algebra}.

To translate the results into terms of the renormalized disorder, we can
perform the replacements
\begin{eqnarray}
 \overline{B}''_{ab} &=& \tilde{B}''_{ab}/(1 + 4 I_2 \tilde{B}''_{ab}) \\
 B''(2 T I_1) &=& \tilde{B}''(0)/(1 + 4 I_2 \tilde{B}''(0)) \\
 \overline{B}'''_{ab} &=& \tilde{B}'''_{ab}/(1 + 4 I_2 \tilde{B}''_{ab})^3 \\
 B'''(2 T I_1) &=& \tilde{B}'''(0)/(1 + 4 I_2 \tilde{B}''(0))^3
\ ,
\end{eqnarray} 
since to the same accuracy we can use in all above expressions the
dominant order or the exact one.

The ``simple'' part of $P_2\ \tr\ \ln$ thus gives:
\begin{eqnarray}
 \delta^{(\ind{simple})} \tilde B (x) &=& \frac{8 T}{N} \int_p I_3(p)
H_0(p) \tilde B' (x) + \frac{T^{2}}{N} \Bigg\{-\frac{1}{2}\int_{p} \ln
\left( \frac{ 1+ 4\tilde{B}''
(x)[I_{2}-I_{2} (p)]} {1+ 4\tilde{B}''(x)I_{2} } \right) \nonumber \\ 
&&\qquad + 8 I_{2} \int_{p} I_{2} (p) \tilde B' (x) \frac{\tilde B'''
(0)}{(1+4 I_{2}\tilde B'' (0))^{2}(1+4 (I_{2}-I_{2}(q)) \tilde B''
(0))} \Bigg\} \label{simple}
\ ,
\end{eqnarray}
where integrals are defined in Appendix \ref{app:integrals}.

%%%%%%%%%%%%%%%%%%%%%%%%%%%%%%%%%%%%%%%%%%%%%%%%%%%%%%%%%%%%%%%%%%%%%%
%%%%%%%%%%%%%%%%%%%%%%%%%%%%%%%%%%%%%%%%%%%%%%%%%%%%%%%%%%%%%%%%%%%%%%
%%%%%%%%%%%%%%%%%%%%%%%%%%%%%%%%%%%%%%%%%%%%%%%%%%%%%%%%%%%%%%%%%%%%%%
%%%%%%%%%%%%%%%%%%%%%%%%%%%%%%%%%%%%%%%%%%%%%%%%%%%%%%%%%%%%%%%%%%%%%%
%%%%%%%%%%%%%%%%%%%%%%%%%%%%%%%%%%%%%%%%%%%%%%%%%%%%%%%%%%%%%%%%%%%%%%
%%%%%%%%%%%%%%%%%%%%%%%%%%%%%%%%%%%%%%%%%%%%%%%%%%%%%%%%%%%%%%%%%%%%%%
%%%%%%%%%%%%%%%%%%%%%%%%%%%%%%%%%%%%%%%%%%%%%%%%%%%%%%%%%%%%%%%%%%%%%%
%%%%%%%%%%%%%%%%%%%%%%%%%%%%%%%%%%%%%%%%%%%%%%%%%%%%%%%%%%%%%%%%%%%%%%
%%%%%%%%%%%%%%%%%%%%%%%%%%%%%%%%%%%%%%%%%%%%%%%%%%%%%%%%%%%%%%%%%%%%%%
%%%%%%%%%%%%%%%%%%%%%%%%%%%%%%%%%%%%%%%%%%%%%%%%%%%%%%%%%%%%%%%%%%%%%%
%                              Results                               %
%%%%%%%%%%%%%%%%%%%%%%%%%%%%%%%%%%%%%%%%%%%%%%%%%%%%%%%%%%%%%%%%%%%%%%
%%%%%%%%%%%%%%%%%%%%%%%%%%%%%%%%%%%%%%%%%%%%%%%%%%%%%%%%%%%%%%%%%%%%%%
%%%%%%%%%%%%%%%%%%%%%%%%%%%%%%%%%%%%%%%%%%%%%%%%%%%%%%%%%%%%%%%%%%%%%%
%%%%%%%%%%%%%%%%%%%%%%%%%%%%%%%%%%%%%%%%%%%%%%%%%%%%%%%%%%%%%%%%%%%%%%
%%%%%%%%%%%%%%%%%%%%%%%%%%%%%%%%%%%%%%%%%%%%%%%%%%%%%%%%%%%%%%%%%%%%%%
%%%%%%%%%%%%%%%%%%%%%%%%%%%%%%%%%%%%%%%%%%%%%%%%%%%%%%%%%%%%%%%%%%%%%%
%%%%%%%%%%%%%%%%%%%%%%%%%%%%%%%%%%%%%%%%%%%%%%%%%%%%%%%%%%%%%%%%%%%%%%
%%%%%%%%%%%%%%%%%%%%%%%%%%%%%%%%%%%%%%%%%%%%%%%%%%%%%%%%%%%%%%%%%%%%%%
%%%%%%%%%%%%%%%%%%%%%%%%%%%%%%%%%%%%%%%%%%%%%%%%%%%%%%%%%%%%%%%%%%%%%%
%%%%%%%%%%%%%%%%%%%%%%%%%%%%%%%%%%%%%%%%%%%%%%%%%%%%%%%%%%%%%%%%%%%%%%
\section{Results for the effective action}\label{results}

\comment{*** VERSION FINALE POUR PIERRE, KAY est REPASS\'e *************} 

The $1/N$ correction to the 2-replica part of the effective action
in terms of the bare disorder can be written as:
\begin{eqnarray}
P_2 \tilde{U}^1(vv) = - \frac{1}{2 T^2} \sum_{ab} \delta \tilde B (v_{ab}^2)
\end{eqnarray}
with $v^2_{ab}=(v_a-v_b)^2$. It can be read off from sections
\ref{T=0}, \ref{order} and \ref{falseT2}.  We will give the expression
for the correction to the renormalized second cumulant $\delta \tilde
B(x)$ both in terms of the bare disorder and the renormalized one.

\subsection{The effective action as a function of the bare disorder}
We find:
\begin{eqnarray}\label{lf71}
 \delta \tilde B (vv)&=& \frac{1}{N} \Bigg [8
\int_{p}C (p)^{2} \left[H_v(p) v^{2} \right]^{2} \nonumber  \\
&&\qquad + 64 
\int_{p} I_{3}(p) C (p)  H_v(p)^{2} v^{2}
\left[B' (\overline{\chi}_v) -B' (\overline{\chi}_0) \right]\nonumber 
\\
&& \qquad + 128 \int_{p} I_{3}(p)^2  H_v(p)^{2}
     \left[B'(\overline{\chi}_v)-B'(\overline{\chi}_0)\right]^{2} \nonumber \\
&&\qquad + 8 \int_{p}C (p)^{2} v^{2}H_v(p)\left[
B'(\overline{\chi}_v)-B'(\overline{\chi}_0) \right]   \nonumber 
\\
&& \qquad +16 \int_{p} H_v(p) 
\left[B'(\overline{\chi}_v)-B'(\overline{\chi}_0) \right]^{2} I_4 (p)  \Bigg]\nonumber\\
% order T
&+& \frac{T}{N}\Bigg[4 \int_{p} \left[C (p)v^{2}
+ 4 {I_3}(p)\,\left( B'(\overline{\chi}_v ) - B'(\overline{\chi}_0) \right) \right]
 H_v(p) [1+2 I_{2} (p)H_v(p) ] \nonumber \\
&&\qquad -32 \int_{p}{{{I_2}(p)}^2\,
     \left[ C (p) {v^2} + 4\,{I_3}(p)\,\left( B'(\overline{\chi}_v ) 
-B'(\overline{\chi}_{0}) \right)  \right]  H_{v} (p)^{2}} \,B''(\overline{\chi}_0) \nonumber \\
&&
\qquad -16 \int_{p}  {I_3}(p)\,B'(\overline{\chi}_v )\, \left(
1 + 2\,{I_2}(p) {B''(\overline{\chi}_0)} \right) H_{0} (p)
 \Bigg] \nonumber \\
% order T^{2}
&+&  \frac{T^{2}}{N} \Bigg[-\frac{1}{2}\int_{p}{\log [1 -
4\,I_2(p)\,B''(\overline{\chi}_v)]} \nonumber \\ 
&&\qquad + 2\int_{p}  I_{2} (p) H_v (p) [1+2I_{2} (p) B'' (\overline{\chi}_0)]
[1-4I_{2} (p)B'' (\overline{\chi}_0) ]\nonumber \\ 
&& \qquad + 2\int_{p}  I_{2} (p)^{2} H_v (p)^{2} \left[1-4 I_{2}
(p)B'' (\overline{\chi}_0)\right]^{2}\nonumber \\ 
&&\qquad -16I_{2}\int_{p}{  I_{2} (p) B' (\overline{\chi}_v) [ 1+6 I_{2} (p)
H_{0} (p)]B''' (\overline{\chi}_0)} \Bigg]\label{Beff1oN}\\ 
H_v(p)&=&\frac{{B}'' (\overline{\chi}_v)}{1-4I_{2}(p) {B}''
(\overline{\chi}_v)}\ , \label{Hvp} 
\end{eqnarray}
where $v^2$ stands for $v_{ab}^2$ and is the argument of the function. The
integrals $I_n$ and $I_n(q)$ are defined in appendix \ref{app:integrals}
and we recall that:
\begin{eqnarray}
\overline{\chi}_v = v^2 + 2 T I_1 + 4 I_2 (B'(\overline{\chi}_v) - B'(\overline{\chi}_0))
\end{eqnarray}

\subsection{The effective action in terms of the renormalized
disorder}\label{efakren}

To make the conversion to renormalized disorder, we use that on the
r.h.s.\ of (\ref{Beff1oN}) we can express $B$ in terms of the  renormalized
$\tilde B$ at leading order. Using (\ref{lf72}) 
\begin{eqnarray}\label{6.3}
\tilde B' (x) &=& B' (x+2TI_{1}+ 4 I_{2} (\tilde B' (x)-\tilde B' (0)))\\
x&:=& v^{2}
\end{eqnarray}
and differentiating (\ref{6.3}) w.r.t $x$, we obtain
\begin{equation}
\frac{\tilde B'' (x)}{1+4 I_{2} \tilde B'' (x) } = B'' (x+2TI_{1}+ 4
I_{2} (\tilde B' (x)-\tilde B' (0)))
\end{equation}
This allows to rewrite  (\ref{Hvp}) as  (already noted in (\ref{Habp}))
\begin{equation}\label{asdf}
\tilde{H} _{x} (p) : = \frac{\tilde B'' (x)}{1+4 \left(I_{2}-I_{2}
(p) \right) \tilde B'' (x)} =  H_{v} (p)
\end{equation}

\begin{eqnarray}
 \delta \tilde B (x)&=& \frac{1}{N} \Bigg \{ 8
\int_{p}C (p)^{2} \left[\tilde{H} _x(p) x \right]^{2} \nonumber  \\
&&\qquad + 64 
\int_{p} I_{3}(p) C (p)  \tilde{H} _x(p)^{2} x
\left[\tilde B' (x) -\tilde B' (0) \right]\nonumber 
\\
&& \qquad + 128 \int_{p} I_{3}(p)^2  \tilde{H} _x(p)^{2}
     \left[\tilde B'(x)-\tilde B'(0)\right]^{2} \nonumber \\
&&\qquad + 8 \int_{p}C (p)^{2}\, x \,\tilde{H} _x(p)\left[
\tilde B'(x)-\tilde B'(0) \right]  \nonumber 
\\
&& \qquad +16 \int_{p} \tilde{H} _x(p) 
\left[\tilde B'(x)-\tilde B'(0) \right]^{2} I_4 (p)\Bigg \}\nonumber \\ 
% order T
&+& \frac{T}{N}\Bigg\{4 \int_{p} \left[C (p)x
+ 4 {I_3}(p) \left( \tilde B'(x ) - \tilde B'(0) \right) \right]
 \tilde{H} _x(p) [1+2 I_{2} (p)\tilde{H} _x(p) ] \nonumber \\
&&\qquad -32 \int_{p}{{{I_2}(p)}^2\,
     \left[ C (p) x + 4\,{I_3}(p)\,\left( \tilde B'(x ) -\tilde
B'(0) \right)  \right]  \tilde{H} _{x} (p)^{2}} \frac{\tilde B'' (0)}{1+4
I_{2}\tilde B'' (0)} \nonumber \\  
&& \qquad -16 
\int_{p} {I_3}(p) \tilde B'(x ) \left( 1 + 2\,{I_2}(p) 
\frac{\tilde B'' (0)}{1+4 I_{2}\tilde B'' (0)} \right) \tilde{H} _{0} (p)
 \Bigg\} \nonumber \\
% order T^{2}
&+& \frac{T^{2}}{N} \Bigg\{-\frac{1}{2}\int_{p} \ln \left( \frac{ 1+
4\tilde{B}''
(x)[I_{2}-I_{2} (p)]} {1+ 4\tilde{B}''(x)I_{2} } \right) \nonumber \\ 
&&\qquad + 2\int_{p} I_{2} (p) \tilde{H} _x (p) \left[1+2I_{2} (p) \frac{\tilde
B'' (0)}{1+4 I_{2}\tilde B'' (0)} \right] \left[1-4I_{2}
(p)\frac{\tilde B'' (0)}{1+4
I_{2}\tilde B'' (0)}  \right]\nonumber \\ 
&& \qquad + 2\int_{p}  I_{2} (p)^{2} \tilde{H} _x (p)^{2} \left[1-4 I_{2}
(p) \frac{\tilde B'' (0)}{1+4
I_{2}\tilde B'' (0)}   \right]^{2}\nonumber \\ 
&&\qquad -16I_{2}\int_{p} I_{2} (p) \tilde B' (x) [ 1+6 I_{2} (p)
\tilde{H} _{0} (p)]\frac{\tilde B''' (0)}{(1+4 I_{2}\tilde B'' (0))^{3}}
\Bigg\} \label{deltaBren1} 
\end{eqnarray}

\subsection{Expression in terms of rescaled dimensionless quantities}
\label{effresc}

As in Ref. \cite{LeDoussalWiese2003a}, we define the dimensionless
function $b$ of the dimensionless argument $z$ through
\begin{equation}\label{changeofvar}
b(z) = 4 A_d\, m^{ 4 \zeta - \epsilon} \tilde B(z m^{-2 \zeta } ) \ .
\end{equation}
where $\epsilon=4-d$ and $A_d = 2 (4 \pi)^{-d/2} \Gamma ( 3 -
\frac{d}{2})$.  This prefactor has been chosen to simplify the
leading-order $\beta$-function. For $d<4$ and $T=0$ all integrals
which will appear here and below are convergent in the limit of UV
cutoff $\Lambda \to \infty$ that we consider, e.g.\ in that limit $I_2
= A_d m^{-\epsilon}/\epsilon$. The rescaling factor $\zeta$ is for now
unspecified, but at the fixed point it will yield the roughness
exponent.

In terms of rescaled quantities, one finds that the result of the
$1/N$ expansion for the second cumulant of the disorder can be
rewritten as
\begin{eqnarray}
 b_m(x) = b^0_m(x) + \frac{1}{N} b^1[b^0_m](x)\ ,
\end{eqnarray}
where $b^0_m(x)$ is the solution to dominant order, obtained
previously for an arbitrary bare disorder. We have made apparent the
dependence on the mass $m$ (and raise the indices whenever necessary).
The expression for the $O(1/N)$ correction $b_1[b](x)$ can be obtained
from the last subsection upon rescaling and reads (for $d<4$):
\begin{eqnarray}
b_1[b](x) &=& \frac{1}{ A_d} \int_{p} \Bigg \{
2  c(p)^{2} \left[{h} _x(p) x \right]^{2} 
+ 4 i_{3}(p) c(p)  {h}_x(p)^{2} x
\left[b' (x) - b' (0) \right]\nonumber 
\\
&& \hspace{8ex} + 2  i_{3}(p)^2  {h} _x(p)^{2}
     \left[b'(x)-b'(0)\right]^{2} + 2  c (p)^{2}\, x \,{h} _x(p)\left[
b '(x)-b '(0) \right]  \nonumber 
\\
&& \hspace{8ex} +  {h} _x(p) 
\left[b '(x)-b '(0) \right]^{2} i_4 (p)\Bigg \}\nonumber \\ 
% order T
&+&  \frac{\epsilon T_{m}}{A_d} \int_{p} \Bigg\{ \left[c(p) x
+  {i_3}(p) \left( b '(x ) - b '(0) \right) \right]
 {h} _x(p) \left[1+\frac{1}{2} i_{2} (p){h} _x(p) \right] \nonumber \\
&&\hspace{9ex} -\frac{1}{2} {{{i_2}(p)}^2\,
     \left[ c (p) x + {i_3}(p)\,\left( b '(x ) - b'(0) \right)  \right]  
{h} _{x} (p)^{2}} \frac{b '' (0)}{1+
i_{2} b '' (0)} \nonumber \\  
&& \hspace{9ex} -
 {i_3}(p) b '(x ) \left( 1 + \frac{1}{2}\,{i_2}(p) 
\frac{b '' (0)}{1+ i_{2} b '' (0)} \right) {h} _{0} (p)
 \Bigg\} \nonumber \\
% order T^{2}
&+& \frac{\epsilon^{2} T_m^{2}}{4 A_d} \int_{p} \Bigg\{-\frac{1}{2}
\ln \left( \frac{ 1+  b''
(x)[i_{2}-i_{2} (p)]} {1+  b''(x)i_{2} } \right) \nonumber \\ 
&&\hspace{11ex} + \frac{1}{2} i_{2} (p) {h} _x (p) \left[1+\frac{1}{2}
i_{2} (p) \frac{ b'' (0)}{1+ i_{2} b '' (0)} \right] \left[1-i_{2}
(p)\frac{b '' (0)}{1+
i_{2}b '' (0)}  \right] \nonumber \\
&&\hspace{11ex} + \frac{1}{8} i_{2} (p)^{2} {h} _x (p)^{2} \left[1-
i_{2} (p) \frac{b '' (0)}{1+
i_{2}b '' (0)}   \right]^{2}\nonumber \\ 
&&\hspace{11ex} - i_{2} i_{2} (p) b ' (x) \left[ 1+\frac{3}{2} i_{2}
(p) {h} _{0} (p)\right]\frac{b ''' (0)}{(1+ i_{2} b '' (0))^{3}}
\Bigg\} \label{deltaBren2}
\end{eqnarray}  
where one has defined 
\begin{eqnarray}\label{DEF:hx(p)}
h_x(p)&=& \frac{b''(x)}{1+ [i_2-i_2(p)] b''(x)}
\end{eqnarray} 
as well as the rescaled temperature:
\begin{eqnarray}
 T_m & := &  \frac{4 
A_{d} m^{\theta}}{\epsilon} T
\end{eqnarray}
where $\theta=d-2+2 \zeta$ is the energy fluctuation exponent, and the
rescaled integrals are denoted by small letters:
\begin{eqnarray}\label{DEFin}
i_n(p)&:=& \frac{1}{ A_{d}} I_{n} (p)\ts_{m=1} \equiv \frac{1}{A_{d}}\int
\frac{\rmd^{d}p}{( 2 \pi)^d} \dots \\
c (p)&:=& \frac{1}{1+p^{2}} \label{DEFi0}
\end{eqnarray}
with, e.g $i_2=i_2(p=0)=1/\epsilon$. In the expression above we have
kept the order of the diagrams from section \ref{T=0} ff.. Note that
explicit $\Lambda$ dependence can be reinstated in (\ref{deltaBren2})
by restricting all rescaled momentum integrals by $\Lambda/m$ as upper
cutoff, and is necessary in $T>0$ integrals (since they are usually UV
divergent).

Regrouping terms, this result can be rewritten in a more compact form:
\begin{eqnarray} \label{b1compact}
 b_1(x) &=& 2 x^2 g_1(a_x) + 2 x (b'(x)-b'(0)) g_2(a_x) + 2
(b'(x)-b'(0))^2 g_3(a_x) \\
&&+ \epsilon T_m \big[ x ( g_4(a_x) + a_0 g_5(a_x) ) +
(b'(x)-b'(0)) ( g_6(a_x) + a_0 g_7(a_x) ) \nonumber  \\
&&\hphantom{+ \epsilon T_m \big[}+
b'(x) ( ( g_8(a_0) + a_0 g_9(a_0) ) \big] \nonumber  \\
&&+ (\epsilon T_m)^2 \big [ g_{10}(a_x) + a_0 g_ {11}(a_x) + a_0^2 g_ {12}(a_x)
+ b'(x) \alpha (\gamma + a_0 g_{13}(a_0))  \big ] \nonumber 
\end{eqnarray}
with
\begin{equation}\label{axetcdef}
a_x = \frac{b''(x)}{1+ \frac{b''(x)}{\epsilon}} \ , \quad h_x(p) =
\frac{a_x}{1 - a_x i_2(p)} \ , \quad \alpha = \frac{b'''(0)}{(1+
\frac{b''(0)}{\epsilon})^3}\ , \quad \bar \alpha =\frac{ b''''
(0)/\epsilon }{[1+b'' (0)/\epsilon ]^{2}} - \frac{2 b'''
(0)^{2}/\epsilon^{2}}{[1+b'' (0)/\epsilon ]^{3}}
\end{equation}
and
\begin{eqnarray}\label{d=3.1}
&& g_1(a_x) =  \frac{1}{A_{d}} \int_{p} c (p)^{2} h_{x} (p)^{2} \\
&& g_2(a_x) = \frac{1}{A_{d}} \int_{p} [ 2 c (p) i_{3} (p) h_{x}(p)^{2}
+ c (p)^{2}  h_{x}(p)  ]  \\
&& g_3(a_x) = \frac{1}{A_{d}} \int_{p} [ \frac{1}{2} i_{4} (p) h_{x} (p)
+ i_{3}(p)^{2} h_{x} (p)^{2} ]\ .
\end{eqnarray}
The other functions $g_i(a)$ and $\gamma$, which characterize 
non-zero temperature, are given in Appendix \ref{app:detailsbeta}.

Finally, note that the results (\ref{deltaBren2}) and (\ref{b1compact})
given above are for the choice $\zeta=0$. For a non-zero $\zeta =
\zeta_0 + \frac{1}{N} \zeta_1 + ...$ the result for $b_1[b](x)$ is
identical to (\ref{deltaBren2}), (\ref{b1compact}) above, up to 
the trivial linear rescaling term
\begin{eqnarray}\label{b1resc}
 b_1[b](x) \to b_1[b](x) - (2 \zeta_1 x b'(x) + 4 \zeta_1 b(x) ) \ln
 (m) \ .
\end{eqnarray}

%%%%%%%%%%%%%%%%%%%%%%%%%%%%%%%%%%%%%%%%%%%%%%%%%%%%%%%%%%%%%%%%%%%%%%
%%%%%%%%%%%%%%%%%%%%%%%%%%%%%%%%%%%%%%%%%%%%%%%%%%%%%%%%%%%%%%%%%%%%%%
%%%%%%%%%%%%%%%%%%%%%%%%%%%%%%%%%%%%%%%%%%%%%%%%%%%%%%%%%%%%%%%%%%%%%%
%%%%%%%%%%%%%%%%%%%%%%%%%%%%%%%%%%%%%%%%%%%%%%%%%%%%%%%%%%%%%%%%%%%%%%
%%%%%%%%%%%%%%%%%%%%%%%%%%%%%%%%%%%%%%%%%%%%%%%%%%%%%%%%%%%%%%%%%%%%%%
%%%%%%%%%%%%%%%%%%%%%%%%%%%%%%%%%%%%%%%%%%%%%%%%%%%%%%%%%%%%%%%%%%%%%%
%%%%%%%%%%%%%%%%%%%%%%%%%%%%%%%%%%%%%%%%%%%%%%%%%%%%%%%%%%%%%%%%%%%%%%
%%%%%%%%%%%%%%%%%%%%%%%%%%%%%%%%%%%%%%%%%%%%%%%%%%%%%%%%%%%%%%%%%%%%%%
%%%%%%%%%%%%%%%%%%%%%%%%%%%%%%%%%%%%%%%%%%%%%%%%%%%%%%%%%%%%%%%%%%%%%%
%%%%%%%%%%%%%%%%%%%%%%%%%%%%%%%%%%%%%%%%%%%%%%%%%%%%%%%%%%%%%%%%%%%%%%
%                       The beta-function                            %
%%%%%%%%%%%%%%%%%%%%%%%%%%%%%%%%%%%%%%%%%%%%%%%%%%%%%%%%%%%%%%%%%%%%%%
%%%%%%%%%%%%%%%%%%%%%%%%%%%%%%%%%%%%%%%%%%%%%%%%%%%%%%%%%%%%%%%%%%%%%%
%%%%%%%%%%%%%%%%%%%%%%%%%%%%%%%%%%%%%%%%%%%%%%%%%%%%%%%%%%%%%%%%%%%%%%
%%%%%%%%%%%%%%%%%%%%%%%%%%%%%%%%%%%%%%%%%%%%%%%%%%%%%%%%%%%%%%%%%%%%%%
%%%%%%%%%%%%%%%%%%%%%%%%%%%%%%%%%%%%%%%%%%%%%%%%%%%%%%%%%%%%%%%%%%%%%%
%%%%%%%%%%%%%%%%%%%%%%%%%%%%%%%%%%%%%%%%%%%%%%%%%%%%%%%%%%%%%%%%%%%%%%
%%%%%%%%%%%%%%%%%%%%%%%%%%%%%%%%%%%%%%%%%%%%%%%%%%%%%%%%%%%%%%%%%%%%%%
%%%%%%%%%%%%%%%%%%%%%%%%%%%%%%%%%%%%%%%%%%%%%%%%%%%%%%%%%%%%%%%%%%%%%%
%%%%%%%%%%%%%%%%%%%%%%%%%%%%%%%%%%%%%%%%%%%%%%%%%%%%%%%%%%%%%%%%%%%%%%
%%%%%%%%%%%%%%%%%%%%%%%%%%%%%%%%%%%%%%%%%%%%%%%%%%%%%%%%%%%%%%%%%%%%%%

\section{The $\beta$-function at order $1/N$}
\label{s:beta1overN}
\comment{*** VERSION FINALE POUR PIERRE, KAY est REPASS\'e *************}  

We are now ready to obtain the flow equation of the dimensionless
disorder, i.e.\ compute the $\beta$-function. Here we give the most
direct method to do so and give the results. A second method, closer
in spirit to the diagrammatic approach and which yields more compact
expressions has also been devised. Since it is rather involved it is
detailed in appendix \ref{a:betaKay}.

Our goal in the present paper is as follows.  The dimensionless
disorder $b(x)$ depends on the IR cutoff $m$, and a priori, also on
the UV cutoff $\Lambda$.  To obtain a FRG flow equation we want to
express:
\begin{eqnarray}\label{betaooinb0}
- m \partial_m b(x)&=& \beta [b] (x) = \beta_0 [b] (x) + \frac{1}{N}
  \beta_1 [b] (x) + \cdots
\end{eqnarray}
in terms of $b(x)$, at fixed $\Lambda$.  Furthermore, we are
interested in the behavior of the resulting expression when
$m/\Lambda$ becomes very small, which we hope can be made independent
of $\Lambda/m$, if necessary with appropriate redefinitions of
parameters.

To compute $\beta [b] (x)$ in the $1/N$ expansion we write
\begin{equation}\label{schema}
b(x) = b_{0}(x) + \frac{1}{N} b_{1}[b_{0}](x) +
O\left({\frac{1}{N^{2}}} \right)\ ,
\end{equation}
where $b_{0}(x)$ is the dimensionless disorder at leading order.  The
corresponding $\beta$-function was derived in
Ref.\ \cite{LeDoussalWiese2003a}.  It can be recovered by inserting
(\ref{changeofvar}) into (\ref{beta finite T N=oo}) using $m \partial_m
I_1=-2m^2 I_2$ and the above value of $I_2$ (for $\Lambda
=\infty$). In the variables $b$ it reads
\begin{eqnarray}\label{betaooinb}
- m \partial_m b_{0} (x)&=& \beta_{0} [b_{0}] (x)\nonumber \\
\beta_{0}[b] (x) &=& (\epsilon - 4 \zeta) b(x) + 2 \zeta x b'(x) + \frac{1}{2} b' (x)^{2}-b' (x) b' (0) + T_m \frac{b' (x)}{1
+\frac{b'' (0)}{\epsilon }} \ ,
\end{eqnarray}
where we recall $T_m = 4 T
A_{d} m^{\theta}/\epsilon$. Thus the $\beta$-function at leading order
has a simple and well-defined $\Lambda =\infty$ limit.

We now turn to the next order correction.

\subsection{The $\beta$-function at $T=0$}
We first detail the $T=0$ limit, which has a well-defined $\Lambda =\infty$ 
limit. At order $1/N$, deriving (\ref{schema}) w.r.t.\ $m$, we obtain
\begin{eqnarray}\label{schembet}
-m \partial_{m} b &=& \beta_{0}[b_{0}] +\frac{1}{N} \beta_{0}[b_{0}]\frac{\rmd
b_{1}[b_{0}]}{\rmd b_{0}}   + O\!\left(\frac{1}{N^{2}} \right)\ .
\end{eqnarray}
Replacing $b_{0}$ by $b$, using (\ref{schema}), we obtain the
$\beta$-function at
order $1/N$
\begin{eqnarray}\label{lf72a}
\beta [b] &=& \beta_{0} \! \left[b-\frac{1}{N}b_{1}[b] \right] 
+\frac{1}{N} \beta_{0}[b]\frac{\rmd 
b_{1}[b]}{\rmd b}   + O\!\left(\frac{1}{N^{2}} \right) \nonumber \\
&=& \beta_{0}[b] + \frac{1}{N}\left\{   \beta_{0}[b]\frac{\rmd 
b_{1}[b]}{\rmd b}  -  b_{1}[b]\frac{\rmd 
\beta _{0}[b]}{\rmd b} \right\}  + O\!\left(\frac{1}{N^{2}} \right) \ .
\end{eqnarray}
This expression is still symbolic. The derivative $\frac{\rmd}{\rmd b}
b_{1}[b]$ e.g.\ is the sum over the derivatives w.r.t.\ all
derivatives of $b$, and in the above expression is multiplied by the
$\beta$-function of the corresponding derivative, obtained by deriving
(\ref{betaooinb}) w.r.t.\ $x$. Since at $T=0$, $b_{1}[b](x)$ depends
only on $b'(x)-b'(0)$ and $b''(x)$, (\ref{lf72a}) gives:
\begin{eqnarray} \label{betasum}
 \beta[b]&=& \frac{\delta b_1[b](x)}{\delta b''[x]} \beta_0[b]''(x) +
\frac{\delta b_1[b](x)}{\delta(b'[x]-b'(0))} (\beta_0[b]'(x)-\beta_0[b]'(0)) 
\nonumber 
\\
&& - \Big\{ \epsilon b_1[b](x) + b_1[b]'(x) \left[ b'(x) -
b'(0)\right] - b'(x) b_1[b]'(0) \Big\}
\end{eqnarray}
Inserting the expression (\ref{deltaBren2}) with $T=0$ one finds:
\begin{eqnarray}\label{beta1overN} 
\beta ({b} (x)) &=& \epsilon{b} (x) + \half {b}' (x)^{2}-
{b}' (x) {b}' (0) + \frac{1}{N}  
\frac{1}{ A_{d}}\int\frac{\rmd ^{d}p}{(2\pi )^{d}}    \nonumber \\
%term 1:
&&\bigg\{ 2x{c(p)}^2{h_x(p)}^2[   x\epsilon  + 2x\epsilon h_x(p){i_2}(p)
+ 2b'(0) -  2b'(x) ]  \nonumber \\
%term 2:
&& + 4c(p){h_x(p)}^2{i_3}(p)[   b'(x) - b'(0) ] [  2x\epsilon +
2x\epsilon h_x(p){i_2}(p)+ b'(0) - b'(x) ] \nonumber \\
%term 3
&& + 2\epsilon {h_x(p)}^2[   3 +
  2h_x(p){i_2}(p) ] {{i_3}(p)}^2{[ b'(0) - b'(x) ]}^2 \nonumber \\
%term4
&&+ 2{c(p)}^2h_x(p)[ b'(x) - b'(0) ] [x\epsilon (1 + i_{2} (p) h_x(p))
+ b'(0) - b'(x) ] \nonumber \\
%term5
&& + \epsilon h_x(p)[ 2 + h_x(p){i_2}(p) ] {i_4}(p){[ b'(0) - b'(x)
]}^2
\bigg\}\ .
\end{eqnarray} 
A more compact expression can be found if one uses (\ref{b1compact})
as a starting point.  The first term on the l.h.s.\ of (\ref{betasum})
is replaced by $\frac{\delta b_1}{\delta a_x} (- m \partial_m^0 a_x)$
and one uses
\begin{eqnarray}
- m \partial_m^0 a_x = \epsilon a_x + \big[b'(x)-b'(0)\big] b''' (x) (
  a_x)^{2}= \epsilon a_x + \big[b'(x)-b'(0)\big] \partial_{x} a_x\ .
\end{eqnarray}
One finds a form rather similar to (\ref{b1compact}):
\begin{eqnarray} 
 - m \partial_m b &=& \epsilon b + \frac{1}{2} b'^2 - b' b'(0)\nonumber  \\
&& + \frac{1}{N} \bigg [ 2 x^2 \tilde g_{1} (a) + 2 x (b'(x)-b'(0))
\tilde g_{2} (a) + 2 (b'(x)-b'(0))^2 \tilde g_{3} (a) \bigg]
\label{betatildeg}
\end{eqnarray}
with 
\begin{eqnarray}\label{tilde g1}
\tilde g_{1} (a) &=&  \epsilon (a g_1'(a) - g_1(a))  \\
\tilde g_{2} (a) &=&  \epsilon a g_2'(a) - 2 g_1(a)   \\
\tilde g_{3} (a) &=& \epsilon ( g_3(a) + a g_3'(a)) - g_2(a) \ .
\label{tilde g3}
\end{eqnarray}
All terms proportional to $\partial_{x}a_{x}$ have canceled. Also note
that we have used analyticity in the derivation.  Issues related to
the non-analytic regime will be discussed in a subsequent publication.

\subsection{The $\beta$-function at non-zero temperature}
At non-zero temperature the $\beta $-function to order $1/N$ is
 independent of the UV cutoff $\Lambda$
only for $d<2$. Its expression is more complicated and we give here
only its form, for $d<2$. The derivation and explicit expressions for
the functions are given in Appendix \ref{app:detailsbeta}, together
with some comments about $d>2$.
\begin{eqnarray}
\beta_1[b](x) &=& \beta_1^{T=0}[b](x) + T_m \left[ x ( \tilde g_4(a_x) + a_0 
\tilde g_5(a_x) ) +
b'(x) ( \tilde g_6(a_x) + a_0 
\tilde g_7(a_x) + \tilde g_8(a_0) )\right] \nonumber \\
&& + T_m^2 \left[  \tilde g_{10}(a_x) + a_0 
\tilde g_{11}(a_x) + a_0^2 
\tilde g_{12}(a_x)  + x a'_0 (\epsilon - a_0) g_5(a_x)
+ \alpha b'(x)  \left(  \epsilon  g_7(a_x) + \phi(a_0) \right)
 \right] \nonumber \\
&& + T_m^3 \left[ b'(x) ( \psi(a_0) \alpha ^2 + \tilde \psi(a_0) \bar \alpha )
+ \alpha \epsilon^{2} ( g_{11}(a_x) + 2 a_0 g_{12}(a_x) )\right]
\label{finiteTbetaexpr}
\end{eqnarray}
where the $T=0$ expression $\beta_1^{T=0}[b](x)$ was given above.

\section{Conclusion}

\comment{**** KAY est REPASS\'E***}

In this article we have computed the effective action of the field
theory of random manifolds at large $N$.  The 2-replica part of this
quantity is what is needed to compute the renormalized disorder to
order $1/N$.  Although similar in spirit (one must compute the
determinant of fluctuations around a $N=\infty$ saddle point) the
problem solved here is much more complex than for the standard $1/N$
expansion (say in the $\phi^{4}$-model), first because one needs to
perform the calculation of fluctuations around the saddle point at
fixed averaged field value, second because this involves four-replica
matrices. It does however not involve spontaneous replica-symmetry
breaking of the Parisi type, but rather some type of simpler explicit
(vector) symmetry breaking.
  
To handle such additional difficulties we have introduced in this
paper two complementary methods. The first one is graphical and uses a
diagrammatics which is able to handle both the $O(N)$ and the replica
indices. In this diagrammatics the zero-temperature diagrams are
reasonably easy to compute. Much more subtle are the finite-$T$
diagrams. Interestingly, only order $T$ and $T^2$ are found to be
non-vanishing to this order in $1/N$. It is even simpler for $d=0$,
where the result for the $\beta$-function is polynomial. (This can be
derived by using Bogoliubov's subtraction operator.) The second method
uses an algebraic formula for the determinant of fluctuations around
the saddle point.  The algebra of the four-replica matrices is worked
out and one uses an expansion in number of free replica sums to
compute all components iteratively in a given order of this expansion.
Since we only need the two-replica part, this is a finite calculation,
although rather tedious and has to be performed with Mathematica. The
two methods are complementary, since any forgotten diagram of the
graphical method can be traced to some term in the algebraic result,
and vice versa. They are also, to our knowledge, new, henceforth the
detailed exposition.

Having obtained the effective action, we rewrote it in terms of the
dimensionless renormalized disorder. By varying with respect to the
infrared cutoff, we obtained the $\beta$-function to order $1/N$. We
noted that this $\beta$-function is UV finite at $T=0$.

It is important to note that the derivation was made, strictly
speaking, using an analytic action.  This is familiar for $N=\infty$,
where the same strategy was applied successfully: Although the
derivation was done in the analytic regime, the $\beta$-function could
then be continued to the non-analytic one. This was done via a
careful analysis of the solution when it reaches the Larkin scale. A
similar analysis will be performed in a forthcoming publication, together
with a comparison to the two-loop result, and a detailed analysis of the
physical consequences of the FRG flow derived here.

\subsection*{Acknowledgements} It is a pleasure to thank the KITP at
UC Santa Barbara, where part of this work was done.
K.W.\ greatly acknowledges financial support from the German National
Science Foundation DFG through Heisenberg grant Wi1932/1-1.

\appendix

\section{The large-$N$ formalism for the effective action}
\label{largeNgeneral}

\comment{*** VERSION FINALE POUR PIERRE, KAY est REPASS\'e *************} 

Before studying specific models, in the following sections, we first
present schematically the framework of the large-$N$ calculations. 

\subsection{General properties of the $1/N$ expansion}
The general problem can be formulated as follows. We want to compute the
effective action $\Gamma[u]$ defined as the Legendre transform
of ${\cal W}[J] = \ln {\cal Z}[J]$ ($J$ being the source
field conjugated to $u$), in the case where the
partition function can be written as:
\begin{eqnarray}
 {\cal Z}[J] = \int {\cal D} [\psi] \rme^{- N S[\psi,j]} \label{ZJ}
\end{eqnarray}
where $j=J/\sqrt{N}$ is the rescaled source and $\psi$ is some
auxiliary field (or a set of such fields).
Here all space coordinates and indices are
suppressed and integrals and sums implicit, in order
to exhibit the structure more clearly.

The first step is to write ${\cal W}[J]$ in an  $1/N$
expansion using the standard saddle point method. One finds:
\begin{eqnarray}
 {\cal W}[J] &=&  N \tilde W[j] \\
  \tilde W[j] &=& W^0[j] + \frac{1}{N} W^1[j] + \frac{1}{N^2} W^2[j] +
  \dots  \ ,
\label{WN} 
\end{eqnarray}
where the $j$-dependent saddle-point value $\psi_j$ of the auxiliary
field is solution of
\begin{eqnarray}
 S'_\psi[\psi_j,j] = 0 \label{SP}\ .
\end{eqnarray}
The expansion yields
\begin{eqnarray}
 W^0[j] &=& - S[\psi_j,j] \label{W0} \\
 W^1[j] &=& - \frac{1}{2} \tr  \ln S''_{\psi \psi} [\psi_j,j] \label{W1} \\
 W^2[j] &=& - \frac{1}{4!} S''''_{abcd}[\psi_j,j] \left<\phi_a \phi_b
\phi_c \phi_d\right>_{(S'')^{-1}} + \frac{1}{3!^2 2 }
S'''_{abc}[\psi_j,j] S'''_{efg}[\psi_j,j] \left<\phi_a \phi_b \phi_c
\phi_e
 \phi_f \phi_g\right>_{(S'')^{-1}}\ . \qquad 
\end{eqnarray}
More generally, the $W^n(j)$ are obtained from the loop expansion of
the field theory:
\begin{equation}
\ln Z[j] = N W^0[j]
+ \ln \int {\cal D} [\phi] \exp\left(- \frac{1}{2} \phi S''[\psi_j,j]
  \phi - \sum_{p=3}^{\infty} \frac{N^{1-\frac{p}{2}}}{p!}
\partial_{\psi_{a_1}} \dots \partial_{\psi_{a_p }}S[\psi_j,j]
\phi_{a_1} ...\phi_{a_p} \right) \ .
\end{equation}
In these formula, the indices $a$, $b$, \dots  summarize all spatial
coordinates, indices etc.\ of the field.

The effective action is then defined as:
\begin{eqnarray}
 \Gamma[u] &=& N \tilde{\Gamma}[v=u/\sqrt{N}] \\
 \tilde{\Gamma}[v] &=& v j_v - \tilde W[j_v] \ , \qquad v = \tilde W'[j_v] 
\end{eqnarray}
The equation for $j_v$ can formally be inverted into an expansion in
$1/N$ using (\ref{WN}), 
\begin{eqnarray}
 j_v = j^0_v + \frac{1}{N} j^1_v + \dots \ ,
\end{eqnarray}
which yields in turn the expansion for the effective action:
\begin{eqnarray}
  \tilde{\Gamma}[v] = \sum_{p=0}^{\infty} N^{-p} \tilde{\Gamma}^p[v]\ .
\end{eqnarray}
One finds that the leading order is simply the Legendre transform
of $W^0[j]$, 
\begin{eqnarray}
 \tilde{\Gamma}^0[v] = v j_v^0 - W^0[j_v^0] \ , \qquad v =
 (W^0)'[j_v^0] \ ,
\end{eqnarray}
where here and below $(W^n)'[j]=\partial_j W^n[j]$. 
Since one has $\partial_v \tilde{\Gamma}^0[v] = j_v^0$, it 
satisfies the self-consistent equation
\begin{eqnarray}
\tilde{\Gamma}^0[v] = v \partial_v \tilde{\Gamma}^0[v] 
- W^0(\partial_v \tilde{\Gamma}^0[v] )\ ,
\end{eqnarray}
and one has the usual relation between the second derivative matrices
$(W^0)''[j_v^0] = \left[ \frac{\delta \tilde{\Gamma}^0[v]}{\delta v
\delta v}\right]^{-1}$.

In this paper we use the result for the next order:
\begin{eqnarray}
\tilde{\Gamma}^1[v] = - W^1[j_v^0]  \label{gamma1}
\end{eqnarray}
For completeness we give the two next orders:
\begin{eqnarray}
 \tilde{\Gamma}^2[v] &=& - W^2[j_v^0] 
+ \frac{1}{2} (W^1)'[j_v^0] (\partial_j^2 (W^0)''[j_v^0]^{-1}
  (W^1)'[j_v^0]\nonumber 
 \\
& =& - W^2(j_v^0) + \frac{1}{2} \frac{\delta \tilde \Gamma^1 }{\delta
v} \left(\frac{\delta ^2 \tilde \Gamma^0}{\delta v \delta
v}\right)^{\!\!-1}
\frac{\delta \tilde \Gamma^1 }{\delta v}   \\
 \tilde{\Gamma}^3[v] &=& - W^3[j_v^0] 
+ (W^1)'[j_v^0] (W^0)''[j_v^0]^{-1} (W^2)'[j_v^0] \nonumber \\
&& - \frac{1}{2}
(W^1)'[j_v^0] (W^0)''[j_v^0]^{-1} (W^1)''[j_v^0] (W^0)''[j_v^0]^{-1}
(W^0)''[j_v^0]^{-1} (W^1)'[j_v^0] \nonumber 
\\
&& + \frac{1}{6} (W^0)'''[j_v^0] (W^0)''[j_v^0]^{-3} 
(W^1)'[j_v^0]^3 
\end{eqnarray}
where the following graphical rules allow to restore correctly all
index contractions and spatial integrals implicit in the schematic
notation above.  Denote $-\Gamma^n$ by a square with $n$ inside and
$W^n$ by a circle with a $n$ inside.  One treats the circled 1,2,3 etc
as vertices and considers all tree graphs.  A line is a propagator
$W_0''[j_v^0]^{-1}$ and is thus not a vertex (it is summed in a
line). The sum of these numbers is just the order.  The formula are
equivalent to:

\medskip

\centerline{\fig{12cm}{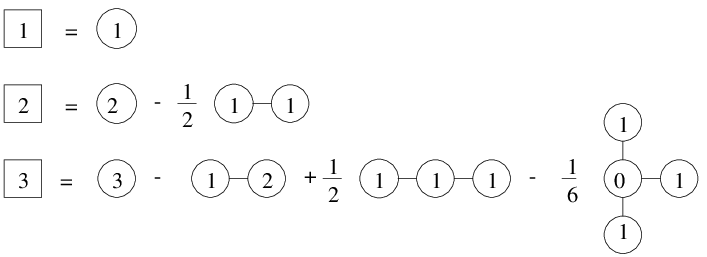}} \noindent This has, including the
combinatorial factors (from expanding the exponential and number of
choices), an immediate interpretation as a perturbative expansion with
all 1PI graphs subtracted.  In the last line, the 1-2 graph comes with
$1/2!$ in expanding the exponential, and adding the two possibilities,
the $1-1-1$ comes with 1/3!, but then there is a 3 for selecting the
middle one, and the last graph has a 1/3! from expanding the
exponential.  There is also a relative minus sign from each vertex
added.  To justify this graphical method, one simply recalls that quite
generally correlation functions, i.e.\ ${\cal W}[J]$ are given as the
sum of possible tree diagram made with $\Gamma[u]$ vertices. Since the
Legendre transform is involutive, the same is true for $\Gamma[u]$ in
terms of ${\cal W}[J]$ vertices. Thus to write $\tilde \Gamma^n[v]$
one must simply insert the proper orders in $1/N$ at each vertex, in
all possible ways, so as to match the total order.

Thus all the $\tilde{\Gamma}^p[v]$ can all be expressed as functions
of the $W^p(j_v^0)$. Inserting the results (\ref{W0}), (\ref{W1}), a.s.o.\ 
from the saddle point expansion, one obtains the $\tilde{\Gamma}^p[v]$
explicitly in terms of derivatives of the $S$ functional in
(\ref{ZJ}).

\subsection{Summary of main result}

\comment{***Kay rechecked****}

Before we detail the calculations in a more pedagogical
way for some specific models below, we first summarize here 
in compact notations the main result for the two lowest
orders of the $1/N$ expansion, with a generalization to
a bilocal bare action. 

Let us consider a $N$-component field 
theory whose action functional can
be written as:
\begin{eqnarray}
 S(\phi) &=& \frac{1}{2} \phi G^{-1} \phi + N S_{\mathrm{int}}[
 \psi_{xy} ]
 \\ \psi_{xy} &=& \frac{1}{N} \phi_x \cdot \phi_y
\ ,
\end{eqnarray}
where $S_{\mathrm{int}}$ is a functional of the bilocal field
$\psi_{xy}$ (which is also a bi-index matrix, if the field $\phi$
carries other indices (e.g.\ $a,b,\dots $).  Then its effective action can
be written as:
\begin{eqnarray}
 \Gamma[\phi] = \frac{1}{2} \phi G^{-1} \phi + N \Gamma^0[ \psi_{xy}]
+   \Gamma^1[ \psi_{xy}] + \dotsb 
\ ,
\end{eqnarray}
where $\Gamma^0$ is also a functional of a bilocal field and satisfies
the self-consistent equation:
\begin{eqnarray}
 \frac{\delta \Gamma^0}{\delta \psi_{zt}}[ \psi_{xy} ] &=& 
\frac{\delta S_{\mathrm{int}}}{\delta \psi_{zt}}[ \psi_{xy} + 
G[ \psi ]_{xy} ] \\
 G[ \psi ]_{xy} &=& \left\{  G^{-1} + 2 \frac{\delta \Gamma^0}{\delta
\psi}[ \psi ] \right\}^{-1}_{xy}
\end{eqnarray}
and 
\begin{eqnarray}
 \Gamma^1[\psi_{xy} ] = \frac{1}{2} \tr  \ln \Big[ 1_{zt,z't'} &+& 2
\frac{\delta^2 S_{int}}{\delta \psi_{zt} \delta \psi_{u v} }[
\psi_{xy} + G[ \psi ]_{xy} ] \times 
\nonumber  \\
& &\Big ( G[ \psi ]_{u t'} G[\psi]_{v z'} + \psi_{u t'}
G[ \psi ]_{v z'} + \psi_{v z'} G[ \psi ]_{u t'} \Big) \Big]
\ ,\qquad \qquad 
\end{eqnarray}
where $1_{xyab,ztcd}=\delta_{xz} \delta_{yt} \delta_{ac} \delta_{bd}$.

\section{Toy model} \label{toy} For pedagogical clarity, we will give
all details for the simpler case of the toy model.
\comment{***Kay est repass\'e sur cet appendix B***}

\subsection{Model and effective action to leading order}
We study the following $O(N)$ toy model, defined by the partition sum:
\begin{eqnarray}
 {\cal Z}[J] &=& \int {\cal D} [u]\, \rme^{ - S[u] + \sqrt{N} j u} =
 \int {\cal D}[u]\, \rme^{ - S[u,j]
} = \int {\cal D}[u]\,{\cal D} [\chi]\, {\cal D}[ \lambda]\, \rme^{ -
S[u,\chi,\lambda,j] } \nonumber \\
&=& \int {\cal D}
[\chi]\, {\cal D} [\lambda]\, \rme^{ - N S[\chi,\lambda,j] } 
\label{startoy} \\ 
S[u,j] &=& \int_{x} \frac{1}{2} (\nabla u)^2 + \frac{1}{2} m^2 u^2 + N
V\left(\frac{u^2}{N}\right) - \sqrt{N} j u \\ 
 S[u,\chi,\lambda,j] &=& S(u) -\int_{x} \left[\frac{1}{2} \lambda (N
 \chi - u^2)
 + \sqrt{N} j u  \right] \\ 
 S[\chi,\lambda,j] &=& \frac{1}{2} \tr \ln ( - \nabla^2 + m^2 +
\lambda) + \int_{x}\left[ V(\chi) - \frac{1}{2} \lambda \chi \right] -
\frac{1}{2}\int_{xy} j_{x} (- \nabla^2 + m^2 + \lambda)^{-1}_{xy}
j_{y} \ ,\qquad
\end{eqnarray}
where $\lambda(x)$ and $\chi(x)$ are local fields (the factor $i$ 
has been absorbed in the field $\lambda$). The above
expression (\ref{startoy}) is thus of the form (\ref{ZJ})
where $\psi=(\chi,\lambda)$ is a set of two auxiliary fields.

The saddle point equation (\ref{SP}) thus read:
\begin{eqnarray}
 \chi_j(x) &=& (- \nabla^2 + m^2 + \lambda_j)^{-1}_{xx} 
+ j_{y'} (- \nabla^2 + m^2 + \lambda_j)^{-1}_{y' x} (- \nabla^2 + m^2 +
\lambda_j)^{-1}_{x y}  j_y \\
 \lambda_j(x) &=& 2 V'( \chi_j(x) ) \label{sptoy}
\ .
\end{eqnarray}
From Appendix A one finds the dominant order:
\begin{eqnarray}
 W^0[j] = - S[\chi_j,\lambda_j,j] 
\end{eqnarray}
Using the saddle point equation one finds:
\begin{eqnarray}
 (W^0)'[j] = - \partial_j S[\chi_j,\lambda_j,j] = (- \nabla^2 + m^2 + \lambda_j)^{-1} j 
\end{eqnarray}
Thus one obtains: 
\begin{eqnarray}
 \Gamma(u) &=& N \tilde{\Gamma}(u/\sqrt{N}) \ ,\qquad v= u/\sqrt{N}  \\
 j_v &=& (- \nabla^2 + m^2 + \lambda) v \\
 \tilde \Gamma[v] &=& j_v v + S[\chi_{j_v},\lambda_{j_v},j_v] \nonumber \\
& =&  \frac{1}{2} v (- \nabla^2 + m^2 + \lambda) v 
+ \frac{1}{2} \tr  \ln ( - \nabla^2 + m^2 + \lambda_{j_v}) + N V(\chi_{j_v}) 
- \frac{1}{2} \lambda_{j_v} \chi_{j_v} 
\ .
\end{eqnarray}
One finds:
\begin{eqnarray}
 \tilde{\Gamma}(v) &=& \frac{1}{2} \tr  \ln G_v^{-1} +
\int_x  \frac{1}{2} (\nabla v)^2 + \frac{1}{2} m^2 v^2 - \frac{1}{2}
(\nabla^2 G_v)_{xx}
+ V(v_x^2 + (G_v)_{xx})  \\
 (G_v^{-1})_{xx'} &=& [ - \nabla_x^2 + m^2 + 2 V'(v_x^2 + (G_v)_{xx} )
 ]
\delta_{xx'} 
\ .
\end{eqnarray}
This can be simplified further as the complicated part of the
effective action is just a constant. First one notes that
$\tilde{\Gamma}(v) = \tilde{\Gamma}[G_v, v]$ satisfies:
\begin{eqnarray}
 \partial_{G} \tilde{\Gamma}[G, v]|_{G=G_v} = 0
\ .
\end{eqnarray}
As a consequence one sees that:
\begin{eqnarray}
 \frac{\delta  \tilde{\Gamma}[v]}{\delta v^i_x} =
(- \nabla^2 + m^2) v^i_x + 2 v^i_x V'\left(v_x^2 + (G_v)_{xx}\right)
\end{eqnarray}
The effective action (per unit volume $\Omega$) for a uniform
configuration of the field $v_x=v$ reads:
\begin{eqnarray}
 \frac{1}{\Omega} \tilde{\Gamma}(v) &=& \frac{1}{2} m^2 v^2 + V(v^2 +
G_v) + \frac{1}{2}
\int_q  \ln(q^2 + m^2 + 2 V'(v^2 + G_v) ) + \frac{q^2 + m^2 }{q^2 + 
m^2 + 2 V'(v^2 + G_v) }   \qquad\quad \\
 G_v &=& \int_q \frac{1}{q^2 + m^2 + 2 V'(v^2 + G_v) } 
\ ,
\end{eqnarray}
which can also be written as
\begin{equation}
\frac{1}{\Omega} \tilde{\Gamma}(v) = \frac{1}{2} m^2 v^2 + V(v^2 +
G_v) - G_v V'(v^2 + G_v)
+ \frac{1}{2}
\int_q  \ln(q^2 + m^2 + 2 V'(v^2 + G_v) ) 
\ .
\end{equation}
Here one can write:
\begin{eqnarray}
 \tilde{\Gamma}(v) &=& \tilde{\Gamma}[G_v, v^2] \\ \partial_{G}
 \tilde{\Gamma}[G, v^2]|_{G=G_v} &=& 0
\ .
\end{eqnarray}
For uniform configurations the effective action is simply a function
of $v^2$ such that
\begin{equation}
\half m^{2}+\tilde V' (v^{2}) = \frac{1}{\Omega} \frac{\rmd
\tilde{\Gamma}[v]}{\rmd v^2} = \frac{1}{2} m^2 + V'(v^2 + G_v)
\end{equation}
i.e.\ it satisfies the self-consistent equation
\begin{equation}\label{B24}
 \tilde V' (v^{2})=  V'\left(v^2 + \int_q \frac{1}{q^2 + 2
 \tilde V' (v^{2}) } \right) \ .
\end{equation}
This defines a renormalized potential $\tilde V(v^2)$, whose RG flow
is studied in Appendix I of \cite{LeDoussalWiese2003b}.

\subsection{$1/N$ corrections to the effective action}
To next order one has, from (\ref{W1}) :
\begin{eqnarray}
 W^1[j] = - \frac{1}{2} \tr  \ln S''[\chi_j,\lambda_j,j] 
\end{eqnarray}
where $\chi_j$, $\lambda_j$ are taken at their saddle point 
values (\ref{sptoy}). The symmetric matrix of second derivatives 
reads:
\begin{eqnarray}
 (S''_{\chi \chi})_{xy} &=& V''(\chi(x)) \delta_{xy} \quad , \quad (S''_{\chi \lambda})_{xy} = - \frac{1}{2} \delta_{xy} \\
 (S''_{\lambda \lambda})_{xy} &=& - \frac{1}{2} G_{xy} G_{yx} 
- (j G)_x G_{xy} (G j)_y  \\
 G_{xy} &=& (- \nabla^2 + m^2 + \lambda)^{-1}_{xy}
\ .
\end{eqnarray}
It can be put in the form:
\begin{eqnarray}
 S'' = - \frac{1}{2} \left(\begin{array}{cc}
A & 1\\
1 & B
\end{array} \right)\ .
\end{eqnarray}
Its inverse reads
\begin{eqnarray}
 (S'')^{-1} &=& - 2 
\left( \begin{array}{cc}
- B (1 - A B)^{-1} &   (1- BA)^{-1}\\
(1- A B)^{-1} & - A (1 - B A)^{-1 }
\end{array} \right)\\
 A_{xy} &=& - 2 V''(\chi(x)) \delta_{xy} \\
 B_{xy} &=& G_{xy} G_{yx} + 2 (j_v G)_x G_{xy} (G j)_y = G_{xy} G_{yx}
 + 2 v_x G_{xy} v_y\ ,
\end{eqnarray}
where the last equality holds only for $j=j^0_v$. 
Note that $A$ and $B$ do not commute (unless fields are uniform).

The result for the effective action is thus, from (\ref{gamma1})
\begin{eqnarray}
 \tilde{\Gamma}^1(v) &=& \frac{1}{2} \tr  \ln S''(\chi_{j_v^0} , \lambda_{j_v^0} , j_v^0) \\
 j^0_v &=& G^{-1} v \\
 G_{xy} &=& (- \nabla^2 + m^2 + \lambda_{j_v^0})^{-1}_{xy}
\end{eqnarray}
A more explicit form can be given for a uniform field configuration
$v_x=v$, with in that case:
\begin{eqnarray}
 A_{xy} &=& - 2 V''(\chi) \delta_{xy} \ , 
\qquad \chi=\chi_{j_v^0} = v^2 + G_{xx} \\
 B_{xy} &=& (G_v^{xy})^2 + 2 v^2 G_v^{xy} \\
 (S'')^{-1}(q) &=& \frac{2}{1 + 2 V''(v^2 + \int_k G_v(k)) [\Pi_v (q)
 + 2 v^2 G(q)] }
\left(\begin{array}{cc}
\Pi_v(q) + 2 v^2 G(q) & - 1 \\
-1 & - 2 V''(v^2 + \int_k G_v(k))
\end{array} \right)\nonumber \\
 \\
 \Pi_v(q) &=& \int_k G_v(q-k) G_v(k) \\
 G_v(k) &=& \frac{1}{k^2 + m^2 + 2 V'(v^2 + \int_k G_v(k))}
\end{eqnarray}
yielding finally
\begin{equation}\label{B.45}
 \tilde{\Gamma}^1(v) = \frac{1}{2} \int_q \ln\left(1 - A(q) B(q)\right) =
\frac{1}{2} \int_q \ln\left(1 + 2 V''\left(v^2 + \int_k
G_{v}(k)\right) \left( \Pi_v(q) + 2 v^2 G_q(v) \right) \right)
\end{equation}
up to a constant.

\section{Calculation of order $1/N$ for the random manifold}
\label{app:Details random manifold}
\comment{**** Kay est repass\'e*******}

We now sketch the derivation of $\tilde \Gamma^1$ for the
case of the random manifold.

It was shown in Ref. \cite{LeDoussalWiese2003b} that the partition sum
(\ref{lf19}) can be put in the form:
\begin{eqnarray}\label{lf85}
{\cal Z}[J] &=& \int D \chi D\lambda e^{ - N S[\chi,\lambda,j] } \\
 S[\chi,\lambda,j] &=& \frac{1}{2} \tr \ln ( C^{-1} + i
\lambda ) + \int_x  U(\chi(x)) - \frac{i}{2} \lambda^{ab}(x)
\chi^{ab}(x) \nonumber \\
&& - \frac{1}{2} \int_{x,x'} j_a(x) (C^{-1} + i \lambda)^{-1}_{a x, b
x'} j_b(x') \ , \qquad
\end{eqnarray}
where the inversion and trace are performed in both replica space and
spatial coordinate space. It has again the form (\ref{ZJ}) where
$\psi=(\chi^{ab}(x),\lambda^{ab}(x))$ is a set of 2-replica-matrix
auxiliary fields. The saddle-point equation (\ref{SP}) reads (see
Ref.\ \cite{LeDoussalWiese2003b}):
\begin{eqnarray}
  \chi_j^{ab}(x) &=& (G_j)_{ax,bx} + (G_j :j)_{a x} \cdot (G_j :j)_{b
x}\qquad  \\ 
 i \lambda_j^{ab}(x) &=& 2 \partial_{ab} U (\chi_j(x)) \\
 G_j^{-1} &=& C^{-1} + i \lambda_j \label{spW} \ ,
\end{eqnarray}
where $G_j$ is a matrix with both replica indices and spatial
coordinates and inversion is carried out for both. Here and below,
replica indices are raised whenever explicit dependency is given,
e.g.\ $\chi_{ab} \equiv \chi_j^{ab}$. The notation for the
$N$-component vector $(G:j)^i_{b x} = \sum_c \int_y G_{b x,c y}
j^i_c(y)$ is a shorthand for a matrix product, and everywhere we
denote by
\begin{equation}\label{lf24}
 \partial_{ab} U(\phi) := \partial_{\phi_{ab}} U(\phi)
\end{equation}
the simple derivative of the function $U(\phi)$ with respect
to its matrix argument $\phi_{ab}$. 

The dominant order, $W^0[j]$ and $\tilde \Gamma^0[v]$, was computed in
Ref.\ \cite{LeDoussalWiese2003b}, thus here we study only the next
order. It is given by:
\begin{eqnarray}
 \tilde{\Gamma}^1[v] &=& \frac{1}{2} \tr  \ln S''[\chi_{j^0}, 
\lambda_{j^0},j^0]   \\
 j^0_{ax} &=& (G_v^{-1}  : v)_{ax} \\
 (G_v^{-1})_{ax,by} &=& (C^{-1})_{x,y} \delta_{ab}
+ 2 \partial_{ab} U(\chi_v(x)) \delta^d(x-y)\ .
\end{eqnarray}
Note that when computing the fluctuations around the saddle point
we consider $\chi_{ab}$ and $\chi_{ba}$ as independently fluctuating fields,
symmetry being restored at the saddle point (it is also possible to perform
the calculation with symmetric matrices only). 

The matrix of second derivatives can again be put in the form (see the
previous section):
\begin{eqnarray}
 S'' &=&
=- \frac{1}{2} \left( {A \atop i1} {i 1 \atop -B} \right)\ ,
\end{eqnarray}
 and one can show that:
\begin{eqnarray} 
 \tr  \ln S'' &=& \tr  \ln (1 - A B) \\
 A_{abx, cdy} &=& - 2 \partial_{\chi_{ab}}  \partial_{\chi_{cd} }
 U(\chi(x)) \delta_{xy} \\
 B_{abx, cdy} &=& G_{a x, d y} G_{c y, b x} + (j : G)_{c y}  G_{a x, d
 y} (G : j)_{b x}  + (j : G)_{a x}  G_{bx, c y} (G: j)_{d y} + M_{ab,cd}
 \nonumber \\
&=& G_{a x, d y} G_{c y, b x} + v_{b x} G_{a x, d y} v_{c y} + v_{a x}
G_{b x, c y} v_{d y} + M_{ab,cd}
\end{eqnarray}
and note that $A$ and $B$ do not commute (unless fields are
uniform). The last equality is valid only when $j=j^0_v$ is inserted. 

For uniform fields one finds:
\begin{eqnarray}
 \tilde{\Gamma}^1[v] &=& \frac{1}{2} \int_q \ln(1 - A(v) B_q(v)) \\
 A(v)_{ab,cd} &=& - 2 \partial_{\chi_{ab}}  \partial_{\chi_{cd} } U(\chi_v) \\
\label{C35}
 B_q(v)_{ab,cd} &=& \int_k G_v^{a d}(k) G_v^{c b}(q-k) + v_b v_c G_v^{a d}(q) 
+ v_a v_d G_v^{b c}(q) \\
 G_v^{ab}(k) &=& [C(k)^{-1} + i \lambda_v ]^{-1}_{ab}  \\
 \chi_v^{cd} &=& v_c v_d + \int_k G_v^{cd}(k) \\
 i \lambda_v^{ab} &=& 2 \partial_{ab} U(\chi_v)  \ .
\end{eqnarray}
We note that the above matrices are {\em not} a priori symmetric in
$a\leftrightarrow b$ or in $c\leftrightarrow d$ since we have chosen
the representation where all components of the fields fluctuate
freely. The alert reader will notice later (when coming back to the
main text) that in fact the final matrix $A B_q$ will possess such a
symmetry.

\comment{
*** JAIMERAIS GARDER CA EN REMARQUE A LA FIN.. 
****PIERRE A NETTOYER!!!!!!!!****

**** Kay says: Pierre, if you want to keep this, please explain, and
     put a proper notation. Gamma has now additional argument,
     w.r.t. which you do or do not do a Legendre???  ****
Let us add a remark on a more general definition of the effective action
$\Gamma[q,v]$ with an additional source $- h_{ab} \chi_{ab}$ 
(Attention: Factor etc. wrong. *** KAY RECHECK....) 
We compute $\Gamma[q,v]$ with a source $- h_{ab} \chi_{ab}$
to dominant order..

\begin{eqnarray}
 \partial_{v} \Gamma &=& (m^2/2T + U'(q) - \partial_q \Gamma(q,v)) v \\
 q &=& v^2 + \int_k ( (k^2 + m^2)/2T +  U'(q) - \partial_q \Gamma(q,v))^{-1}
\end{eqnarray}
$q=\chi$. We cannot use the saddle-point equation obtained by
 taking a derivative w.r.t.\ $\chi$,
but we can use the other one. CLARIFY.. 
}

%%%%%%%%%%%%%%%%%%%%%%%%%%%%%%%%%%%%%%%%%%%%%%%%%%%%%%%%%%%%%%%%%%%%%%
%%%%%%%%%%%%%%%%%%%%%%%%%%%%%%%%%%%%%%%%%%%%%%%%%%%%%%%%%%%%%%%%%%%%%%
%%%%%%%%%%%%%%%%%%%%%%%%%%%%%%%%%%%%%%%%%%%%%%%%%%%%%%%%%%%%%%%%%%%%%%
%%%%%%%%%%%%%%%%%%%%%%%%%%%%%%%%%%%%%%%%%%%%%%%%%%%%%%%%%%%%%%%%%%%%%%
%%%%%%%%%%%%%%%%%%%%%%%%%%%%%%%%%%%%%%%%%%%%%%%%%%%%%%%%%%%%%%%%%%%%%%
%%%%%%%%%%%%%%%%%%%%%%%%%%%%%%%%%%%%%%%%%%%%%%%%%%%%%%%%%%%%%%%%%%%%%%
%%%%%%%%%%%%%%%%%%%%%%%%%%%%%%%%%%%%%%%%%%%%%%%%%%%%%%%%%%%%%%%%%%%%%%
%%%%%%%%%%%%%%%%%%%%%%%%%%%%%%%%%%%%%%%%%%%%%%%%%%%%%%%%%%%%%%%%%%%%%%
%%%%%%%%%%%%%%%%%%%%%%%%%%%%%%%%%%%%%%%%%%%%%%%%%%%%%%%%%%%%%%%%%%%%%%
%%%%%%%%%%%%%%%%%%%%%%%%%%%%%%%%%%%%%%%%%%%%%%%%%%%%%%%%%%%%%%%%%%%%%%
%                  The excluded replica-formalism                    %
%%%%%%%%%%%%%%%%%%%%%%%%%%%%%%%%%%%%%%%%%%%%%%%%%%%%%%%%%%%%%%%%%%%%%%
%%%%%%%%%%%%%%%%%%%%%%%%%%%%%%%%%%%%%%%%%%%%%%%%%%%%%%%%%%%%%%%%%%%%%%
%%%%%%%%%%%%%%%%%%%%%%%%%%%%%%%%%%%%%%%%%%%%%%%%%%%%%%%%%%%%%%%%%%%%%%
%%%%%%%%%%%%%%%%%%%%%%%%%%%%%%%%%%%%%%%%%%%%%%%%%%%%%%%%%%%%%%%%%%%%%%
%%%%%%%%%%%%%%%%%%%%%%%%%%%%%%%%%%%%%%%%%%%%%%%%%%%%%%%%%%%%%%%%%%%%%%
%%%%%%%%%%%%%%%%%%%%%%%%%%%%%%%%%%%%%%%%%%%%%%%%%%%%%%%%%%%%%%%%%%%%%%
%%%%%%%%%%%%%%%%%%%%%%%%%%%%%%%%%%%%%%%%%%%%%%%%%%%%%%%%%%%%%%%%%%%%%%
%%%%%%%%%%%%%%%%%%%%%%%%%%%%%%%%%%%%%%%%%%%%%%%%%%%%%%%%%%%%%%%%%%%%%%
%%%%%%%%%%%%%%%%%%%%%%%%%%%%%%%%%%%%%%%%%%%%%%%%%%%%%%%%%%%%%%%%%%%%%%
%%%%%%%%%%%%%%%%%%%%%%%%%%%%%%%%%%%%%%%%%%%%%%%%%%%%%%%%%%%%%%%%%%%%%%
\section{The excluded replica-formalism}\label{excludedrepformalism}

\comment{*** KAY est repasse, c'est maintenant Pierre qui DOIT REPASSER
************}

In this section, we calculate the terms at order $T$ and $T^{2}$ using
the excluded replica-formalism. This gives an independent derivation
of the moon-shaped building blocks $\textdiagram{halfmoon}$ and
$\textdiagram{moon}$ in section \ref{buildingblocks}.

We start by recalling that corrections at finite temperature are more
difficult to obtain. The simplest example is given in the next
appendix \ref{2looptadpole}. Here we want to understand this by making
two contractions between $B (u_{a} (x) -u_{b} (x))$ and $B (u_{a} (y)
-u_{b} (y))$. Be the first contraction (focusing on the
replica-structure)
\begin{equation}\label{lf47}
\diagram{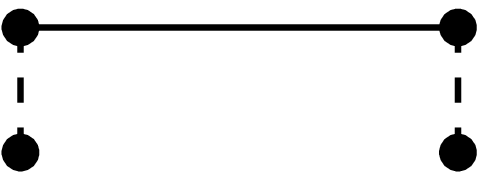} \ .
\end{equation}
The following possible contractions are
\begin{equation} \label{e2}
\diagram{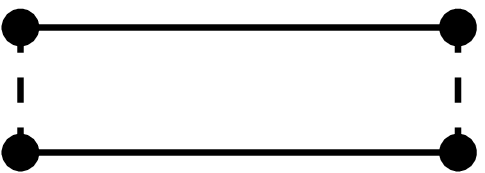} +\diagram{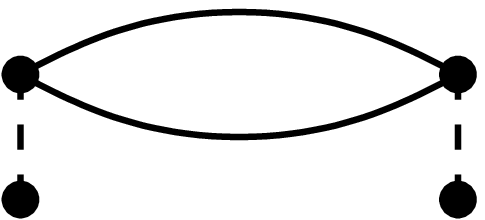}-
\diagram{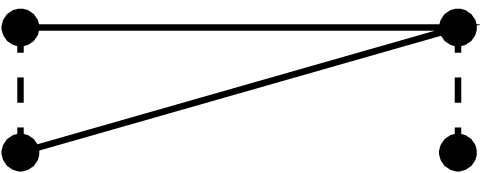}- \diagram{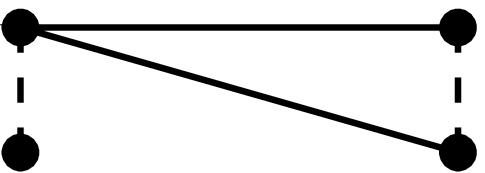}\ .
\end{equation}
At zero temperature, we have found in section \ref{T=0} that 
only the first term contributes. This enabled us to do the calculation
at $T=0$. 

The general case is more complicated, since there are now three more
possible contractions, one may draw, i.e.\ 4 for each $B''$. This
looks discouraging. Two ways out of the dilemma can be thought of.
The first, called {\em recursive construction} or {\em successive
construction}, tries to add one more link to a chain of $B''$. The
decisive simplification then is that whenever arriving at e.g.
\begin{equation}\label{lf48}
\diagram{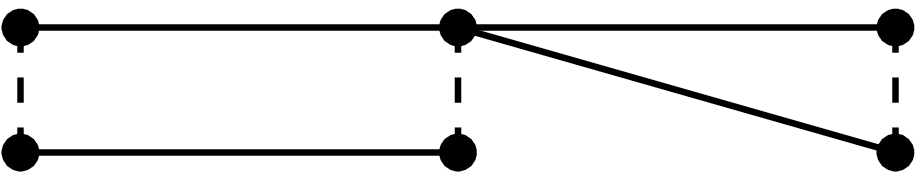}
\end{equation}
no more links can be added to the right, since the rightmost vertex is
$B''_{aa}$, which is a constant. This procedure has however one
{\em crucial deficiency}: Contractions are non-commutative! Let us
illustrate this on the  diagram 
\begin{equation}\label{lf49}
\diagram{e2}\ .
\end{equation}
If one first does the two leftmost contractions, then the two
rightmost contractions are not possible. On the other hand, when doing
first the two rightmost contractions, the two leftmost are
possible! Working with this formalism, i.e.\ using
its implied simplifications, one has to number the lines. This is
pretty awkward, since one wants to be able to interpret diagrams as
such, without having to number lines. This is especially important
here, since one wants to recognize the chains which give rise to
$H_{ab} (p)$. Also note that problems arise at finite temperature,
since there one has  additional lines to add, and not all lines are
needed to get a 2-replica term, i.e.~there are lines which can be
discarded, and still the diagram would be a 2-replica term.
A very instructive example is the simplest 2-loop order diagram at
finite temperature, which is derived in appendix
\ref{2looptadpole}.
One sees that diagrams can be grouped differently to cancel, and that
these different cancellations correspond to different paths in the
contractions.

One way out of the dilemma is the {\em excluded replica approach},
which at 2-loop order has as its descendent the sloop-algorithm
\cite{LeDoussalWieseChauve2002}. Here we explain the {\em excluded
replica approach}. We start by writing the trivial relation
\begin{equation}\label{lf50}
\sum_{a,b}B_{ab} = \sum_{a\neq b}B_{ab} + \sum_{a}B_{aa} \ .
\end{equation}
The first term on the r.h.s.\ is the excluded replica disorder, the
second a constant. When performing contractions, 
the second term does not contribute. Thus one can perform perturbation
theory instead with $\sum_{ab}B_{ab}$ with $\sum_{a\neq b}B_{ab}$. The
big advantage is that the last two terms in (\ref{e2}) do not
contribute.  The backdraw is that the final result is a sum over
excluded replicas, which has to be projected onto the 2-replica term,
and one may have more terms at intermediate steps. The final
projection can be done formally, by replacing $B_{ab}\to B_{ab}
(1-\delta_{ab})$, expanding and then collecting the 2-replica
contributions.

We now introduce the excluded replica-formalism. 
Recalling that
\begin{equation}\label{lf58}
\stackrel{{p} }{\longrightarrow}
{ _a \atop ^b}\diagram{Bsummed}{ _a \atop ^b} = 
\frac{1}{1-4 I_{2}(p)\, B''(\chi_{ab})} \ { _a \atop ^b}
\diagram{Bpp}{ _a \atop ^b} = \frac{1}{2} \left[(u_{a}-u_{b})^{2}
\right] H_{ab}(p) 
 \left[(u_{a}-u_{b})^{2}\right]
\ .
\end{equation}
We then proceed as follows: Whenever we have a long chain with some
replica conserving lines (as above) and with doubled lines (or single
lines when inserting vertices), we will always sum the double-line
part, resulting in $H_{ab}$. We then have to sum explicitly the rest,
and project it onto the 2-replica-contribution (in general). For
illustration, we start with the open chain. This is, say when fixing
one end to have replicas $a$ and $b$
\begin{equation}\label{D7}
\stackrel{\vec{p} }{\longrightarrow} { _a \atop ^b}\diagram{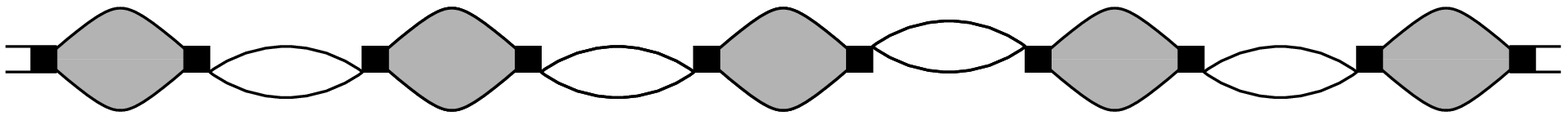} 
\ .
\end{equation}
We furthermore observe that the combinatorial weight for either having a
replica-conserving double line ($\diagram{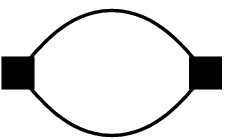}$), or a doubled
line up ($\diagram{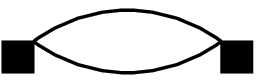}$) or down
($\diagram{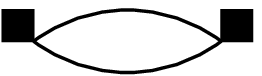}$), is the same. We do not draw crossed
lines; by definition they are incorporated into the combinatorial
factor.  The above has to be projected onto the 2-replica sum. Writing
any $H_{cd} (p)$ with excluded indices as $H_{cd} (p) (1-\delta
_{cd})$, and multiplying all terms, one always has to take the term
with the $\delta _{cd}$, and (\ref{D7}), summed over all lengths of
the chain, is
\begin{eqnarray}\label{lf59}
\sum_{\mbox{\footnotesize all lengths}}
\stackrel{\vec{p} }{\longrightarrow} { _a \atop ^b}\diagram{openchain}
& =&H_{ab} (p) \sum_{n=0}^{\infty } (- 4 I_{2} (p) H_{aa})^{n}\nn\\
&=& H_{ab} (p) \frac{1}{1+4 I_{2} (p)H_{aa} (p)} \nonumber \\
&=& H_{ab} (p) \left(1-4 I_{2} (p)B''_{aa} \right)\ .
\end{eqnarray}
This object is well known: It is the chain times the moon-diagram, see
table \ref{f:buildingblocks}. We remark the important point that,
even though we have an infinity of diagrams to sum up, in the
projection onto 2 replicas (or since we are already fixing two
replicas, maybe we should better say onto one replica) all but 1
non-trivial term vanish. (Also remark that we have been a bit sloppy
to amputate the last $u^{2}$ legs. Otherwise, that is not so easy to
represent.)

We need another intermediate result. Summing all diagrams which
conserve the lower index, we obtain
\begin{equation}\label{S1}
\diagram{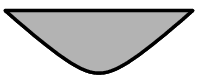} := \diagram{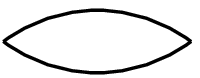}+\dots +
\diagram{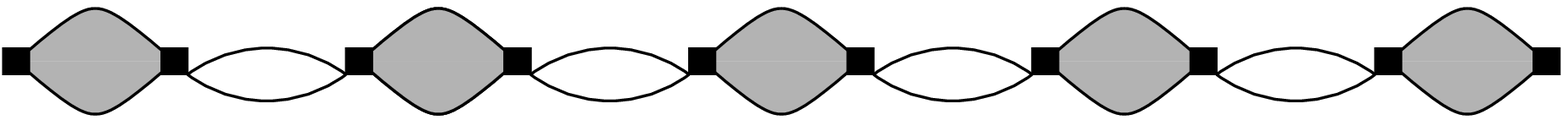}+ \dots
\ .
\end{equation}
We want this sum $\diagram{lowerhalfchain}$, projected onto 1 replica,
$\diagram{lowerhalfchain}_{\!\!\!1}$. Using the same procedure as
above, we find 
\begin{eqnarray} \label{S2}
{\cal S} (p):= \diagram{lowerhalfchain}_{\!\!\!1\ } &=& 2 I_{2} (p)
 \sum _{n=0}^{\infty} (-2 I_{2} (p) H_{aa})^{n} \nonumber \\
&=&  2 I_{2} (p)\frac{1}{1+2 I_{2} (p)H_{aa}}\nonumber \\
&=&
 2 I_{2} (p) \frac{1-4 I_{2} (p)B''_{aa}}{1-2 I_{2} (p)B''_{aa}}
\ .
\end{eqnarray}
The 2-replica contribution is 
\begin{equation}\label{S3}
\diagram{lowerhalfchain}_{\!\!\!2\ }
=\diagram{lowerhalfchain}_{\!\!\!1\ }\times
\diagram{Bsummed}_{\!\!\!2\ }\times \diagram{lowerhalfchain}_{\!\!\!1\
} = {\cal S} (p) H_{ab} (p) {\cal S} (p)
\ .
\end{equation}
This half-chain  is very practical, since when  inserted into a more
complicated diagram, no further restrictions apply. In a diagram
involving this half-chain and some  ``rest'', there are two  
2-replica contributions: The half-chain projected onto the
1-replica part times the rest projected onto 2 replicas (taking care
of the replica-conserving line of the half-chain) or the 2-replica
part of the half-chain times the 1-replica-part of the rest.

We continue on the order $T^{2}$-term. We have the following terms
before projection:
\begin{equation}\label{tbp}
\diagram{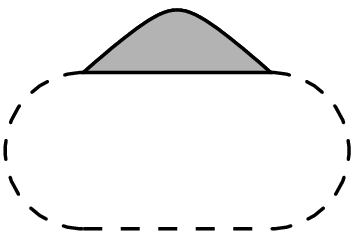}+
\diagram{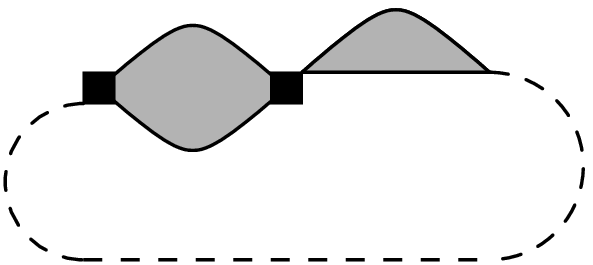}+\frac{1}{2}\diagram{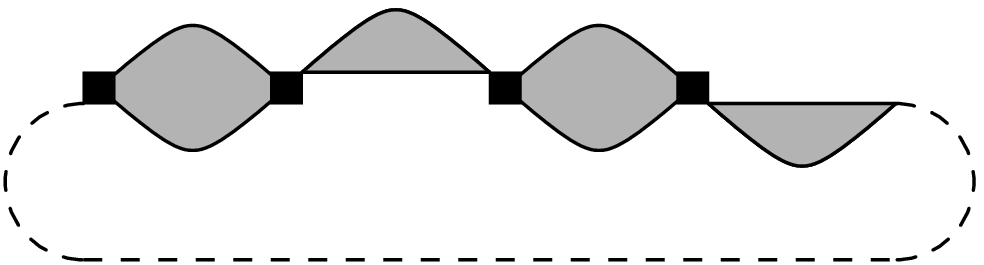}+\frac{1}{3}
\diagram{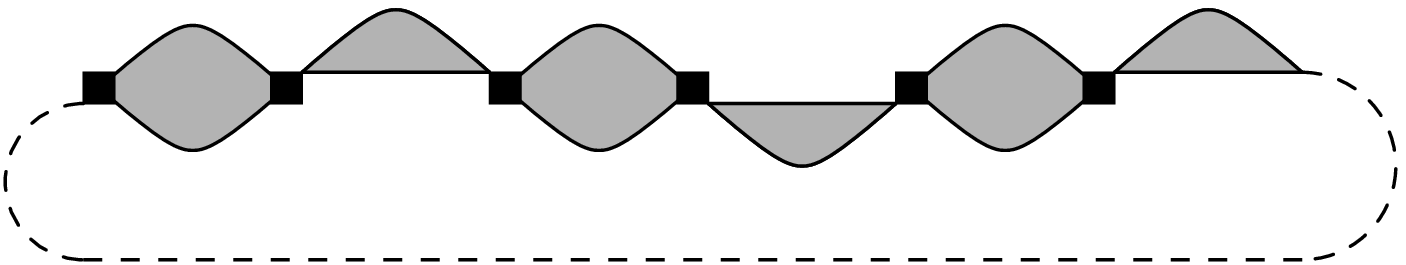}+\ldots
\ .
\end{equation}
A dashed line denotes the 
identity; one would like to print this on a torus.
The first term is very symbolic, since it is not a geometric sum but 
a log, and there should be at least one vertex.
The second is 0 (coinciding replicas in the chain)
\begin{equation}\label{lf60}
\diagram{logchain1} = 0\ .
\end{equation}
We now have the choice of where to project onto the 2-replica
term. The rest has to be projected onto the 1-replica term.  We first
project one of the half-chains onto a 2-replica-term, using
(\ref{S3}). The 2-replica term is shaded in dark grey. The rest
consequently has to be projected onto the 1-replica terms, shaded in
light grey.  We obtain the class ${\cal C}_{1}$:
\begin{eqnarray}
{\cal C}_{1}&=&\diagram{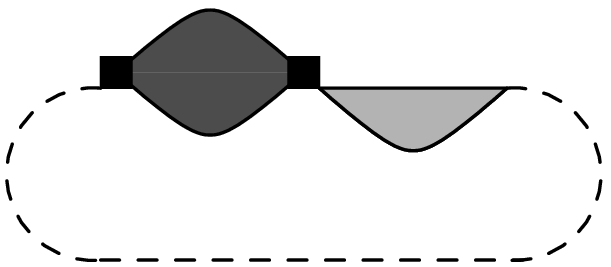}+\diagram{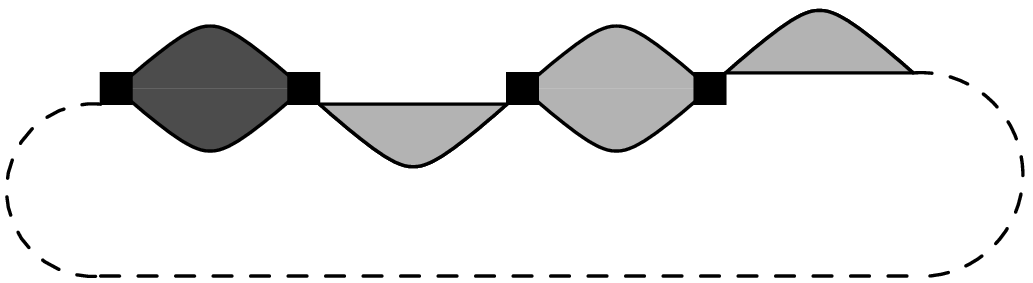}+
\diagram{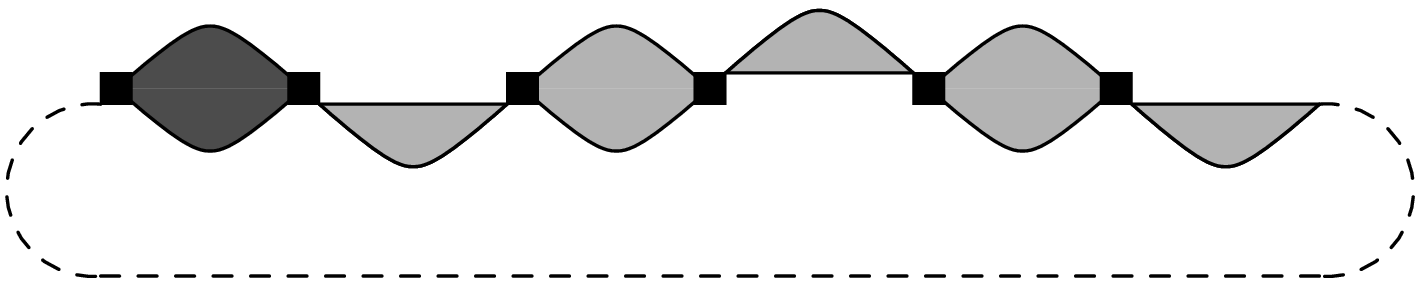}\nonumber \\ 
&&\quad + \diagram{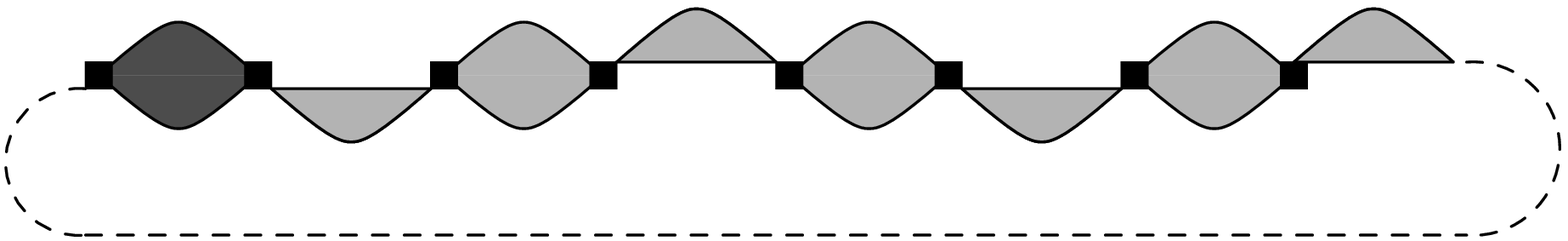} + \dots \label{C1}
\ .
\end{eqnarray}
Denote by $n$ the number of $H_{ab}$ in each diagram, which have lines
starting at different replicas. Each such $H_{ab}$ contributes a factor of
$(1-\delta_{ab})$, for a  product of 
\begin{equation}\label{topo}
\sum _{i_{1},i_{2},i_{n}}(1-\delta _{i_{1}i_{2}}) (1-\delta
_{i_{2}i_{3}})\dots (1-\delta _{i_{n}i_{1}})
\ .
\end{equation}
We need all indices to be restricted to be identical. 
The terms which lead to that restriction are 
either the product of all $\delta$'s (1 term) or any one of the $\delta
$'s left out.  This gives another $n$ terms, but with a different
sign, for a total of 
\begin{eqnarray} \label{lf61}
{\cal C}_{1}= {\cal S} (p) H_{ab} (p)\sum _{n=0}^{\infty} (1-n)
\left[- H_{aa} (p) {\cal S} (p)\right]^{n} \ ,
\end{eqnarray}
where it is important to note that also the first term in (\ref{C1})
for $n=0$ is correctly given by the above formula.  We note the
auxiliary sum %(Mathematica-proof)
\begin{equation}\label{lf62}
\sum _{n=0}^{\infty} (1-n) (-x)^{n} = \frac{1+2x}{( 1+x)^{2}}
\ .
\end{equation}
We give the intermediate results
\begin{eqnarray}\label{jkl}
\frac{1}{1+H_{aa} (p){\cal S}(p) } &=& 1-2 I_{2} (p)B''_{aa}\\ \label{im2}
\frac{{\cal S}(p) }{1+H_{aa} (p){\cal S}(p) } &=& 2 I_{2} (p) ( 1-4
I_{2} (p)B''_{aa})\\ 
1+ 2 H_{aa} (p){\cal S}(p) &=&\frac{1+2 I_{2} (p)B''_{aa}}{1-2 I_{2}
(p)B''_{aa}} \label{im3}
\end{eqnarray}
the above sum is %(M-proof)
\begin{eqnarray}\label{lf63}
{\cal C}_{1}= 2 I_{2} (p) H_{ab} (p) (1+2I_{2} (p) B''_{aa}) (1-4I_{2}
(p)B''_{aa}) \ .
\end{eqnarray} 
This reproduces diagram (\ref{1oNT2b}).

We now turn to the diagrams where the half-chain is always projected
onto the 1-replica contribution, whereas the non-trivial terms come
from omitting a $\delta $ belonging to one of the $H_{ab}$ with lines
entering into different replicas. With the same shading for 1- and
2-replica terms as above, this is 
\begin{eqnarray}\label{lf64}
{\cal C}_{2}=\diagram{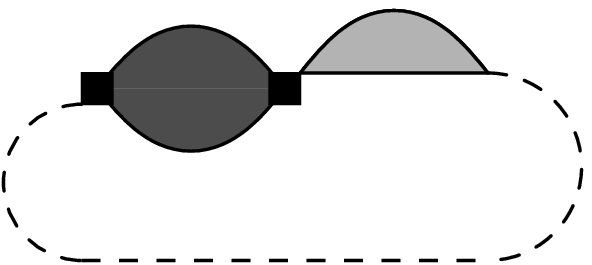}+
\diagram{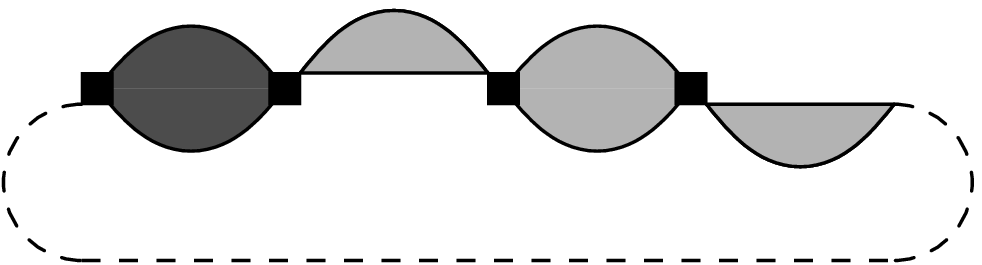}+
\diagram{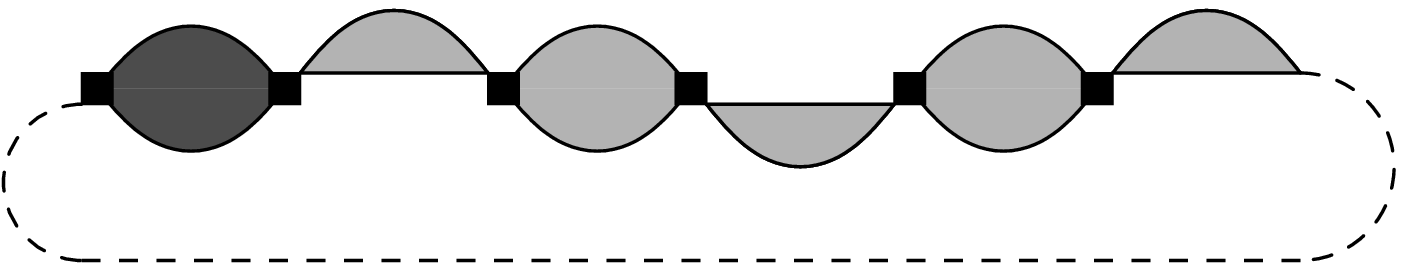}+\ldots \label{4.74}
\end{eqnarray}
%\begin{eqnarray}
%{\cal C}_{2}=\diagram{logchain1}+\frac{1}{2}
%\diagram{logchain2}+\frac{1}{3}
%\diagram{logchain3}+\ldots
%\end{eqnarray}
%projected onto replicas. 
%Note the change in combinatorics, induced by the choice of the
%2-replica term. 
The first term is 0, since it necessarily has crossed indices:
\begin{equation} \label{lf65}
\diagram{logchain1black}=0\ .
\end{equation}
Each $H_{ab}$ which is projected onto a single replica again
contributes a factor of $(1-\delta_{ab})$ Starting from the second
diagram, we have to leave out exactly 2 $\delta $'s in
\begin{equation}\label{lf66}
\frac{1}{n}( 1-\delta _{i_{1}i_{2}}) (1-\delta
_{i_{2}i_{3}})\dots (1-\delta _{i_{n}i_{1}})\ .
\end{equation}
The first has already been left out in plotting (\ref{4.74}),
accounting for the  factor of $n$. Leaving out two $\delta$'s leads to
a factor of $n(n-1)/2$, since they are indistinguishable. 
The result contains two factors of  $H_{ab}$; the remaining factors
are all  $H_{aa}$:
\begin{equation}\label{lf67}
{\cal C}_{2} = \frac{1}2  H_{ab} (p)^{2}{\cal S}(p) ^{2} \sum
_{n=2}^{\infty } (n-1) [-H_{aa} (p){\cal S}(p) ]^{n-2}
\ .
\end{equation}
With the sum 
\begin{equation}\label{lf68}
\sum_{n=2}^{\infty } (n-1) ( -x)^{n-2} = \frac{1}{(1+x)^{2}}\ ,
\end{equation}
and using (\ref{S2}) and (\ref{jkl})  we obtain
\begin{equation}\label{lf69}
{\cal C}_{2} = 2 I_{2} (p)^{2} H_{ab} (p)^{2} \left(1-4 I_{2}
(p) B''_{aa}\right)^{2} \ . 
\end{equation}
This reproduces the term (\ref{1oNT2c}).

We now turn to the diagrams of order $T$.  First of all, note that the
following two diagrams have no contributions proportional to
$B''_{aa}$:
\begin{equation}\label{5.1}
\diagram{1oNT1a} \qquad \mbox{and}\qquad \diagram{1oNT1b}
\ .
\end{equation}
The reason is that they have no doubled line, as in 
\begin{equation}\label{5.2}
\diagram{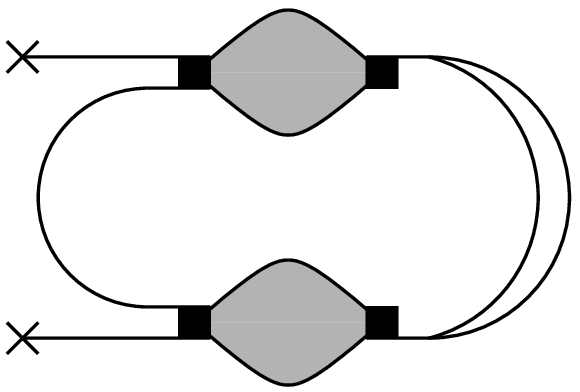} \qquad \mbox{and}\qquad \diagram{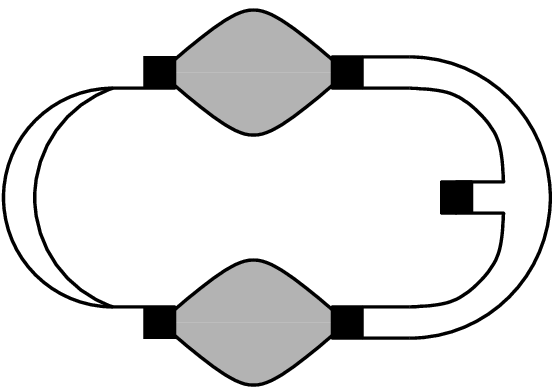}
\ .
\end{equation}
Note that both diagrams necessitate a replica-conserving double line
(i.e.$\diagram{doubledlineup}$ or
$\diagram{doubledlinedown}$). Otherwise they vanish. The
replica-conserving double line can be more generally replaced by all
chains, which start and end with a double line, and which conserve the
index running through.    The
chains with one index entering and one index exiting are
\begin{equation}\label{5.3}
\diagram{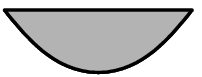}+ \diagram{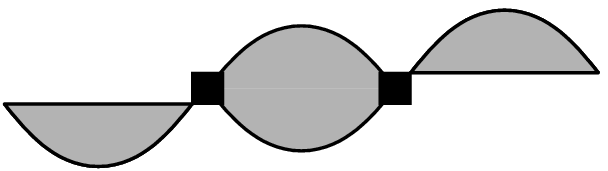} + \diagram{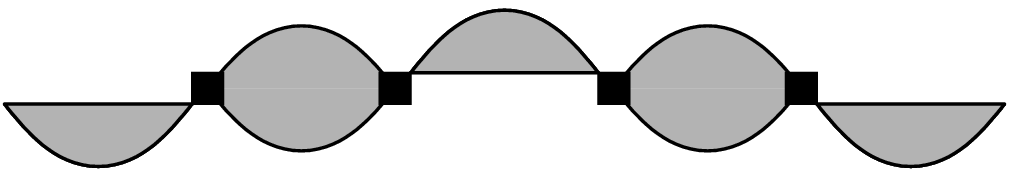} + \diagram{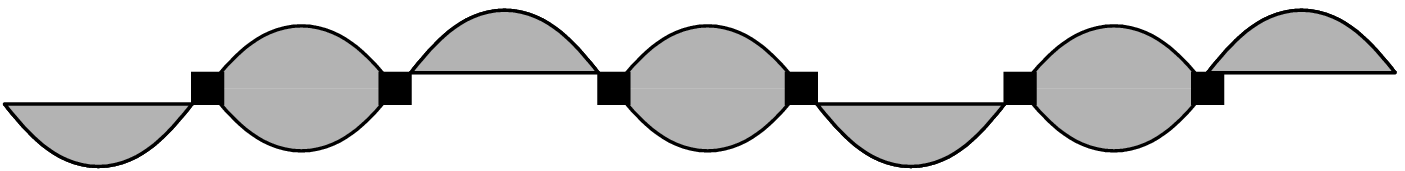} + \dots 
\ .
\end{equation}
Each of these has the form (with $n+1$ half-chains)
\begin{equation}\label{D31}
{\cal S} (p) \prod_{j=1}^{n}
\left((1-\delta_{i_{j}i_{j+1}}){\cal S} (p)
H_{i_{j}i_{j+1}} (p) \right)
\ .
\end{equation}
The only replica-conserving term is obtained by using
$\delta_{i_{j}i_{j+1}} $  in {\em each} factor. The projection
onto the 1-replica term is denoted $\cal R$ (for replica-conserving)
\begin{equation}\label{5.5}
{\cal R} = \diagram{rcc1}_{\!\!\!1}+ \diagram{rcc2}_{\!\!\!1} +
\diagram{rcc3}_{\!\!\!1} + \diagram{rcc4}_{\!\!\!1} + \dots
\end{equation}
and is evaluated as
\begin{equation}\label{5.6}
{\cal R} = S (p) \sum_{n =0}^{\infty} \left[- {\cal S}(p)
H_{aa}(p)\right]^{n } = \frac{{\cal S }(p)}{1+{\cal S} (p)H_{aa} (p)} \
= 2 I_{2} (p)\left(1-4 I_{2} (p) B''_{aa} \right)\ ,
\end{equation}
where the last identity can be found in (\ref{im2}). This
reestablishes the two factors of $(1-4 I_{2}B''_{aa})$ in diagram
(\ref{5.2}).
Note that $\cal R$ is $\diagram{moon}$ introduced in section
\ref{buildingblocks}.

We can also give an equivalent interpretation of (\ref{lf63}). It is
$H_{ab}$ times summed half-chains, but since the indices are forced to
be equal at the end, we can drop one of the $\delta$'s in
(\ref{D31}). Instead of (\ref{5.6}) this is
\begin{eqnarray}\label{5.7}
{\cal T} &=&{\cal S} (p) \sum_{n =0}^{\infty} \left[- {\cal S}(p)
H_{aa}(p)\right]^{n } (1-n ) \nonumber \\
&=& 2 I_{2} (p) \left(1+2 I_2
(p) B''_{aa} \right)\left(1-4 I_{2} (p)
B''_{aa} \right)\ ,
\end{eqnarray}
where the combinatorial factor $(1-n)$ is due to the fact, that one
can drop one of the $\delta$'s. We have also used the simplifications
of equations (\ref{lf62}) ff.  $\cal T$ is nothing but
$\diagram{halfmoon}$, introduced in section \ref{buildingblocks}.

%%%%%%%%%%%%%%%%%%%%%%%%%%%%%%%%%%%%%%%%%%%%%%%%%%%%%%%%%%%%%%%%%%%%%%
%%%%%%%%%%%%%%%%%%%%%%%%%%%%%%%%%%%%%%%%%%%%%%%%%%%%%%%%%%%%%%%%%%%%%%
%%%%%%%%%%%%%%%%%%%%%%%%%%%%%%%%%%%%%%%%%%%%%%%%%%%%%%%%%%%%%%%%%%%%%%
%%%%%%%%%%%%%%%%%%%%%%%%%%%%%%%%%%%%%%%%%%%%%%%%%%%%%%%%%%%%%%%%%%%%%%

\section{The 2-loop diagram with a tadpole and graphical
interpretation of perturbation theory}\label{2looptadpole}
\comment{***KAy finale***}
For simplicity of notations, this calculation is done for a
1-component field $u$. We also note $R (u):=B (u^{2})$. We here
calculate the 2-loop diagrams at finite temperature. This shows how
the naive rules one uses at zero temperature can be misinterpreted.

All diagrams have two vertices $R$, two lines between these two $R$'s,
and a tadpole attached to one of the $R$'s:  
\begin{equation}\label{lf11}
\fig{0.5\textwidth}{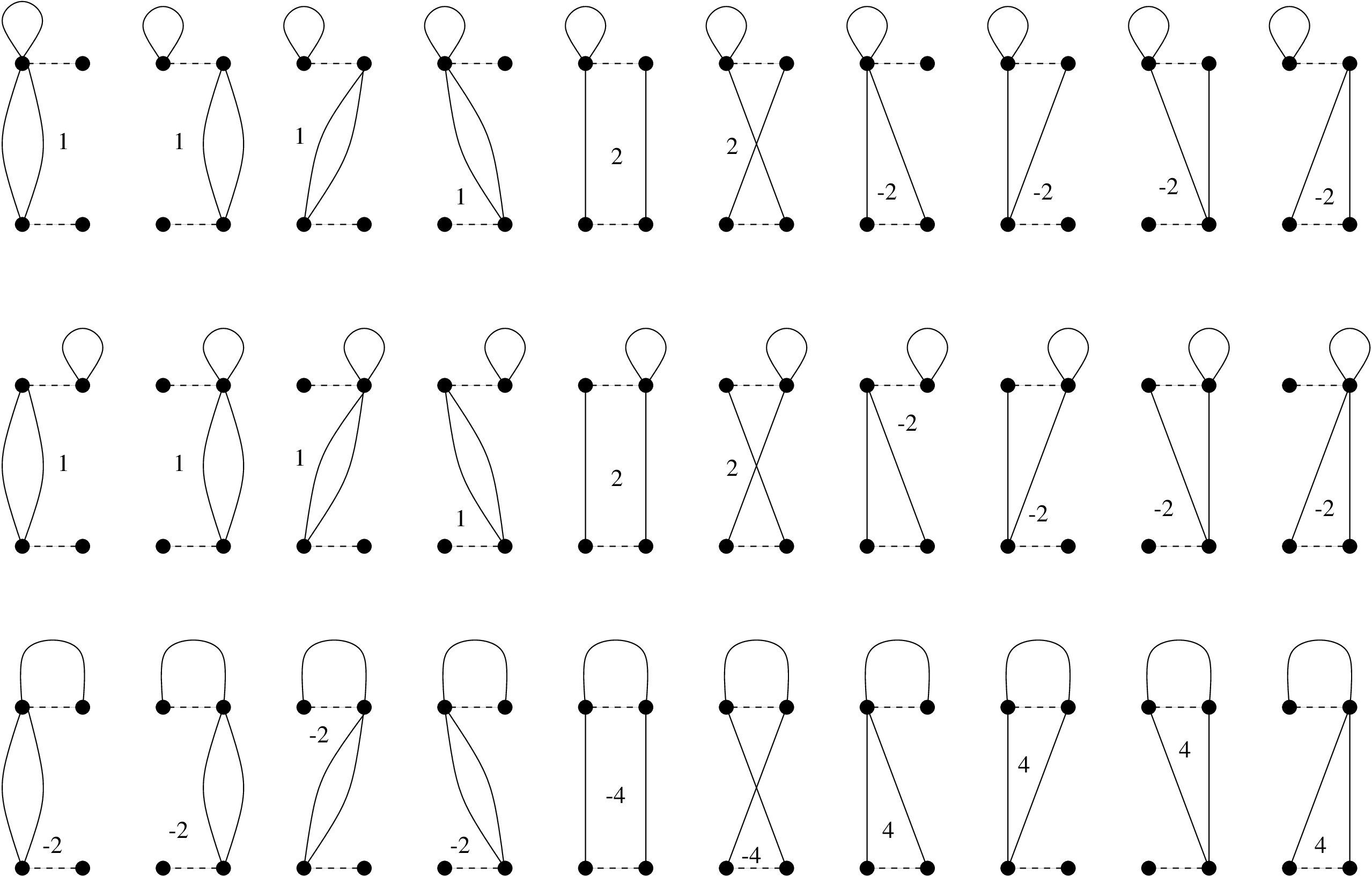}
\ .
\end{equation}
This is a graphical representation of $R''''_{ab}R''_{cd} (\delta
_{aa}+\delta _{bb}-2\delta _{ab})\left(\delta _{ac}\delta _{bd}+\delta
_{ad}\delta _{bc} \right)^{2}$, together with combinatorial
factors. Projecting onto 2-replica-terms only gives:
\begin{equation}\label{lf12}
\fig{0.3\textwidth}{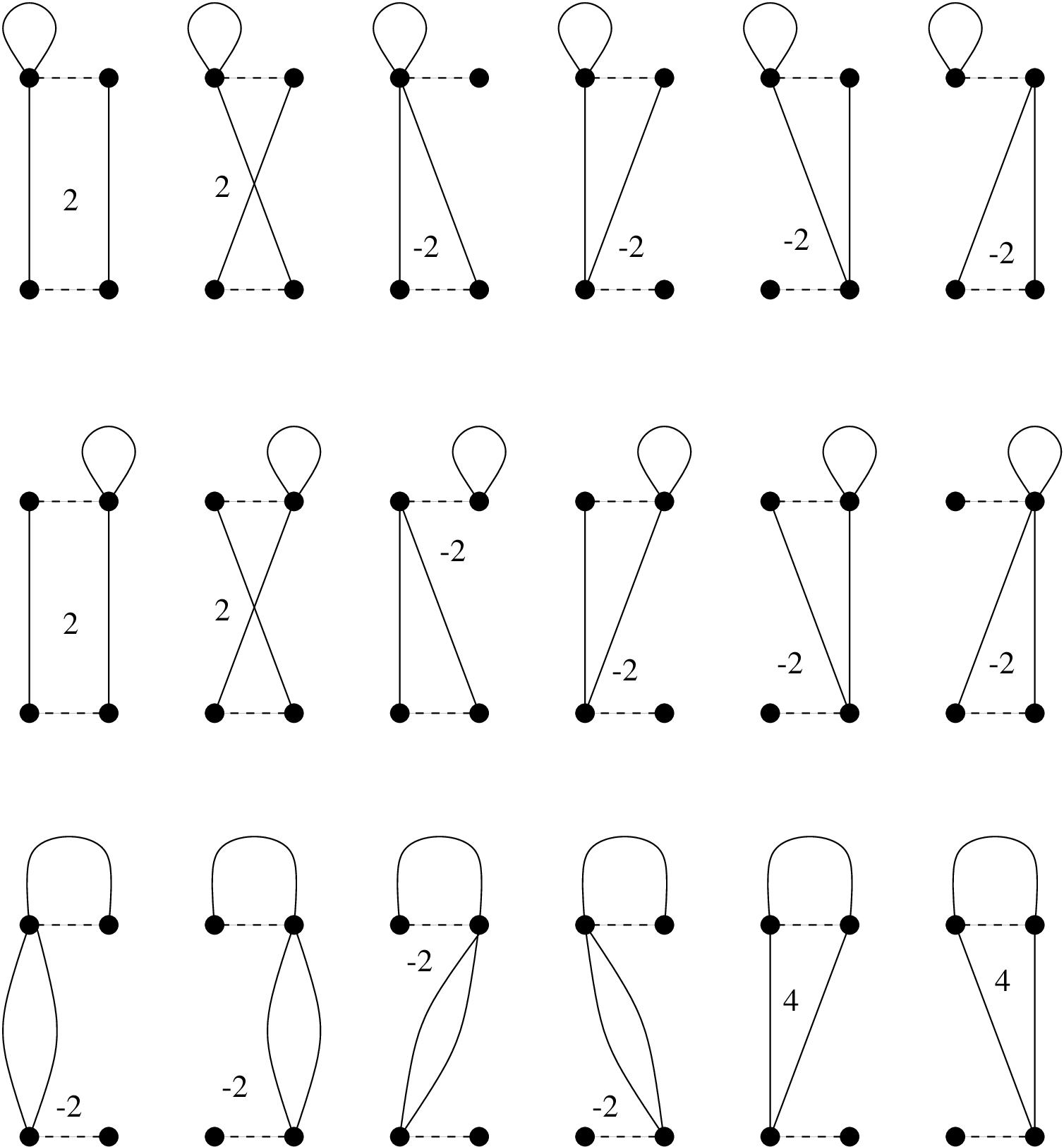}
\ .
\end{equation}
The interesting diagrams proportional to $R''''(0)$ are:
\begin{equation}\label{lf13}
\fig{0.3\textwidth}{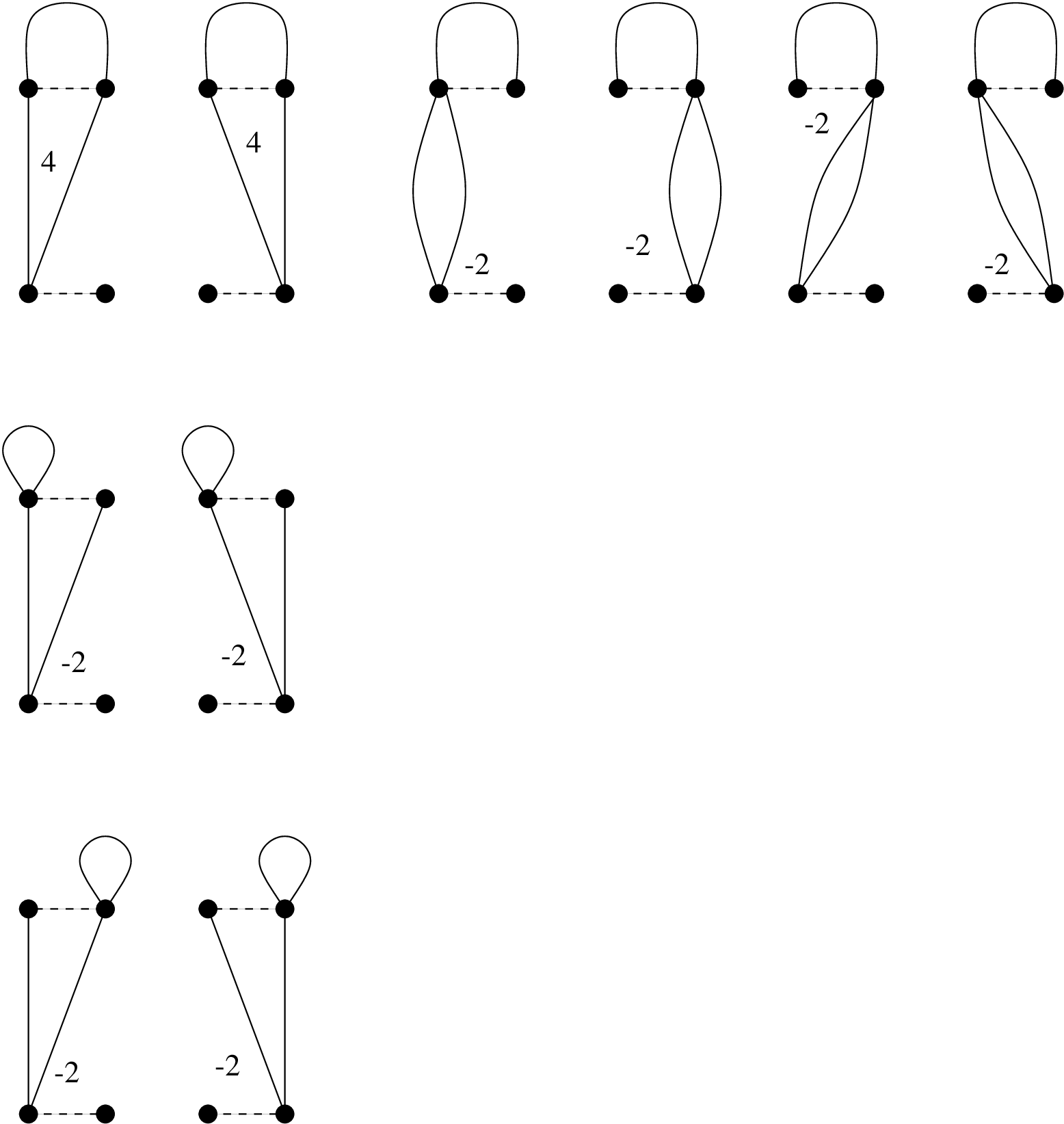}
\ .
\end{equation}
One sees that one has cancellation either in one column or in one line.
One can interpret this as follows: Making first contractions between
the two $R$'s, identifying the replicas of the upper $R$, one can say
that no tadpole can be added to the upper $R$. This is the cancellation
in the first column. Equivalently, one can first draw the tadpole. If
one does this connecting replicas $a$ and $b$ on the upper $R$, then
one can say that it is no longer possible to add correlators
connecting the two $R$'s. This is the cancellation in the first line,
as opposed to the first column.

%%%%%%%%%%%%%%%%%%%%%%%%%%%%%%%%%%%%%%%%%%%%%%%%%%%%%%%%%%%%%%%%%%%%%%
%%%%%%%%%%%%%%%%%%%%%%%%%%%%%%%%%%%%%%%%%%%%%%%%%%%%%%%%%%%%%%%%%%%%%%
%%%%%%%%%%%%%%%%%%%%%%%%%%%%%%%%%%%%%%%%%%%%%%%%%%%%%%%%%%%%%%%%%%%%%%
%%%%%%%%%%%%%%%%%%%%%%%%%%%%%%%%%%%%%%%%%%%%%%%%%%%%%%%%%%%%%%%%%%%%%%
%%%%%%%%%%%%%%%%%%%%%%%%%%%%%%%%%%%%%%%%%%%%%%%%%%%%%%%%%%%%%%%%%%%%%%
%%%%%%%%%%%%%%%%%%%%%%%%%%%%%%%%%%%%%%%%%%%%%%%%%%%%%%%%%%%%%%%%%%%%%%
%%%%%%%%%%%%%%%%%%%%%%%%%%%%%%%%%%%%%%%%%%%%%%%%%%%%%%%%%%%%%%%%%%%%%%
%%%%%%%%%%%%%%%%%%%%%%%%%%%%%%%%%%%%%%%%%%%%%%%%%%%%%%%%%%%%%%%%%%%%%%
%%%%%%%%%%%%%%%%%%%%%%%%%%%%%%%%%%%%%%%%%%%%%%%%%%%%%%%%%%%%%%%%%%%%%%
%%%%%%%%%%%%%%%%%%%%%%%%%%%%%%%%%%%%%%%%%%%%%%%%%%%%%%%%%%%%%%%%%%%%%%
%                            Algebra                          %
%%%%%%%%%%%%%%%%%%%%%%%%%%%%%%%%%%%%%%%%%%%%%%%%%%%%%%%%%%%%%%%%%%%%%%
%%%%%%%%%%%%%%%%%%%%%%%%%%%%%%%%%%%%%%%%%%%%%%%%%%%%%%%%%%%%%%%%%%%%%%
%%%%%%%%%%%%%%%%%%%%%%%%%%%%%%%%%%%%%%%%%%%%%%%%%%%%%%%%%%%%%%%%%%%%%%
%%%%%%%%%%%%%%%%%%%%%%%%%%%%%%%%%%%%%%%%%%%%%%%%%%%%%%%%%%%%%%%%%%%%%%
%%%%%%%%%%%%%%%%%%%%%%%%%%%%%%%%%%%%%%%%%%%%%%%%%%%%%%%%%%%%%%%%%%%%%%
%%%%%%%%%%%%%%%%%%%%%%%%%%%%%%%%%%%%%%%%%%%%%%%%%%%%%%%%%%%%%%%%%%%%%%
%%%%%%%%%%%%%%%%%%%%%%%%%%%%%%%%%%%%%%%%%%%%%%%%%%%%%%%%%%%%%%%%%%%%%%
%%%%%%%%%%%%%%%%%%%%%%%%%%%%%%%%%%%%%%%%%%%%%%%%%%%%%%%%%%%%%%%%%%%%%%
%%%%%%%%%%%%%%%%%%%%%%%%%%%%%%%%%%%%%%%%%%%%%%%%%%%%%%%%%%%%%%%%%%%%%%
%%%%%%%%%%%%%%%%%%%%%%%%%%%%%%%%%%%%%%%%%%%%%%%%%%%%%%%%%%%%%%%%%%%%%%

\section{Details of the algebraic method}
\comment{Kay finished**********************}
\label{app:Algebra}

\subsection{Algebra of 4-replica matrices}
The 4-replica matrices needed in the $tr \ln$ calculation
of Section \ref{s:algebramethod} can be parameterized as
in (\ref{para1}) and (\ref{para2}). They form an algebra, and
to solve Eq. (\ref{eqmat}) one needs to write explicitly the
components of an arbitrary product of such matrices.

Let us consider two matrices $M^1$ and $M^2$ parameterized
respectively by $(x^i_a,y^i_{ab},z^i_{ab},t^i_{abc},
u^i_{ab},v^i_{abc},w^i_{abc}$, $g^i_{abcd})$ for $i=1,2$. Then the
product $M_{ab,cd} = \sum_{ef} M^1_{ab,ef} M^2_{ef,cd}$
is itself parameterized by 
$(x_a,y_{ab},z_{ab},t_{abc},$ $
u_{ab},v_{abc},  w_{abc},g_{abcd})$ and one finds:
\begin{eqnarray}
  x_a &=& x^1_a  x^2_a + 2 y^1_{aa} x^2_a  + 2 x^1_a z^2_{aa} + 2
\sum_f y^1_{af} z^2_{af} \label{algebra} \\ 
  y_{ad}&=&x^1_a y^2_{ad} + 2 x^1_a t^2_{aad} + 2
y^1_{aa} y^2_{ad} + 2 \sum_f y^1_{af} t^2_{afd} \nonumber  \\ 
  z_{ab} &=& z^1_{ab} x^2_a + 2 t^1_{aba} x^2_a + 2 z^1_{ab} z^2_{aa} +
2 \sum_f t^1_{abf} z^2_{af} \nonumber \\ 
  t_{abd} &=& z^1_{ab} y^2_{ad} + 2 t^1_{aba} y^2_{ad} + 2 z^1_{ab}
t^2_{aad} + 2 \sum_f t^1_{abf} t^2_{afd} \nonumber \\ 
  u_{ac}&=& x^1_a u^2_{ac} + u^1_{ac} x^2_c + v^1_{acc} x^2_c + x^1_a
w^2_{aac} + 2 y^1_{ac} z^2_{ca} 
+ 2 y^1_{aa} u^2_{ac} + 2 u^1_{ac} z^2_{cc} \nonumber \\
&&  +  \sum_f \left[ 2 y^1_{af} w^2_{afc} + 2 v^1_{acf} z^2_{cf} 
+ u^1_{af} u^2_{fc} + v^1_{aff} u^2_{fc} 
+ u^1_{af} w^2_{ffc} \right] + \sum_{ef} v^1_{aef} w^2_{efc} \nonumber \\
  v_{acd} &=& x^1_a v^2_{acd} + x^1_a g^2_{aacd} + 2 y^1_{ac} t^2_{cad} + 2 y^1_{ad} t^2_{dac} + 
u^1_{ac} y^2_{cd} + u^1_{ad} y^2_{dc} + 2 y^1_{aa} v^2_{acd} + v^1_{acc} y^2_{cd} 
+ v^1_{add} y^2_{dc} \nonumber \\
 &&  +
 2 u^1_{ac} t^2_{ccd} + 2 u^1_{ad} t^2_{ddc} \nonumber \\
&&  + \sum_f \left[ 2 y^1_{af} g^2_{afcd} + 
2  v^1_{acf} t^2_{cfd} + 2  v^1_{adf} t^2_{dfc} +
 u^1_{af} v^2_{fcd} +  u^1_{af} g^2_{ffcd} + v^1_{aff} v^2_{fcd} \right]
+ \sum_{ef} v^1_{aef} g^2_{efcd}   \nonumber  \\
  w_{abc}&=& w^1_{abc} x^2_c + g^1_{abcc} x^2_c + 2 t^1_{abc} z^2_{ca} + 2 t^1_{bac} z^2_{cb} +
z^1_{ab} u^2_{ac} + z^1_{ba} u^2_{bc} + z^1_{ab} w^2_{aac} + z^1_{ba} w^2_{bbc} +
2 w^1_{abc} z^2_{cc} \nonumber \\
 && + 2 t^1_{aba} u^2_{ac} + 2 t^1_{bab} u^2_{bc}
\nonumber \\
&&  + \sum_f \left[ 2 g^1_{abcf} z^2_{cf} 
2 t^1_{abf} w^2_{afc} + 2 t^1_{baf} w^2_{bfc} + w^1_{abf} u^2_{fc} 
+ g^1_{abff} u^2_{fc} + w^1_{abf} w^2_{ffc} \right] +
\sum_{ef} g^1_{abef} w^2_{efc}  \nonumber \\ 
  g_{abcd} &=& w^1_{abc} y^2_{cd} + w^1_{abd} y^2_{dc} + g^1_{abcc} y^2_{cd} +  g^1_{abdd} y^2_{dc} +
z^1_{ab} v^2_{acd} + z^1_{ba} v^2_{bcd} + z^1_{ab} g^2_{aacd} + z^1_{ba} g^2_{bbcd} 
\nonumber \\
&&  + 2 t^1_{abc} t^2_{cad} 
+ 2 t^1_{bac} t^2_{cbd} 
+ 2 t^1_{abd} t^2_{dac} + 2 t^1_{bad} t^2_{dbc} + 2 t^1_{aba}  v^2_{acd} + 2 t^1_{bab}  v^2_{bcd} +
2 w^1_{abc} t^2_{ccd} + 2 w^1_{abd} t^2_{ddc}  \nonumber \\
&&  + \sum_f \left[ 2 t^1_{abf} g^2_{afcd} + 2 t^1_{baf} g^2_{bfcd} +
2 g^1_{abcf} t^2_{cfd} + 2  g^1_{abdf} t^2_{dfc} + w^1_{abf} v^2_{fcd} + 
 g^1_{abff} v^2_{fcd} +  w^1_{abf} g^2_{ffcd}\right]  \nonumber \\
&&+ \sum_{ef} g^1_{abef} g^2_{efcd} 
\ , \nonumber
\end{eqnarray}
where we have made replica sums explicit.

Using these multiplication rules one can rewrite Eq. (\ref{eqmat}) in
terms of a set of nonlinear equations for the components of
$M^\lambda$ in terms of the components of $\overline{\cal
M}$. Unfortunately no closed solution seemed possible (except in some
very special cases). The next step is thus to expand each component in
number of replica sums as in (\ref{expsum}). This results in a
hierarchy of equations for components with increasing number of
replica sum. For instance the zero-sum components behave under
multiplication as in (\ref{algebra}), dropping all terms with replica
sums, and so on.  These equations can be solved iteratively, as
discussed in the text.  For this, one needs the zero- and one-sum
components of the matrix $\overline{\cal M}$, the calculation of which
we now detail.

\subsection{Calculation of the matrix $M$} We start by computing the
matrix $M$ in (\ref{start}). As in the following subsections there are
two stages. First make all Kronecker-deltas explicit, then expand each
term in the number of replica sums. As discussed in the text, all
intermediate free sums over two or more replicas can be dropped.

A straightforward calculation from (\ref{bpot}), for a model
with only a bare second cumulant, gives:
\begin{eqnarray}
M_{ab,cd} &=& (2 T \partial_{\chi_{ab}}  \partial_{\chi_{cd}} U(\chi))|_{\chi=\chi_v} 
\label{mM} \\
& =& - \frac{2}{T} ( \delta_{abcd} \sum_e B''_{ae} +
\delta_{ab} \delta_{cd} B''_{ac} -  (\delta_{acd} + \delta_{bcd}) B''_{ab} 
- (\delta_{abc} + \delta_{abd}) B''_{cd}
+ (\delta_{ac} \delta_{bd} + \delta_{bc} \delta_{ad}) B''_{ab} )
\nonumber 
\end{eqnarray}
where  $B''_{ab}=B''(\tilde{\chi}_v^{ab})$, and 
$\tilde{\chi}_v^{ab}=\chi_v^{aa}+\chi_v^{bb}- 2 \chi_v^{ab}$; 
$\chi_v^{ab}$ is given by its saddle point value 
(\ref{chiv}). Note that since one takes everything at the
saddle point at the end, which is symmetric in $a,b$, all
expressions resulting from two derivatives 
are symmetric: $B''_{ab} = B''_{ba}$ (even if one chooses the
fluctuating fields a priori non-symmetric). 

The matrix $\tilde{\chi}_v^{ab}$ still contains 
explicit Kronecker-deltas. As in the main text, one writes
\begin{equation}
 \tilde{\chi}_v^{ab} = \tilde{\chi}_{ab} + \delta_{ab} \tilde \chi_a \ ,
\end{equation}
where $\tilde{\chi}_{ab}$ and $\tilde \chi_a$ contain no Kronecker
delta, and are computed below. Then one sees that
\begin{equation}
 B''(\tilde{\chi}_v^{ab}) = \delta_{ab} \left[
B''(\tilde \chi_{aa} + \tilde \chi_a) - B''(\tilde \chi_{aa}) \right]
+ B''(\tilde \chi_{ab})\ .
\end{equation}
Inserting this form into (\ref{mM}) above one finds that the
contribution of the piece $\delta_{ab} [ B''(\tilde \chi_{aa} + \tilde
\chi_a) - B''(\tilde \chi_{aa}) ]$ cancels exactly and thus one
obtains that $M$ is given by (\ref{mM}) but with now
$B''_{ab}=B''(\tilde{\chi}_{ab})$, i.e.\ the part with no Kronecker
delta.

We can now continue the calculation from (\ref{chiv}) by expanding in
the number of replica sums. First we define:
\begin{eqnarray}
 (- 2 T \partial \tilde{U}^0)_{ab} &=& \delta_{ab} \sum_c U_{ac} - U_{ab} \\
 U_{ab} &=& \frac{2}{T} \tilde{B}'_{ab} + \frac{2}{T^2} \sum_g
\tilde{S}'_{abg} \ , \label{order1}
\end{eqnarray}
where $\tilde{B}'_{ab} = \tilde{B}'(v^2_{ab})$ and similarly
for the three-replica term (which will drop later on). 
From there and (\ref{chiv}) we obtain, dropping all
higher order sums, the expansions:
\begin{eqnarray}
 \tilde{\chi}_{a} &=& - 2  T  I_1 - 2 T  I_2 \sum_e U_{ae} \\
 \tilde{\chi}_{ab} &=& v_{ab}^2 + 2 T  I_1 + 2 T  I_2 \left[U_{ab} 
- \frac{1}{2} (U_{aa} + U_{bb}) 
+ \frac{1}{2} \sum_e (U_{ae}  + U_{be}) \right] \\
&& + 2 T I_3 \left[ \sum_e U_{ab}( U_{ae} + U_{be}) - \sum_e U_{ae}
U_{eb} - \sum_e ( U_{aa} U_{ae} + U_{bb} U_{be}
- \frac{1}{2} U_{ae} U_{ea} - \frac{1}{2} U_{be} U_{eb}) \right]
\ . \nonumber 
\end{eqnarray}
Thus in each  of the $B''_{ab}
=B''(\tilde{\chi}_{ab})$ matrices appearing in (\ref{mM}) the argument
can be Taylor expanded, i.e.\ as
\begin{equation}\label{Mm1}
B''_{ab}= \overline{B}''_{ab} + \overline{B}'''_{ab} \sum_f O_{abf} \ ,
\end{equation}
where we have defined $\overline{B}''_{ab} = B''(\overline{\chi}_{ab})$,
$\overline{B}'''_{ab} = B'''(\overline{\chi}_{ab})$ and 
\begin{eqnarray}
 \overline{\chi}_{ab} &=& v_{ab}^2 
+ 2 T  I_1 + 4 I_2 (\tilde{B}'_{ab} 
- \frac{1}{2} (\tilde{B}'_{aa} + \tilde{B}'_{bb})) \\
 O_{abf} &=& \frac{4}{T} I_2 (\tilde{S}'_{abf} 
- \frac{1}{2} ( \tilde{S}'_{aaf} +  \tilde{S}'_{bbf} )) +
2 I_2 (\tilde{B}'_{af}  + \tilde{B}'_{bf}) \nonumber \\
&&  + \frac{8}{T} I_3  \left[\tilde{B}'_{ab}( \tilde{B}'_{af} 
+ \tilde{B}'_{bf}) - \tilde{B}'_{af} \tilde{B}'_{ef} 
 -  (\tilde{B}'_{aa} \tilde{B}'_{af} + \tilde{B}'_{bb} \tilde{B}'_{bf}  
- \frac{1}{2} \tilde{B}'_{af} \tilde{B}'_{fa} 
- \frac{1}{2} \tilde{B}'_{bf} \tilde{B}'_{fb} ) \right] \qquad \label{Mm2}
\end{eqnarray}
It will turn out below that at the end we will only need $O_{aaf} = 4
I_2 \tilde{B}'_{af}$.  We will not perform this expansion and
replacement now. First, we turn to the calculation of the matrix
$\overline{N}$ and perform the product $M \overline{N}^q$, keeping
$B''_{ab}$ unspecified.

\subsection{Calculation of the matrix $\overline{N}^q$} We now compute
the second matrix, $\overline{N}^q$, expanded up to one free replica
sum.  One has:
\begin{eqnarray}
 \overline{N}^q_{ab,cd} &=& v_a v_d \overline{G}^q_{bc} + v_b v_c
 \overline{G}^q_{ad} + T \overline{\Pi}^q_{ab,cd}  \\
 \overline{G}^q &=& C(q) \delta + \sum_{n \geq 1} C(q)^{n+1} (- 2 T
 \partial \tilde{U}^0)^n \\
 \overline{\Pi}^q_{ab,cd} &=& J^q_{1,1} \delta_{ad} \delta_{bc} +
\sum_{n \geq 1} J^q_{1,n+1} ( \delta_{ad} (- 2 T \partial
\tilde{U}^0)_{bc} + \delta_{bc} (- 2 T \partial \tilde{U}^0)_{ad} )\nonumber 
\\
&& + \sum_{m \geq 1, n \geq 1} J^q_{m+1,n+1} (- 2 T \partial
\tilde{U}^0)^m_{ad}
(- 2 T \partial \tilde{U}^0)^n_{bc} \\
  J^q_{i,j} &=& \int_k \frac{1}{(k^2 + m^2)^i} \frac{1}{((q-k)^2 + m^2)^j}\ . 
\end{eqnarray}
In addition to (\ref{order1}) we also need:
\begin{eqnarray}
&& (- 2 T \partial \tilde{U}^0)^2_{ab} = - \sum_e U_{ab} (U_{ae} + U_{be}) 
   + \sum_c U_{ac} U_{cb} = 
- \frac{4}{T^2} \sum_e \tilde{B}'_{ab} (\tilde{B}'_{ae} + \tilde{B}'_{be}) +  \frac{4}{T^2} \sum_e \tilde{B}'_{ae}  \tilde{B}'_{eb} 
\ . \nonumber
\end{eqnarray}
Since $(- 2 T \partial \tilde{U}^0)^3_{ab}$ etc.\ contains only at least 
2-replica sums, it can be dropped. We define:
\begin{eqnarray}
 \overline{N}^q_{ab,cd} &=& N^q_{ad,bc} + N^q_{bc,ad} \\
 N^q_{ad,bc} &=& \frac{1}{2} T J^q_{1,1} \delta_{ad} \delta_{bc} + C(q) \delta_{ad} v_b v_c 
+ v_a v_d ( C(q)^2  (- 2 T \partial \tilde{U}^0)_{bc}
+ C(q)^3  (- 2 T \partial \tilde{U}^0)^2_{bc}\nonumber   \\
&& 
+ T \delta_{ad} ( J^q_{1,2} (- 2 T \partial \tilde{U}^0)_{bc} 
+ J^q_{1,3} (- 2 T \partial \tilde{U}^0)^2_{bc} )
+ \frac{1}{2} T J^q_{2,2} (- 2 T \partial \tilde{U}^0)_{ad} 
(- 2 T \partial \tilde{U}^0)_{bc} \nonumber \\ 
&&
+ T J^q_{2,3} (- 2 T \partial \tilde{U}^0)_{ad} (- 2 T \partial
  \tilde{U}^0)^2_{bc} )
\ .
\end{eqnarray}
We  obtain:
\begin{eqnarray}
 N^q_{ad,bc} &=& \delta_{ad} \delta_{bc} L^1_{ab} + \delta_{ad}
 P^1_{a,bc} + \delta_{bc} P^2_{b,ad} - U_{bc} v_a v_d C(q)^2 \nonumber \\ &&
 + \frac{1}{2} T J_{2,2}^q U_{ad} U_{bc} +
(v_a v_d C(q)^3 - T J^q_{2,3} U_{ad} ) \sum_f [ U_{bf} U_{fc} - U_{bc}
(U_{bf}+U_{cf}) ]
\\
 L^1_{ab}  &=& \frac{1}{2} T J^q_{1,1} + T J^q_{1,2} \sum_f U_{b f}  \\
 P^1_{a,bc} &=&  C(q) v_b v_c -
T J^q_{1,2} U_{bc} 
+ T J^q_{1,3}  \sum_f [ U_{bf} U_{cf} - U_{bc} (U_{bf} + U_{cf}) ]
- \frac{1}{2} T J^q_{2,2} U_{bc} \sum_f U_{af} \\
 P^2_{b,ad} &=& v_a v_d  C(q)^2  \sum_f U_{b f} 
- \frac{1}{2} T J^q_{2,2} U_{ad} \sum_f U_{b f} 
\ .
\end{eqnarray}
This yields:
\begin{eqnarray}
 \overline{N}^q_{ab,cd} &=& \delta_{ad} \delta_{bc} L^q_{ab} 
+  \delta_{ad} Q^q_{a,bc} + \delta_{bc} Q^q_{b,ad} 
- \frac{2}{T} (\tilde{B}'_{bc} v_a v_d + \tilde{B}'_{ad} v_b v_c) C(q)^2 
+ \frac{4}{T}  J_{2,2}^q \tilde{B}'_{ad} \tilde{B}'_{bc} 
\nonumber 
\\
&& +
\frac{4}{T^2} (v_a v_d C(q)^3 - 2 J^q_{2,3} \tilde{B}'_{ad} ) \sum_f [
\tilde{B}'_{bf} \tilde{B}'_{fc}  
- \tilde{B}'_{bc} (\tilde{B}'_{bf}+\tilde{B}'_{cf}) ] \nonumber \\
&& +
\frac{4}{T^2} (v_b v_c C(q)^3 - 2 J^q_{2,3} \tilde{B}'_{bc} ) \sum_f [
\tilde{B}'_{af} \tilde{B}'_{fd}  
- \tilde{B}'_{ad} (\tilde{B}'_{af}+\tilde{B}'_{df}) ]\nonumber   \\
 L^q_{ab}  &=& T J^q_{1,1} + 2 J^q_{1,2} \sum_f (\tilde{B}'_{a f} + 
\tilde{B}'_{b f})   \\ 
 Q^q_{a,bc} &=&
v_b v_c ( C(q) + \frac{2}{T} C(q)^2  \sum_f \tilde{B}'_{a f} ) 
- \frac{2}{T} \tilde{B}'_{bc} ( T J^q_{1,2} 
+ 2 J^q_{2,2} \sum_f \tilde{B}'_{af}   ) \nonumber \\
&& + \frac{4}{T} J^q_{1,3}  \sum_f [ \tilde{B}'_{bf} \tilde{B}'_{cf} 
- \tilde{B}'_{bc} (\tilde{B}'_{bf} + \tilde{B}'_{cf}) ]
\ .
\end{eqnarray}

\subsection{Final calculation of the matrix ${\cal M}$}
We now perform the matrix multiplication
\begin{eqnarray}
 {\cal M}^q_{ab,cd} = \sum_{ef}  M_{ab,ef}  \overline{N}^q_{ef,cd} = 
 - \frac{2}{T} [ B''_{ab} {\cal N}^q_{ab,cd} - \delta_{ab} \sum_g  B''_{ag} 
{\cal N}^q_{ag,cd} ]
\ ,
\end{eqnarray}
where we have defined:
\begin{eqnarray}
 {\cal N}^q_{ab,cd} &=& (\overline{N}^q_{ab,cd} +
 \overline{N}^q_{ba,cd}
- \overline{N}^q_{aa,cd} - \overline{N}^q_{bb,cd})
= L^q_{ab} (\delta_{ad} \delta_{bc} + \delta_{bd} \delta_{ac}) -
L^q_{aa} \delta_{acd}
- L^q_{bb} \delta_{bcd} \nonumber  \\
&&
+  \delta_{ad} Q^q_{a,bc} + \delta_{bd} Q^q_{b,ac} - \delta_{ad}
   Q^q_{a,ac} - \delta_{bd} Q^q_{b,bc}
+ \delta_{bc} Q^q_{b,ad} + \delta_{ac} Q^q_{a,bd} - \delta_{ac}
  Q^q_{a,ad} - \delta_{bc} Q^q_{b,bd}
\nonumber  \\
&& - (U_{bc} v_a v_d + U_{ad} v_b v_c + U_{ac} v_b v_d + U_{bd} v_a
v_c
- U_{ac} v_a v_d - U_{ad} v_a v_c -U_{bc} v_b v_d - U_{bd} v_b v_c  ) I^q_2 
\nonumber  \\
&& + T J_{2,2}^q ( U_{ad} U_{bc} + U_{bd} U_{ac} - U_{ad} U_{ac} -
U_{bd} U_{bc} ) \nonumber \\ && + (v_a v_d C(q)^3 - T J^q_{2,3} U_{ad}
) \sum_f [ U_{bf} U_{fc} - U_{bc}
(U_{bf}+U_{cf}) ]\nonumber \\
&& + (v_b v_c C(q)^3 - T J^q_{2,3} U_{bc} ) \sum_f [ U_{af} U_{fd} -
U_{ad}
 (U_{af}+U_{df}) ]  \nonumber  \\ 
&& + (v_b v_d C(q)^3 - T J^q_{2,3} U_{bd} ) \sum_f [ U_{af} U_{fc} -
U_{ac} (U_{af}+U_{cf}) ] \nonumber \\
&& +
(v_a v_c C(q)^3 - T J^q_{2,3} U_{ac} ) \sum_f [ U_{bf} U_{fd} - U_{bd} (U_{bf}+U_{df}) ]  \nonumber  \\
&& - (v_a v_d C(q)^3 - T J^q_{2,3} U_{ad} ) \sum_f [ U_{af}
U_{fc}- U_{ac} (U_{af}+U_{cf}) ]\nonumber \\
&&  -
(v_a v_c C(q)^3 - T J^q_{2,3} U_{ac} ) \sum_f [ U_{af} U_{fd} - U_{ad} (U_{af}+U_{df}) ]  \nonumber  \\
&& - (v_b v_d C(q)^3 - T J^q_{2,3} U_{bd} ) \sum_f [ U_{bf} U_{fc} -
U_{bc} (U_{bf}+U_{cf}) ] \nonumber \\
&& - (v_b v_c C(q)^3 - T J^q_{2,3} U_{bc} )
\sum_f [ U_{bf} U_{fd} - U_{bd} (U_{bf}+U_{df}) ] \nonumber \ .
\end{eqnarray}
Performing the matrix product yields
 the parameterization of the matrix $\overline{\cal M}$, where we have
 not yet fully
expanded in sums, as (the $q$-dependence is implicit):
\begin{eqnarray}
 \alpha_{ab} &=& - \frac{2}{T} \frac{1}{1 - \frac{4}{T} B''_{ab} L_{ab}} \\
 x_a &=& \alpha_{aa} \sum_f B''_{af} L_{aa} \nonumber   \\
 y_{ad}  &=& - \alpha_{aa} B''_{ad} L_{ad} + \alpha_{aa} \sum_f B''_{af} (Q_{a,ad} - Q_{a,fd} ) \nonumber  \\
 z_{ab} &=&  - \alpha_{ab} B''_{ab} L_{aa} \nonumber  \\
 t_{a b d} &=&  \alpha_{ab} B''_{ab}  (Q_{a,bd} - Q_{a,ad} ) \nonumber  \\
 u_{ac} &=& \alpha_{aa} B''_{ac} L_{cc} \nonumber  \\
 v_{acd} &=& \alpha_{aa} \Big[ B''_{ad} (Q_{d,dc} -Q_{d,ac}) + B''_{ac}
 (Q_{c,cd} - Q_{c,ad})\nonumber \\
&& 
- \sum_f B''_{af}  T J_{2,2}^q ( U_{ad} U_{fc} + U_{fd} U_{ac} - U_{ad} U_{ac} - U_{fd} U_{fc} ) \nonumber  \\
&& + \sum_f B''_{af} (U_{fc} v_a v_d + U_{ad} v_f v_c + U_{ac} v_f v_d
+ U_{fd} v_a v_c\nonumber \\
&&\hphantom{+ \sum_f B''_{af} (}
- U_{ac} v_a v_d - U_{ad} v_a v_c -U_{fc} v_f v_d - U_{fd} v_f v_c  ) I^q_2 \Big] \nonumber \nonumber  \\
 w_{abc} &=& 0 \nonumber \\ 
g_{abcd} &=& \alpha_{ab} B''_{ab} \Big\{ -
 (U_{bc} v_a v_d + U_{ad} v_b v_c + U_{ac} v_b v_d + U_{bd} v_a v_c 
\nonumber \\ 
&& \hphantom{\alpha_{ab} B''_{ab} \Big\{} \qquad 
- U_{ac} v_a v_d - U_{ad} v_a v_c -U_{bc} v_b v_d - U_{bd} v_b v_c  ) I^q_2 
\nonumber  \\
&& \hphantom{\alpha_{ab} B''_{ab} \Big\{}+ T J_{2,2}^q ( U_{ad} U_{bc} + U_{bd} U_{ac} - U_{ad} U_{ac} -
U_{bd} U_{bc} ) \nonumber \\ 
 &&\hphantom{\alpha_{ab} B''_{ab} \Big\{} + (v_a v_d I^q_3 - T
 J^q_{2,3} U_{ad}
) \sum_f [ U_{bf} U_{fc} - U_{bc} (U_{bf}+U_{cf}) ] \nonumber \\
&&\hphantom{\alpha_{ab} B''_{ab} \Big\{} +
(v_b v_c I^q_3 - T J^q_{2,3} U_{bc} ) \sum_f [ U_{af} U_{fd} - U_{ad} (U_{af}+U_{df}) ]  \nonumber \\
&& \hphantom{\alpha_{ab} B''_{ab} \Big\{}+ (v_b v_d I^q_3 - T
J^q_{2,3} U_{bd} ) \sum_f [ U_{af} U_{fc} -
U_{ac} (U_{af}+U_{cf}) ]\nonumber \\
&& \hphantom{\alpha_{ab} B''_{ab} \Big\{} +
 (v_a v_c I^q_3 - T J^q_{2,3} U_{ac} ) \sum_f [ U_{bf} U_{fd} - U_{bd} (U_{bf}+U_{df}) ]  \nonumber \\
&& \hphantom{\alpha_{ab} B''_{ab} \Big\{}- (v_a v_d I^q_3 - T J^q_{2,3} U_{ad} ) \sum_f [ U_{af} U_{fc} -
U_{ac} (U_{af}+U_{cf}) ]\nonumber \\
&& \hphantom{\alpha_{ab} B''_{ab} \Big\{}-
 (v_a v_c I^q_3 - T J^q_{2,3} U_{ac} ) \sum_f [ U_{af} U_{fd} - U_{ad} (U_{af}+U_{df}) ]  \nonumber\\
&& \hphantom{\alpha_{ab} B''_{ab} \Big\{}- (v_b v_d I^q_3 - T J^q_{2,3} U_{bd} ) \sum_f [ U_{bf} U_{fc} -
U_{bc} (U_{bf}+U_{cf}) ] \nonumber \\
&& \hphantom{\alpha_{ab} B''_{ab} \Big\{}- (v_b v_c I^q_3 - T
J^q_{2,3} U_{bc} ) \sum_f [ U_{bf} U_{fd} - U_{bd} (U_{bf}+U_{df}) ]
\Big\} \nonumber \ .
\end{eqnarray}

We now finish the expansion in sums, using (\ref{Mm1}), (\ref{Mm2})
and defining the notations:
\begin{eqnarray}
 A^q_{ab} &=& \frac{2}{1 - 4 \overline{B}''_{ab} J^q_{1,1} } \\
 A^q_{aa} &=& \frac{2}{1 - 4 B''(2 T I_1) J^q_{1,1} } \\
 \alpha_{ab} &=& - \frac{1}{T} [ A^q_{ab} + 2 (A^q_{ab})^2
 (\frac{2}{T} J^q_{1,2}
\overline{B}''_{ab} \sum_f ( \tilde{B}'_{af} + \tilde{B}'_{bf})  
+ J^q_{1,1} \overline{B}'''_{ab} \sum_f O_{abf} ) ] \\
 L_{ab} &=& T J^q_{1,1} + 2 J^q_{1,2} \sum_f ( \tilde{B}'_{af} +
 \tilde{B}'_{bf})
\end{eqnarray}h The result is given in the text in section
\ref{s:algebramethod}, as well as:
\begin{eqnarray}
 P_1 y_{ad} &=& 2 J^q_{1,1} (A^q_{aa})^2 (\frac{4}{T} J^q_{1,2}
 \overline{B}''_{aa}
\sum_f  \tilde{B}'_{af} +
J^q_{1,1} \overline{B}'''_{aa} \sum_f O_{aaf} ) \overline{B}''_{ad} 
+ \frac{2}{T} A_{aa}  J^q_{1,2} \overline{B}''_{ad} \sum_f ( \tilde{B}'_{af}
 + \tilde{B}'_{df} ) \nonumber 
\\
&& + J^q_{1,1} A^q_{aa} \overline{B}'''_{ad} \sum_f O_{adf} -
\frac{2}{T} 
\frac{C(q)}{(1 - 
4 B''(2 T I_1) J^q_{1,1})} \sum_f \overline{B}''_{af} (v_a - v_f) v_d   \\ 
 P_1 z_{ab} &=& 2 J^q_{1,1} (A^q_{ab})^2 (\frac{2}{T} J^q_{1,2}
 \overline{B}''_{ab}
\sum_f ( \tilde{B}'_{af} +\tilde{B}'_{bf} ) +
J^q_{1,1} \overline{B}'''_{ab} \sum_f O_{abf} ) \overline{B}''_{ab} 
\nonumber 
\\
&& + \frac{4}{T} A_{ab}  J^q_{1,2} b''_{ab} \sum_f \tilde{B}'_{af} 
+ J^q_{1,1} A^q_{ab} \overline{B}'''_{ab} \sum_f O_{abf}\ . 
\end{eqnarray}

\section{More remarks on the graphical method}
\label{s:moregraphics}
\comment{**** KAY est REPASSe,**********}
\subsection{Diagrammatics}\label{a:K:diagrammatics}
\begin{figure}[t]
\centerline{\fig{0.9\textwidth}{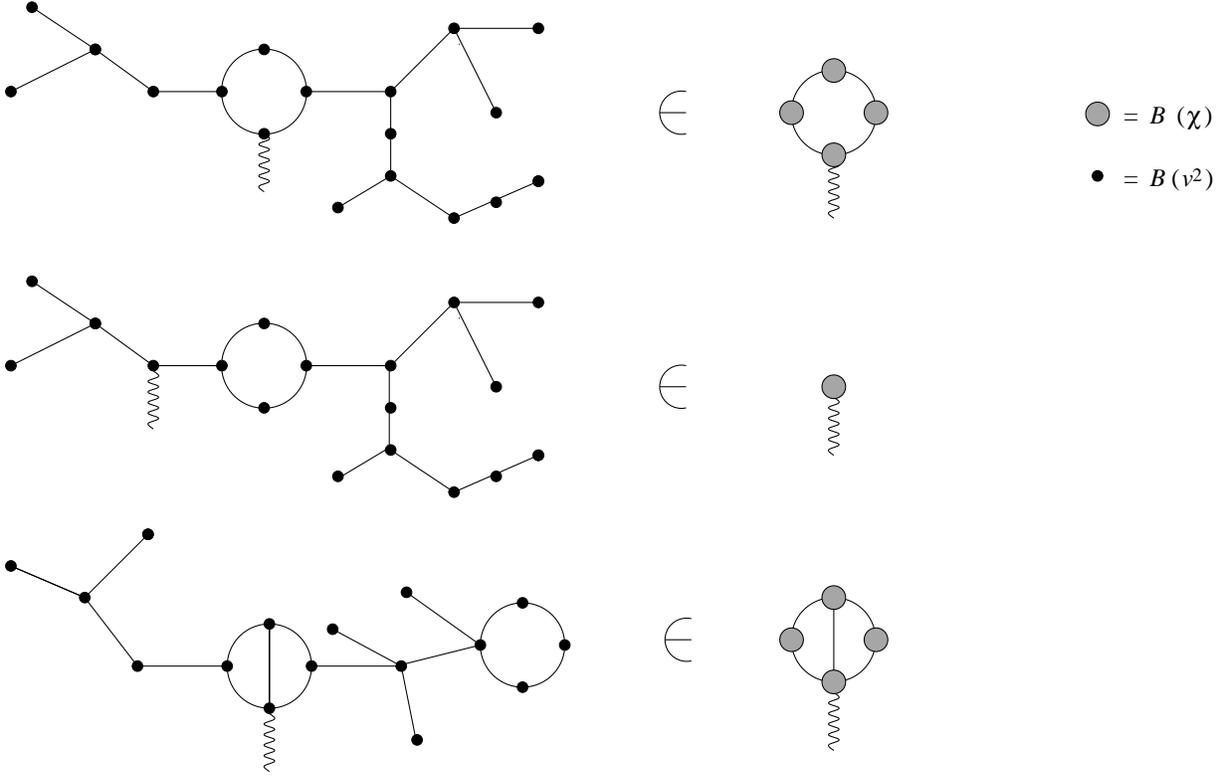}}
\caption{Some typical $1/N$-diagrams and the classes they belong
to. The wiggly line indicates the derivative, the black circle is a
$\tilde{B}$ (or derivative), the grey circle a $B$ (or
derivative). We have restrained from drawing a contribution due to the
explicit $v$-dependence. These terms do not play a role in the
argument, and are only tedious supplements one has to keep track of.}
\label{1overNdiagrams}
\end{figure}
The diagrammatic $1/N$ expansion can be constructed by using the 

\leftline{\underline {Theorem:}}
\begin{eqnarray}\label{tildeBallorders}
\tilde{B}' (v^{2}) &=&\left[ \frac{\partial }{\partial \chi }+ \frac{\partial }{\partial v^{2} } \right] \left(B (\chi
)+\frac{1}{N} B^{(1)} (\chi (v),v^{2} )
+\frac{1}{N^{2}} B^{{(2)}} (\chi (v),v^{2} ) + \dots  \right) \\
B^{(1)} (\chi )&=& \sum \mbox{all 1PI diagrams with 1 loop}\\
B^{(2)} (\chi )&=& \sum \mbox{all 1PI diagrams with 2 loops}\\
\dots &=& \dots \\
\label{tildeBallorderschi}
\chi (v)&=& v^{2} + 2 T I_{1} + 4 I_{2} [\p_{\chi} {B} (\chi (
v^{2}),v^{2})-\p_{\chi} {B} (\chi ( 0 ), 0) ] \ .
\end{eqnarray}
Some explication and precisions are in order: 1-particle-irreducible
diagrams (1PI) are w.r.t.\ lines being correlators $\left<vv\right>$,
and vertices being $B^{(n)} (\chi )$.  The r.h.s.\ of
\Eq{tildeBallorders} are diagrams made out of bare vertices. We have
separated the $\chi $-dependence from the explicit $v$-dependence: The
latter are $v$'s which are connected with a line. These are the terms
in our $1/N$-calculation, which explicitly contain $v$. Note that $v$'s
always pair.  Side-chains only come from the fact that finally one
inserts $\chi $. Note that $\chi $ as defined here is an object which
contains terms at all orders in $1/N$.  The diagrams are 1PI, a fact
which is important for the order $1/N^{2}$. It means that $B^{(2)}$
does {\em not} contain the diagram made out of 2 closed loops,
connected by a single line.

\leftline{\underline{Proof:}} Draw a collection of diagrams
contributing to $\tilde{B}$ (see figure \ref{1overNdiagrams}). This
drawing contains vertices made out of derivatives of $B(v^2)$ (not $B
(\chi )$ -- we have drawn the completely expanded diagram). Now derive
that object with respect to $v^{2}$, giving a couple of terms. Any of
these terms singles out one $B$, namely the one derived.

This $B$ may be part of a tree, by which we mean that either it is a
point or by cutting off one of the attached legs, the diagram will
fall apart. Then it is contained in the first term on the r.h.s.\ of
\Eq{tildeBallorders}, since any attachment which can be made to it in
the form of a tree, is taking care of by choosing the above given
$\chi $. Note that for this to be true, $\chi $ has to be exactly the
object given above, i.e.\ on the r.h.s.\ of \Eq{tildeBallorderschi}
there has to be the full $\tilde{B}$ to all orders in $1/N$. In the
diagrammatic language this is clear: Having a higher-order diagram and
taking the derivative at one of the tree-like vertices, this diagram
may still contain an arbitrary loop somewhere attached to the tree.

\begin{figure}[t]
\centerline{\fig{\textwidth}{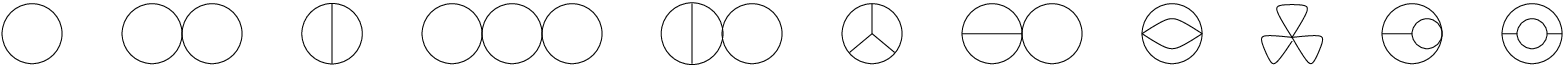}}
\caption{Diagrams at order $1/N$ (first), $1/N^{2}$ (next two) and
$1/N^{3}$ (rest).}
\label{1overNdiagrams2}
\end{figure}

The $B$ which has been derived may as well be part of a closed loop.
By this we mean that when we cut off all parts of the diagram which
can be disconnect from our chosen one by one cut, there remains more
than the vertex itself. This object is of higher connectivity; it can
either be a loop (at order $1/N$); it can be a diagram in the form of
an 8 or a circle to which one has added an additional line between two
arbitrary chosen points on it (at order $1/N^{2}$). Higher order
diagrams are given in figure \ref{1overNdiagrams2}.
%There is still a remark: The procedure given above
%differs form what Pierre and Kay have done up to now. At order $T^{2}$
%there was a diagram proportional to $B'''(\chi(0))$. This diagram is 
%eliminated by the derivative, which is very nice; the procedure used
%to get that diagram in the explicit calculation by Kay was rather 
%weird: Calculating first the loop and then setting the loop to 0, connecting
%to a $\tilde{B}$. I therefore strongly advocate to use the above
%procedure,  in which one never has to calculate these kind of terms. 

\subsection{An alternative derivation of the $\beta$-function ($T=0$)}
\label{a:betaKay} We now give a general derivation of the
$\beta$-function to all orders in $1/N$, following our results of the
last section. This derivation is restricted to $T=0$, since it is rather
complicated at finite $T$. To make the derivation more transparent,
and to avoid having to derive with respect to $v^{2}$ on the r.h.s.\
of \Eq{tildeBallorders}, we introduce the auxiliary function
$\tilde{B} (v^{2},u^{2})$. By $u^{2}$ we shall denote a pair of
background-fields that are non-trivially connected by a line of
propagators, whereas $v^{2}$ denotes the background-field which is
inserted into $B$, but which is not connected to any propagator. Note
that this decomposition is unique; that the paring of $u$'s is
natural, and that deriving with respect to $v^{2}$, but not $u^{2}$
can combinatorically be interpreted as choosing any vertex $B$ and
deriving it once. This object is thus better fitted for calculations
than $\tilde{B} (v^{2})=\tilde{B} (v^{2},u^{2})\ts_{u^2=v^2}$. However
the latter object is of course the only one with a physical meaning.

We now start from a modified version of \Eq{tildeBallorders}, namely
\begin{eqnarray}\label{basicsldkfj}
\p_{v^2}\tilde{B} (v^{2},u^{2})&=& \frac{\p}{\p\chi} \left( B (\chi )
+ \delta B 
(\chi,u^{2} )\right)\\
 \delta B (\chi ,u^{2} ) &=& \delta B[B'(\chi(v)),B''
(\chi (v) ),\dots ,u^{2}] \qquad \nn\\
&=&
\frac{1}{N} {B}^{(1)} [B'(\chi(v)),B'' (\chi (v)),\dots ]+\frac{1}{N^{2}}
{B}^{(2)} [B'(\chi(v)),B'' (\chi (v) ),\dots,u^{2} ] +
\dots\qquad \\
\label{chiequation}
\chi = \chi (v) &=&\chi (v,u^{2})= v^{2} + 4 I_{2}\left(\p_{v^2}\tilde{B}
(v^{2},u^{2})-\p_{v^2}\tilde{B} (0,0) \right)\nonumber \\
&\equiv& v^{2}+ \frac{1}{\epsilon
}\left( \partial _{v^{2}}
\tilde{b} (v^{2},u^{2})-\partial _{v^{2}} {b} (0,0) \right)\ ,
\end{eqnarray}
where we are a little bit sloppy with the notations, suppressing the 
argument $u$ of $\chi (v)$.
We define the dimensionless quantities
\begin{eqnarray}\label{dimlessquant}
{b} (v^{2},u^{2})&=&4 \epsilon I_{2}\tilde{B} (v^{2},u^{2})=4
A_{d} m^{-\epsilon }\tilde{B} (v^{2},u^{2})\\
{b}_{0} (\chi )&=&4 A_{d}m^{-\epsilon }{B} (\chi )\\
\delta {b} (\chi,u^{2} )&=&4 A_{d} m^{-\epsilon }\delta {B} (\chi,u^{2} )\ .
\end{eqnarray}
As in the main text, we use the notation $ i_{n} (p) = \frac{I_{n}
(p)}{A_{d}} = \frac{I_{n} (p)}{\epsilon I_{2}}.$ The
$\beta$-function is
\begin{eqnarray}\label{beta-prelim}
-m \frac{\partial }{\partial m}\partial_{v^{2}} {b}
(v^{2},u^{2}) &=& \epsilon \partial_{v^{2}} {b} (v^{2},u^{2}) -
(4A_{d}m^{-\epsilon } ) \frac{m \partial }{\partial
m}\frac{\p}{\p\chi}
\left( B (\chi ) + \delta B (\chi,u^{2} )\right)\nn\\
&=& \epsilon\partial_{v^{2}} {b} (v^{2},u^{2})+
\frac{\p^{2}}{\partial\chi^{2}} \left( b
(\chi ) + \delta b (\chi,u^{2} )\right)\left(-m \frac{\p \chi }{\partial
m}\right) \nonumber \\
&&+ \epsilon \frac{\p}{\p\lambda}\lts_{\lambda=1}
\left[\frac{1}{\lambda}\frac{\partial }{\partial \chi }\delta b
(\lambda b_0',\lambda b_0'',\dots,u^{2} ) \right]
\ .
\end{eqnarray} 
Note that in the last equation, we have been a little bit sloppy with
the notation. What this means is that having rescaled $B$ to $b_{0}$,
the $m$-dependence of the integrals is canceled. Thus we can evaluate
all integrals at $m^{2}=1$. The derivative w.r.t.\ $\lambda $ is
easily understood as follows: Having a diagram with $n+1$ vertices,
the integrals scale like $m^{-n\epsilon }$.  First, this accounts for
the factor of $\epsilon $. Second, in order to get the right
combinatorial factor of $n$ instead of $n+1$, one has to subtract one
contribution, which is done by the factor of $1/\lambda $ in front of
$\delta b$.

We need two more equations. First, starting from \Eq{chiequation}
and deriving w.r.t.\ $m$, we obtain (exact!)
\begin{equation} 
-m \frac{\p \chi }{\partial m} = \frac{1}{\epsilon } 
\left(-m\frac{\p}{\p m} \right)
\left[ \partial_{v^{2}} {b} (v^{2},u^{2})-
 \partial_{v^{2}}{b}(0,0)\right]
\neq \frac{1}{\epsilon } 
\left(-m\frac{\p}{\p m} \right) \partial_{v^{2}} {b} (v^{2},u^{2})
-\partial_{v^{2}} {b} (0,0)\ ,
\end{equation} where an equality would 
 suppose that due to dimensional reduction $\partial_{m} \tilde{B}'(0)=0$.

Deriving \Eq{chiequation} w.r.t.\ $v^{2}$, we obtain (also exact)
\begin{equation}\label{pchipv2}
\frac{\p \chi }{\p v^{2}} = 1+\frac{1}{\epsilon}
\partial_{v^{2}}^{2} {b} (v^{2},u^{2})\ .
\end{equation}
Deriving  \Eq{basicsldkfj} by $v^{2}$ gives with the help of \Eq{pchipv2}
\begin{eqnarray}
\partial_{v^2}^{2} {b} (v^{2},u^{2}) &=& \frac{\partial^{2} }{\partial \chi ^{2}}
\left[b_{0} (\chi )+\delta b (\chi,u^{2} ) \right] \frac{\p \chi }{\p v^{2}}\\
&=& \frac{\partial^{2} }{\partial \chi ^{2}}
\left[b_{0} (\chi )+\delta b (\chi,u^{2} ) \right] \left( 1+\frac{1}{\epsilon}\partial_{v^2}^{2}  {b}(v^{2},u^{2})\right)
\ .
\end{eqnarray}
Therefrom we infer that (to all orders)
\begin{equation} 
\frac{\partial^{2} }{\partial \chi ^{2}} \left[b_{0} (\chi )+\delta b
(\chi,u^{2} ) \right] = \frac{\partial_{v^2}^2{b}
(v^{2},u^{2})}
{1+\frac{1}{\epsilon }\partial_{v^2}^2{b} (v^{2},u^{2})}\ .
\end{equation}
This equation can also be written as
\begin{equation}\label{lf1}
\frac{1}{\ds\frac{\partial^{2} }{\partial \chi ^{2}} \left[b_{0} (\chi )+\delta b
(\chi,u^{2} ) \right]} = \frac{1}{{\ds\partial_{v^2}^2{b}
(v^{2},u^{2})}}+\frac{1}{\epsilon}
\ .
\end{equation}
This procedure can be repeated to obtain
\begin{eqnarray}\label{lf2} 
\frac{\partial^{3} }{ \partial \chi ^{3}} \left[b_{0} (\chi )+\delta b
(\chi,u^{2} ) \right] &=&  \frac{\partial_{v^2}^3{b}
(v^{2},u^{2})} 
{\left( 1+\frac{1}{\epsilon } \partial_{v^2}^2{b} (v^{2},u^{2})\right)^3}\\
\frac{\partial^{4} }{ \partial \chi ^{4}} \left[b_{0} (\chi )+\delta b
(\chi,u^{2} ) \right] &=&  \frac{\partial_{v^2}^4{b}
(v^{2},u^{2})+\frac{1}{\epsilon}\left(\partial_{v^2}^4{b}
(v^{2},u^{2})\partial_{v^2}^2{b}
(v^{2},u^{2})-3\left[ \partial_{v^2}^3{b}
(v^{2},u^{2})\right]^{2}\right)} 
{\left( 1+\frac{1}{\epsilon } \partial_{v^2}^2{b}
(v^{2},u^{2})\right)^5} \ .\qquad 
\end{eqnarray}
Eliminating $\frac{\p \chi }{\p m }$ and $\frac{\p ^{2}}{\partial \chi
^{2}}[b_{0} (\chi )+\delta b (\chi ,u^{2})]$ from \Eq{beta-prelim}, we obtain
\begin{eqnarray}
-m \frac{\partial }{\partial m}\partial_{v^{2}}{b} (v^{2},u^{2})
&=& \epsilon\partial_{v^{2}} {b} (v^{2},u^{2}) +
\frac{\partial_{v^2}^2{b}
(v^{2},u^{2})}
{1+\frac{1}{\epsilon }\partial_{v^2}^2{b}
(v^{2},u^{2})}\frac{1}{\epsilon} 
\left(-m\frac{\partial }{\partial m} \right)\left(\partial
_{v^{2}}{b} (v^{2},u^{2})-\partial _{v^{2}}{b} (0,0)
\right)\nn\\ 
&&+ \epsilon \frac{\p}{\p\lambda}\lts_{\lambda=1}
\left[\frac{1}{\lambda}\frac{\partial }{\partial \chi }\delta b
(\lambda b_0',\lambda b_0'',\dots ) \right] \ .
\end{eqnarray}
We now take the limit of $v^{2},u^{2}\to 0$. We suppose that
$\left(-m\frac{\partial }{\partial m} \right)\left(\partial
_{v^{2}}{b} (v^{2},u^{2})-\partial _{v^{2}}{b} (0,0)
\right) \to 0 $ in that limit.  Further $\partial_{v^2}^2{b}
(v^{2},u^{2})$ can either remain finite or diverge. However
$\frac{\partial_{v^2}^2{b} (v^{2},u^{2})} {1+\frac{1}{\epsilon
}\partial_{v^2}^2{b} (v^{2},u^{2})}$ remains finite whatever
$\partial_{v^2}^2{b} (v^{2},u^{2})$ will do.  Supposing that
this argument is indeed correct (are there additional IR-divergences?)
the conclusion is that
\begin{eqnarray}
-m \frac{\partial }{\partial m}\partial_{v^{2}}{b} (0,0)
&=& \epsilon\partial_{v^{2}} {b} (0,0) 
+\lim_{u,v\to 0} \epsilon \frac{\p}{\p\lambda}\lts_{\lambda=1}
\left[\frac{1}{\lambda}\frac{\partial }{\partial \chi }\delta b
(\lambda b_0',\lambda b_0'',\dots ) \right] \ .
\end{eqnarray}
%This is equivalent to 
%begin{equation}
%m \frac{\partial }{\partial m}\partial_{v^{2}}{b} (v^{2},u^{2})
%,
%frac{1}{1+\frac{1}{\epsilon }\partial_{v^{2}}{b}(v^{2},u^{2})}
%= \epsilon\partial_{v^{2}} {b} 
%v^{2},u^{2})-\frac{\partial_{v^{2}}{b}(0,0)\partial_{v^{2}}^{2}{b}
%(v^{2},u^{2})}{1+\frac{1}{\epsilon 
%\partial_{v^{2}}^{2}{b} (v^{2},u^{2})} + \epsilon
%\frac{\p}{\p\lambda}\lts_{\lambda=1} 
%left[\frac{1}{\lambda}\frac{\partial }{\partial \chi }\delta b
%\lambda b_0',\lambda b_0'',\dots ) \right] \ .
%end{equation}
The $\beta $-function thus is equivalent to
\begin{eqnarray}\label{beta-tilde-b-prime}
 -m \frac{\partial }{\partial m}\partial_{v^{2}}{b}
(v^{2},u^{2}) &=&\epsilon \partial_{v^{2}}{b}(v^{2},u^{2}) +
[\partial_{v^{2}}{b} (v^{2},u^{2}) -\partial_{v^{2}}{b}
(0,0)]\partial_{v^{2}}^{2}{b}
(v^{2},u^{2})\nonumber \\
& & + \left(\epsilon +\partial_{v^{2}}^{2}{b} (v^{2},u^{2})
\right)\frac{\p}{\p\lambda}\lts_{\lambda=1}
\left[\frac{1}{\lambda}\frac{\partial }{\partial \chi } \delta b
(\lambda b_0',\lambda b_0'',\dots ) \right]\nonumber \\
& &-\partial_{v^{2}}^{2} {b} (v^{2},u^{2}) \lim_{v,u\to
0}\frac{\partial }{\partial \lambda }\lts_{\lambda=1}
\left[\frac{1}{\lambda }\frac{\partial }{\partial \chi} \delta b
(\lambda b_0,\lambda b_0',\dots ) \right] \ .
\end{eqnarray}
Using \Eq{pchipv2}, this can also be written as
\begin{eqnarray}
-m \frac{\partial }{\partial m}\partial_{v^{2}} {b}
(v^{2},u^{2}) &=& \epsilon\partial_{v^{2}} {b}(v^{2},u^{2}) +
[\partial_{v^{2}}{b} (v^{2},u^{2}) -\partial_{v^{2}}{b}
(0,0)]\partial_{v^{2}}^{2}{b} (v^{2},u^{2}) \nn\\
&&+ \epsilon
\frac{\p}{\p\lambda}\lts_{\lambda=1}\frac{\partial }{\partial v^{2}}
\left[\frac{1}{\lambda}\delta b (\lambda b_0',\lambda b_0'',\dots )
\right] \nonumber\\
&&-\partial_{v^{2}}^{2} {b} (v^{2},u^{2}) \lim_{v,u\to 0}\frac{\partial }{\partial \lambda }\lts_{\lambda=1} \left[\frac{1}{\lambda }\frac{\partial }{\partial \chi} \delta b (\lambda b_0,\lambda b_0',\dots ) \right]
\ .
\end{eqnarray}
Integrating the latter equation over $v^{2}$, we obtain 
\begin{eqnarray}\label{beta-tilde-b1}
 -m \frac{\partial }{\partial m}{b} (v^{2},u^{2})
&=& \epsilon {b}(v^{2},u^{2}) +\half  \partial_{v^{2}} {b} (v^{2},u^{2})^{2} -\partial_{v^{2}}{b} (v^{2},u^{2})\partial_{v^{2}}
{b} (0,0) + \epsilon
\frac{\p}{\p\lambda}\lts_{\lambda=1}
\left[\frac{1}{\lambda}\delta b (\lambda b_0',\lambda b_0'',\dots )
\right]\nonumber \\
&& - \partial_{v^{2}} {b} (v^{2},u^{2}) \lim_{v,u\to 0}\frac{\partial }{\partial
\lambda }\lts_{\lambda=1} \left[\frac{1}{\lambda }\frac{\partial
}{\partial \chi} \delta b (\lambda b_0,\lambda b_0',\dots ) \right] \ ,
\end{eqnarray}
which of course has to be read at $u^{2}=v^{2}$.
In a final step, we want to reintroduce proper quantities. 
Noting that 
\begin{equation}
\partial_{v^{2}} b (v^{2},u^{2})\ts_{u=v^2} = {b}' (v^{2}) -
\partial_{v^{2}} \delta b (\chi (v) ,v^{2}) \ ,
\end{equation}
(which does not need $u$), we obtain
\begin{eqnarray}\label{beta-tilde-b2} 
 -m \frac{\partial }{\partial m}{b} (v^{2})
&=& \epsilon {b}(v^{2}) +\half {b}' (v^{2})^{2} -{b}' (v^{2})
{b}' (0)  
- [{b}'(v^{2})-{b}'(0)] \partial_{v^{2}}\delta b (\chi (v)
,v^{2})\nn\\
&& + \partial_{v^{2}}\delta b (\chi ( 0),0) {b}' (v^{2}) +\half
\left[\partial_{v^{2}} \delta {b} (\chi (v), v^{2})\right]^{2}
-\partial_{v^{2}} \delta {b} (\chi (v), v^{2}) \partial_{v^{2}} \delta
{b} (\chi (0), 0)
\nn\\
&&+ \epsilon
\frac{\p}{\p\lambda}\lts_{\lambda=1}
\left[\frac{1}{\lambda}\delta b (\lambda b_0',\lambda b_0'',\dots )
\right]\nonumber \\
&&
 -\left[  {b}' (v^{2}) -
\partial_{v^{2}} b (\chi (v) ,v^{2}) \right] \lim_{v,u\to 0}\frac{\partial }{\partial
\lambda }\lts_{\lambda=1} \left[\frac{1}{\lambda }\frac{\partial
}{\partial \chi} \delta b (\lambda b_0,\lambda b_0',\dots ) \right] 
 \ .
\end{eqnarray}
(Of course
$\partial _{v^{2}}\delta b (\chi (0),0)$ means first to derive and then to
put the arguments to 0.) Also note that  $ \partial_{v^{2}}
\delta {b} (\chi (0), 0)$ is not 0, at least at order $T$.
The $\beta$-function at order $1/N$ therefore is
\begin{eqnarray}\label{beta-tilde-b3}
\ds\!\! -m \frac{\partial }{\partial m}{b} (v^{2})
&=&\ds \epsilon {b}(v^{2}) +\half {b}' (v^{2})^{2} -{b}' (v^{2})
{b}' (0)\nonumber \\
&&\ds+\,\frac{1}{N}\Bigg( \epsilon
\frac{\p}{\p\lambda}\lts_{\lambda=1}
\left[\frac{1}{\lambda}  b^{(1)} (\lambda b_0',\lambda b_0'',\dots )
\right]  
- [{b}'(v^{2}){-}{b}'(0)] \partial_{v^{2}}  b^{(1)} (\chi (v)
,v^{2})\nonumber \\
&&\ds\qquad\quad + {b}'(v^{2})\left\{ \partial_{u^{2}} b^{(1)}
(\chi (0) ,u^{2})\ts_{u=0}- \lim_{v,u\to 0}\frac{\partial }{\partial
\lambda }\lts_{\lambda=1} \left[\frac{1}{\lambda }\frac{\partial
}{\partial \chi} b^{(1)} (\lambda b_0,\lambda b_0',\dots ) \right]
\right\}
\Bigg)\!\!\!\!\!\nonumber \\
&&\ds +\,O\left(\frac1{N^{2}}\right)
\end{eqnarray}
This might better be grouped as
\begin{eqnarray}\label{beta-tilde-b4}
\!\! -m \frac{\partial }{\partial
m}{b} (v^{2}) &=&\ds \epsilon {b}(v^{2}) +\half {b}'
(v^{2})^{2} -{b}' (v^{2})
{b}' (0)\nonumber \\
&&\ds+\,\frac{1}{N}\Bigg( \epsilon
\frac{\p}{\p\lambda}\lts_{\lambda=1}
\left[\frac{1}{\lambda}  b^{(1)} (\lambda b_0',\lambda b_0'',\dots )
\right]  - {b}' (v^{2})  \lim_{v,u\to 0}\frac{\partial }{\partial
\lambda }\lts_{\lambda=1} \left[\frac{1}{\lambda }\frac{\partial
}{\partial \chi} b^{(1)}  (\lambda b_0,\lambda b_0',\dots ) \right]  
\nonumber \\
&&\ds\qquad\quad - [{b}'(v^{2}){-}{b}'(0)]
\partial_{v^{2}} b^{(1)} (\chi (v) ,v^{2}) + {b}'(v^{2})
\partial_{u^{2}} b^{(1)} (\chi (0) ,u^{2})\ts_{u=0}
\Bigg)\!\!\!\!\!\nonumber \\
&&\ds +\,O\left(\frac1{N^{2}}\right)
\ .
\end{eqnarray}
A caveat is in order: The rescaling has to be done on the level of
bare vertices, not on the level of renormalized ones. That would give
a wrong result. However the derivative w.r.t. $v^{2}$ can be taken in
any formulation.

\subsection{The case $d=0$}
As one can see from our final result for the $\beta$-function in
(\ref{beta1overN}), specified to $d=0$,  it is a polynom in $b$ of
finite order, since the denominators present in $h_{x} (p)$, see
(\ref{DEF:hx(p)}) are identical 1. Since this come as quite a
surprise, we show here why this must be so; actually it is a quite
general feature of the $1/N$-expansion of a  renormalizable theory in
$d=0$. 

We start to warm up with the diagram
\begin{equation}
\diagram{1oN3}\ .
\end{equation}
The leading and next-to-leading contributions (in ${b}$) are
\begin{equation}
\diagram{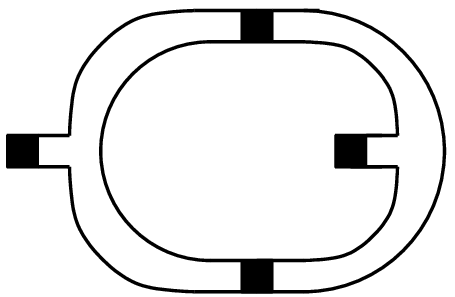} \qquad \diagram{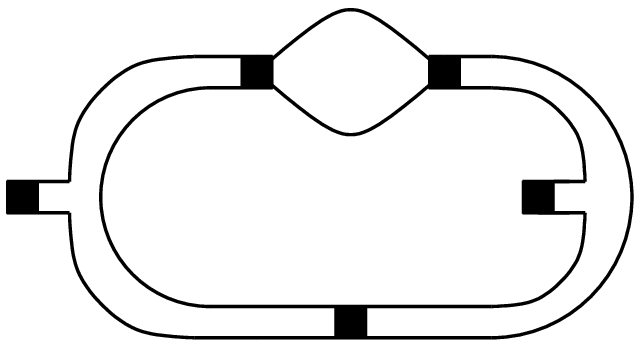} \qquad \diagram{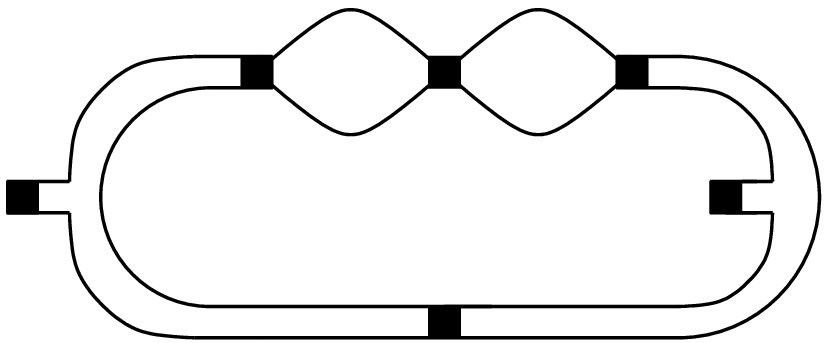} 
\qquad \diagram{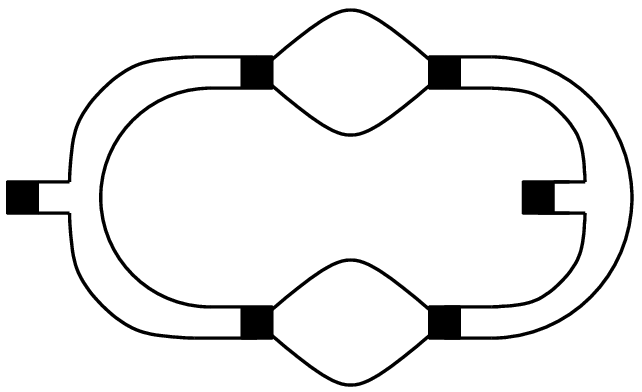}
\end{equation}
Our strategy is to apply Bogoliubov's $\cal R$-operation, see e.g.\
\cite{Zimmermann1969,Hepp1966,BogoliubovParasiuk1957,BergereLam1975}
and to show that only the first two terms contribute. Three remarks
are in order:

1.) The $\cal R$-operation in the context of a $1/N$-expansion is
maybe not entirely natural.  However we have in the above diagrams the
property,  that the terms already encountered at $1/N$ (and thus taken
care of in the $\beta $-function at leading order) are exactly the
iterated 1-loop diagrams, thus the first order in $1/N$.

2.) In order to extract the $\beta $-function from $\cal R$ applied to
a diagram, we only have to derive (w.r.t.\ $m$) the diagrams in the
boxes, since only those are counter-terms.

3.) Applying $\cal R$ to a functional of the bare $b_{0}$ gives the result as a
functional of ${b}$. Thus the contribution $\delta \beta
( {b})$ to the $\beta$-function
is $\delta \beta
( {b})= -m\frac{\partial }{\partial m} {\cal R}[\mbox{diagram} (b_{0})]$.
Now 
\begin{equation}
{\cal R} \diagram{1oN3s1} = \diagram{1oN3s1} -\fbox{\Diagram{1oN3s1}}
\end{equation}
which give as the contribution to $\beta $
\begin{equation}
\delta \beta( {b}) = - 3 \epsilon\  \fbox{\Diagram{1oN3s1}}\ .
\end{equation}
The second diagram gives
\begin{equation}
{\cal R} \diagram{1oN3s2} = \diagram{1oN3s2} -
\diagram{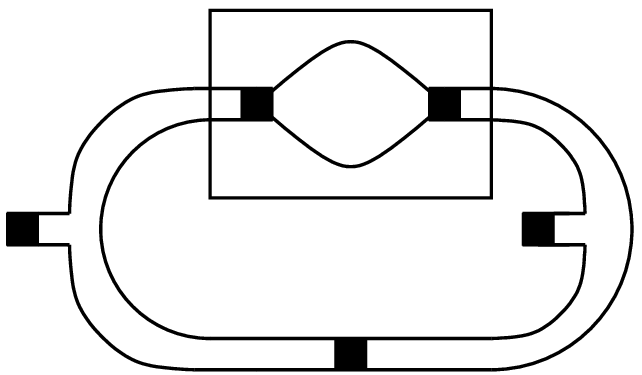}-\fbox{\Diagram{1oN3s2}}+\fbox{\Diagram{1oN3s2B}}
\end{equation}
In $D=0$, the last two diagrams cancel, and 
\begin{equation}
\delta \beta( {b}) = -  \epsilon\  \fbox{\Diagram{1oN3s2B}}\ .
\end{equation}
This gives also the ratio 6 to be found in the explicit formula: The 3
from the first diagram has to be set in relation to the combinatorial
factor of 2 for the second one and a factor of $1/\E $ for the second,
which together give a ratio of $6$; finishing the test.

We now proceed to higher orders: First we remark that for the chain
with $n$ members (here $n=3$) one can show recursively that
\begin{eqnarray}
{\cal R}\left[
\Diagram{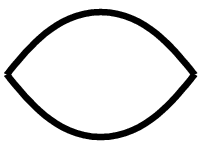}\!\Diagram{1loop}\!\Diagram{1loop}\right] =
\left[\diagram{1loop}-\ \fbox{\Diagram{1loop}}\ \right]^{n}
\end{eqnarray}
Thus having $n$ diagrams in the chain and deriving the $n$-dependence gives
\begin{equation}
-\frac{m\partial }{\partial m} {\cal R}\left[
\Diagram{1loop}\!\Diagram{1loop}\!\Diagram{1loop}\right] = n \E \
\fbox{\Diagram{1loop}}\ \left[\diagram{1loop}-\
\fbox{\Diagram{1loop}}\ \right]^{n-1}
\end{equation}
Of course, in $d=0$ the latter vanishes for $n>1$.

It is now easy to see that only finite order terms can appear, q.e.d.

\section{Details of the calculation of the $\beta$-function at finite $T$}
\label{app:detailsbeta}

\comment{**** KAY est REPASS\'e******}

\subsection{Integrals appearing in the $T>0$ correction to
the effective action}

The following functions have been defined but not given in formula
(\ref{b1compact}) in the main text.
\begin{eqnarray}\label{d=3.1new}
  g_4(a_x) &=& \frac{1}{A_{d}} \int_{p}  c (p) h_{x} (p) + \frac{1}{2} 
c (p) i_2(p) h_{x} (p)^2  \\
  g_5(a_x) &=& - \frac{1}{2} \frac{1}{A_{d}} \int_{p} 
c (p) i_2(p)^2 h_{x} (p)^2 \\
  g_6(a_x) &=& \frac{1}{A_{d}} \int_{p}  i_3(p) h_{x} (p) + \frac{1}{2} 
i_2(p) i_3(p) h_{x} (p)^2  \\
  g_7(a_x) &=& - \frac{1}{2} \frac{1}{A_{d}} \int_{p} 
c (p) i_2(p)^2 i_3(p) h_{x} (p)^2 \\
  g_8(a_0) &=& - \frac{1}{A_{d}} \int_{p} i_3(p) h_{0} (p) \\
  g_9(a_0) &=& - \frac{1}{2} \frac{1}{A_{d}} \int_{p} i_2(p) i_3(p)
  h_{0} (p) \\ 
  g_{10}(a_x) &=& \frac{1}{8 A_{d}} \int_{p} - \ln( 1 - a_x i_2(p))
+ i_2(p) h_x(p) + \frac{1}{4} i_2(p)^2 h_x(p)^2  \\
  g_{11}(a_x) &=& - \frac{1}{16 A_{d}} \int_{p} 
i_2(p)^2 h_x(p) + i_2(p)^3 h_x(p)^2  \\
  g_{12}(a_x) &=& \frac{1}{16 A_{d}} \int_{p}  
- i_2(p)^3 h_x(p) + \frac{1}{2} i_2(p)^4 h_x(p)^2  \\
   \gamma&=& - \frac{1}{4} i_1^2 i_2 \\ 
  g_{13}(a_0) &=& - \frac{3}{8} i_2 \int i_2(p)^2 h_0(p) 
\end{eqnarray}
and we recall that a cutoff $\Lambda/m$ is implicit in all these
rescaled momentum integrals.

These functions are not all independent. Indeed, defining
\begin{equation}
 \gamma _{m,n}(a_x) := \frac{1}{A_{d}} \int_{p} f(p) i_2(p)^n h_x(p)^m
\end{equation}
one easily establishes the recursion relations:
\begin{eqnarray}
 m \gamma_{m+1,n}(a) &=& a^2 \partial_a \gamma_{mn}(a)  \nonumber \\
 \gamma_{m+1,n+1}(a) &=& \frac{1}{a} \gamma_{m+1,n}(a) -
 \gamma_{m,n}(a) \label{rec}\ . 
\end{eqnarray}
They allow to obtain these functions from derivatives of
\begin{equation}
 \gamma(a)= - \frac{1}{A_{d}} \int_{p} \ln( 1 - a i_2(p)) f(p)
\end{equation}
Specializing first to $f(p)=1$, one finds using (\ref{rec})
\begin{eqnarray}
  \gamma_{1,1} (a) &=& a \gamma'(a) \\
  \gamma_{1,2} (a) &=& \gamma'(a) - i_1^{2} \\
  \gamma_{1,3} (a) &=& \frac{1}{a} (\gamma'(a) - i_1^2) - \gamma_{0,2} (a) \\
  \gamma_{2,2} (a) &=& a^2 \gamma''(a) \\
  \gamma_{2,3} (a) &=& a \gamma''(a) - \gamma'(a) + i_1^2 \\
  \gamma_{2,4} (a) &=& \gamma''(a) - \frac{2}{a} (\gamma'(a) - i_1^2) +
\gamma_{0,2} (a)
\end{eqnarray}
Thus the functions $g_{10}(a)$, $g_{11}(a)$, $g_{12}(a)$ and
$g_{13}(a)$ can all be obtained from derivatives of $\gamma(a)$ with
$f(p)=1$. Similarly, $g_{4}(a)$ and $g_{5}(a)$ can be all obtained
from derivatives of $\gamma(a)$ with $f(p)=c(p)$, and similarly for
$g_{6}(a)$, $g_{8}(a)$ and $g_{9}(a)$ with $f(p)=i_3(p)$, $g_7(a)$
with $f(p)=i_3(p) c(p)$.

One may further attempt to relate $\gamma(a)$ for different functions
$f(p)$. Since $\partial_{m^2} I_2 = - 2 I_3$, one can use that
\begin{equation}
 i_3(p) = \frac{1}{4} \epsilon i_2(p) - \frac{1}{4} p i_2'(p) - \frac{1}{4}
\lambda \partial_\lambda i_2(p)
\end{equation}
where $\lambda=\Lambda/m$. Integration by part yields identities such as
\begin{equation}
\int  \frac{\rmd p}{p} p^d (p \partial_p + \lambda \partial_\lambda)
H(i_2(p)) = 
- d \int^\lambda  \frac{\rmd p}{p} p^d  H(i_2(p)) +
\lambda \partial_\lambda \int^\lambda \frac{\rmd p}{p} p^d H(i_2(p)) \
,
\end{equation}
which can be used to relate the integrals.

\subsection{Calculation of the $T>0$ $\beta$-function} Below we
compute $- m \partial_m b(x)$ at fixed $\Lambda/m$, thus we truly compute
 $- (m \partial_m + \Lambda \partial_\Lambda) b(x)$.  It is therefore
useful at $T>0$ only for $d<2$ when all integrals are UV convergent
and the limit $\Lambda/m \to \infty$ can be taken with no further
redefinitions. The calculation of the $\beta$-function for $d\ge 2$
and $T>0$ requires further redefinitions and will eventually be 
detailed elsewhere.

Taking into account all $m$-dependence at $T>0$ in (\ref{b1compact})
one obtains (up to an additive constant):
\begin{eqnarray}
&&\!\!\!\!\!\beta_1[b](x)  = - m \partial_m b(x)   \\
&&\quad =  \frac{\delta b_1}{\delta a_x} \left[ - m \partial_m^0 a_x\right]
+ \frac{\delta b_1}{\delta [b'(x)-b'(0)] } \left[ - m \partial_m^0
(b'(x)-b'(0))\right]
+ \frac{\delta b_1}{\delta a_0} (- m \partial_m^0 a_0) 
+ \frac{\delta b_1}{\delta T_m}(- m \partial_m^0 T_m) \nonumber \\
&&\qquad+ \frac{\delta b_{1}}{\delta \alpha} (-m \partial_m \alpha )
  - \epsilon b_1 - ( b' - b'(0) ) (b_1' - b_1'(0))
 - T_m  \frac{b_1'(x)}{1 + b''(0)/\epsilon} +
T_m  \frac{b'(x)}{(1 + b''(0)/\epsilon)^2} \frac{b_1''(0)}{\epsilon}
\nonumber
\end{eqnarray}
where $b_1$ is given in (\ref{b1compact}). 

One uses that:
\begin{eqnarray}
- m \partial_m^0 a_x &=& \epsilon a_x + [b'(x)-b'(0)] a'_x 
+ \frac{T_m}{1 + b''(0)/\epsilon} a'_x \\
- m \partial_m^0 (b'(x)-b'(0)) &=& [\epsilon + b''(x)] [b'(x)-b'(0)] + 
\frac{T_m}{1 + b''(0)/\epsilon} [b''(x)-b''(0)] \\
 - m \partial_m^0 \alpha &=& \alpha \epsilon +T_{m}
   \left(\frac{\alpha^{2}}{a_{0}-\epsilon} +
   \frac{(a_{0}-\epsilon)^{2}\bar \alpha}\epsilon \right)
%=\epsilon \alpha +
% \frac{T_m}{1 + b''(0)/\epsilon} \left[ \frac{a_0''}{1 + b''(0)/\epsilon}
%- \frac{(a_0')^2}{\epsilon} \right]
\end{eqnarray}
and $- m \partial_m T_m = - \theta T_m$.
After a rather tedious calculation one obtains the form (\ref{finiteTbetaexpr})
 given in the text
with the following definitions (for $d<2$):
\begin{eqnarray}
 \tilde g_4(a) &=& - 4 g_1(a) - \epsilon (\epsilon + \theta)  g_4(a)  
+ \epsilon^2 a g_4'(a) \\
  \tilde g_5(a) &=& \frac{4}{\epsilon} g_1(a) 
- 2 g_2(a) - \epsilon \theta g_5(a) + \epsilon^2 a g_5'(a) \\
 \tilde g_6(a) &=& -2 g_{2} (a)- \epsilon \theta g_{6} (a) +
 \epsilon^{2} a g_{6}' (a)   
- \epsilon g_4(a)   \\
  \tilde g_7(a) &=& \frac{2}{\epsilon} g_2(a) - 4 g_3(a) - \epsilon g_5(a) 
+  \epsilon (\epsilon - \theta) g_7(a) + \epsilon^2 a g_7'(a)  \\
  \tilde g_8(a) &=& 
\frac{a \epsilon^2}{\epsilon-a} 
\left[ g_6(a)+g_8(a)+a(g_7(a)+g_9(a)) \right]
+ \frac{4}{\epsilon} g_1(a)  + \epsilon g_4(a) - \epsilon \theta g_{8} (a) \\
&& +
a \left[- \frac{8}{\epsilon^2} g_1(a) + \frac{4}{\epsilon} g_2(a) +
\epsilon g_5(a) + \epsilon (\epsilon - \theta)  g_9(a)  
 + \epsilon^2 g_8'(a) \right] \\
&& +
a^2 \left[ \frac{4}{\epsilon^3} g_1(a) - \frac{4}{\epsilon^2} g_2(a)
+ \frac{4}{\epsilon} g_3(a) + \epsilon^2 g_9'(a) \right]
\\
 \tilde g_{10}(a) &=& - \epsilon^2 (\epsilon + 2 \theta) g_{10}(a)
- \epsilon  g_4(a) + \epsilon^3 a g_{10}'(a) \\
 \tilde g_{11}(a) &=& - 2 \epsilon^2 \theta g_{11}(a) +  g_4(a) - \epsilon
g_5(a) - \epsilon g_6(a)   + \epsilon^3 a g_{11}'(a) \\
\tilde g_{12}(a) &=& \epsilon^2 (\epsilon - 2 \theta) g_{12}(a) +  g_5(a) - \epsilon
g_7(a)   + \epsilon^{3} a g_{12}'(a) \\
 \phi(a) &=& \frac{\epsilon}{\epsilon -a}\left[ 
{g_6}(a) + a{g_7}(a) + {g_8}(a) + \epsilon {g_9}(a) + {\epsilon }^2
g_{10}'(a) + a{\epsilon }^2 g_{11}'(a) + a^2{\epsilon }^2
g_{12}'(a)+ \epsilon^{3} \gamma + a \epsilon^{3}g_{13} (a) 
\right]\nonumber\!\!\! \\
&& + 2 g_{4}' (a) + \epsilon g_{8}' (a) - 2 \epsilon^{2}\gamma \theta
+ a \left[ \epsilon^{2}( {\epsilon } - 2 \theta ) g_{13}(a) -
\frac2\epsilon {g_4}'(a) + 2 g_5'(a) + 2 g_6'(a) + \epsilon
 g_9'(a)
 \right] \nonumber \\
&& +a^{2}\left[{ \epsilon }^3 g_{13}'(a) - \frac{2g_5'({a})}{\epsilon } +
2 g_7'({a}) \right]
 \\
 \psi(a)&=& {\epsilon}
[ g_{10}''(a) + a g_{11}''(a) + a^2 g_{12}''(a) 
+ \epsilon g_{13}(a) + a \epsilon g_{13}'(a) ] \\
 \tilde \psi(a) &=& 
(\epsilon-a)^2  [  g_{10}'(a) + a g_{11}'(a) + a^2 g_{12}'(a) +
\epsilon \gamma +  \epsilon a g_{13} (a) ]
\end{eqnarray}

\section{Integrals} \label{app:integrals}
\comment{****Pierre recheck all, there were 3 important mistakes***}
\subsection{Definitions}\label{sec:int definitions}
\begin{eqnarray}\label{a:int:1}
I_{n}&:=&\int_{k} \frac1{( k^{2}+m^{2})^{n}}\\
\label{a:int:2}
I_{2}&=&\displaystyle A_{d}
\frac{m^{-\epsilon}}{\epsilon}\\
\label{a:int:3}
A_d &:=& \frac{2 \Gamma ( 3 - d/2) }{(4\pi)^{d/2}}
\ .
\end{eqnarray}
The momentum dependent ones are
\begin{eqnarray}\label{app:I2p}
I_{2}(p) &:=& \int_{k} \frac1{(k+p/2)^{2}+m^{2}}
\frac1{(k-p/2)^{2}+m^{2}}\\\label{app:I3p}
 I_{3}(p) &:=&
\int_{k}\frac{1}{[(k+p/2)^{2}+m^{2}]^{2}}\frac{1}
{(k-p/2)^{2}+m^{2}}\\\label{app:I4p}
\displaystyle I_{4}(p)&:=&
\int_{k}\frac{1}{[(k+p/2)^{2}+m^{2}]^{2}}\frac{1}
{[(k-p/2)^{2}+m^{2}]^{2 }}
\ .
\end{eqnarray}
Dimensionless rescaled variants:
\begin{eqnarray}\label{app:inp}
 i_{n} (p)&:=&\frac{I_{n} (p)}{A_{d}} \Bigg|_{m=1}\\
\label{app:in}
 i_{n}&:=&\frac{I_{n}}{A_{d}}\Bigg|_{m=1}
\ .
\end{eqnarray}

\subsection{Integrals  in fixed dimensions, general formulas}
%\begin{equation}\label{app:i2gen}
%i_{2} (p) = \frac{1}{4-d}\int_{1}^{\infty} \rmd w\, w ^{2-d} (w^{2}+
%(w-1)p^{2})^{d/2-2}
%\end{equation}
%Sometimes it is helpful to restore the dependence on $m$ (but keeping
%the normalization) in order to be able to derive w.r.t.\ it:
%\begin{equation}\label{i2m}
%i_{2}^{m} (p) = \frac{m^{4-d}}{4-d}\int_{1}^{\infty} \rmd w\, w ^{2-d}
%(w^{2}+ (w-1)p^{2}/m^{2})^{d/2-2}
%\end{equation}
%This is useful to get $i_{3} (p)$: 
%\begin{equation}\label{i3gen}
%i_{3}^{m} (p)=-1/2 \frac{\rmd }{\rmd m^{2}} i_{2} (p)
%\end{equation}
%\begin{equation}\label{i4gen}
%i_{4} (p)= \frac{(4-\frac{d}{2})
%(3-\frac{d}{2})}{4-d}\int_{1}^{\infty} \rmd w\,(w-1) w ^{2-d} (w^{2}+
%(w-1)p^{2})^{d/2-4}
%\end{equation}
%\subsubsection{General case: $j_{n,m} (p)$}\label{jnmp}
The general case can be treated as follows:
\begin{equation}\label{ooo}
j_{n,m} (p):=  \frac{1}{A_{d}}\int_{k}
\frac{1}{\left[ (k-\frac{p}{2})^{2}+1\right]^{m}} 
\frac{1}{\left[(k+\frac{p}{2})^{2}+1 \right]^{n}} \ .
\end{equation}
Using the usual Schwinger-parameter representation, this can be
written as
\begin{eqnarray}\label{lf15aaa}
j_{n,m} (p) &=& \frac{1}{\Gamma (n)\Gamma
(m)}\left(\frac{1}{A_{d}}\int_{k} \rme^{-k^2}\right)
\int_{\alpha,\beta >0} \alpha^{n-1}\beta^{m-1}
\left(\alpha +\beta \right)^{-\frac{d}{2}} \rme^{- (\alpha+\beta )} \rme^ {-\frac{\alpha \beta }{\alpha +\beta }p^{2}}\nn \\
&=& \frac{\Gamma (n+m-d/2)}{2 \Gamma (3-d/2)\Gamma (n)\Gamma (m)}
\int_{\beta >0} \frac{\beta^{m-1}}{(1+\beta)^{n+m}} \left[1+
\frac{\beta }{(1+\beta)^{2}}\, p^{2} \right]^{\frac{d}{2}-n-m}\ .
\end{eqnarray}
We make the change of variables $\beta =\frac{s}{1-s}$:
\begin{eqnarray}\label{lf16toc}
j_{n,m} (p) &=& \frac{\Gamma (n+m-d/2)}{2 \Gamma (3-d/2) \Gamma (n)\Gamma
(m)} \int_{0}^{1}\rmd
s\, s^{m-1} (1-s)^{n-1} \left[1+s
(1-s) {p^{2}} \right]^{\frac{d}{2}-m-n}\nn \\ 
&=& \frac{\Gamma (n+m-d/2)}{\Gamma (3-d/2) \Gamma (n)\Gamma (m)2^{n+m}}\nonumber \\ 
&&\times \int_{0}^{1 }\frac{\rmd y}{\sqrt{1{-}y}} \,
\frac{(1{+}\sqrt{1{-}y})^{m-1}
(1{-}\sqrt{1{-}y})^{n-1}+(1{+}\sqrt{1{-}y})^{n-1}
(1{-}\sqrt{1{-}y})^{m-1}}{2} \left[1{+}y \frac{p^{2}}{4 }
\right]^{\frac{d}{2}-m-n} \ ,\nonumber \\
&&
\end{eqnarray}
where we have used another new variable $y=4 s (1-s)$. Note that the
large fraction $\frac{\cdots}{2}$ in the above expression is such that
only integer powers of $(1-y)$ survive.  Some simplifications occur
for $n=m$, and $i_{3} (p)=j_{1,2} (p)$:
\begin{eqnarray}
j_{n,n} (p) &=& \frac{\Gamma (2n-d/2)}{\Gamma (3-d/2)\Gamma
(n)^{2}\,2^{2n} } \int_{0}^{1 }\frac{\rmd y}{\sqrt{1{-}y}} \, y^{n-1}
\left[1{+}y \frac{p^{2}}{4 }
\right]^{\frac{d}{2}-2n}\\
i_{3} (p)= j_{1,2} (p)&=& \frac{1}{8}\int_{0}^{1 }\frac{\rmd
y}{\sqrt{1{-}y}} \, \left[1{+}y \frac{p^{2}}{4 }
\right]^{\frac{d}{2}-3}\ .
\end{eqnarray}

\subsection{$d=0$}\label{app:d=0integrals}
\begin{equation}
i_{n}=i_{n} (p)=1/4 \ , \qquad A_{d}=4\ ,\qquad  \epsilon =4\ .
\end{equation}

\subsection{$d=1$}\label{app:d=1integrals}
\begin{eqnarray}
A_{d}&=& \frac{3}{4} \\
i_{1} &=& \frac{2}{3}\\
i_{{2}} (p)&=& \frac{4}{3 (4+p^{2})} \\
i_{3} (p)&=&  \frac{12+p^{2}}{3 (4+p^{2})^{2}} \\
i_{4} (p)&=& \frac{2( 20 + p^2 ) }{3{( 4 + p^2 ) }^3}
\ .
\end{eqnarray}

\subsection{$d=2$}\label{app:d=2integrals}
\begin{equation}
A_{d}=\frac{1}{2\pi}
\end{equation}
\begin{equation}\label{app:i2:d=2}
i_{2} (p)= \frac{\mbox{arctanh}\left(\frac{|p|
\sqrt{4+p^{2}}}{2+p^{2}} \right)}{|p|\sqrt{4+p^{2}}} = 
\frac{\ln \left({2+p^{2}} +|p|
\sqrt{4+p^{2}} \right)-\ln \left({2+p^{2}}-{|p|
\sqrt{4+p^{2}}} \right)}{2|p| \sqrt{4+p^{2}}} =  \frac{2\, \mbox{arcsinh}
(\frac{|p|}{2})}{|p| \sqrt{4+p^{2}}} 
\end{equation}
\begin{equation}\label{app:i3:d=2}
i_{3} (p)= \frac{1}{8 + 2p^2} +
\frac{2\,\mbox{arcsinh}(\frac{|p|}{2})}{|p|{( 4 + p^2 )
}^{\frac{3}{2}}}
\end{equation}
\begin{equation}\label{app:i4:d=2}
i_{4} (p) = \frac{1}{6}\  {}_{2}F_{1} (2,3,5/2,-p^{2}/4)
\ .
\end{equation}

\subsection{$d=3$}\label{d=3integrals}
\begin{equation}
A_{d}=\frac{1}{8\pi}
\end{equation}
\begin{equation}\label{app:i2:d=3}
i_{2} (p) = 2 \frac{\arctan (\frac{|p|}{2})}{|p|} =
\frac{i}{|p|}\left[\ln (2-i|p|)-\ln (2+i|p|) \right]
\end{equation}
\begin{equation}\label{app:i3:d=3}
i_{3} (p) = \frac{1}{p^{2}+4}
\end{equation}
\begin{equation}\label{app:i4:d=3}
i_{4} (p)=\frac{2}{(p^{2}+4)^{2}}
\ .
\end{equation}

\vfill

\section{Summary of Notation}\label{sec:Notation}
{\renewcommand{\arraystretch}{1.5}\begin{tabular}{|c|c|c|}
\hline 
symbol & definition & defined in equation\\
%%%%%%%%%%%%%%%%%%%%%%%%%%%%%%%%%%%
\hline\hline
$\epsilon$ & $ \epsilon =4-d$ &\\
%%%%%%%%%%%%%%%%%%%%%%%%%%%%%%%%%%%
\hline 
$\zeta$, $\theta$ & $\zeta=$ roughness, $\theta =d-2+2\zeta$ (thermal
exponent) & \\
%%%%%%%%%%%%%%%%%%%%%%%%%%%%%%%%%%%
\hline 
$u (x)$, $v (x)$ & $u (x) = $ field, $\displaystyle v (x)={u (x)}/{\sqrt{N}}$ & \\
%%%%%%%%%%%%%%%%%%%%%%%%%%%%%%%%%%%
\hline \parbox{0in}{\rule{0mm}{6.5ex}} $I_{n}$ & $ \displaystyle
I_{n}:=\int_{k} \frac1{( k^{2}+m^{2})^{n}}$,~~$I_{2}=\displaystyle A_{d}
\frac{m^{-\epsilon}}{\epsilon}$,~~$\displaystyle A_d = \frac{2 \Gamma
( 3 - d/2) }{(4\pi)^{d/2}}$ &
(\ref{defIn}) \\ 
%%%%%%%%%%%%%%%%%%%%%%%%%%%%%%%%%%%
\hline \parbox{0in}{\rule{0mm}{6.5ex}} $I_{2} (p) $ & $\displaystyle
I_{2}(p) := \int_{k} \frac1{(k+p/2)^{2}+m^{2}}
\frac1{(k-p/2)^{2}+m^{2}}$ & (\ref{I2p}) \\ 
%%%%%%%%%%%%%%%%%%%%%%%%%%%%%%%%%%%
\hline \parbox{0in}{\rule{0mm}{6.5ex}} $\displaystyle I_{3}(p)$ &
$\displaystyle I_{3}(p):=
\int_{k}\frac{1}{[(k+p/2)^{2}+m^{2}]^{2}}\frac{1}
{(k-p/2)^{2}+m^{2}} $  & (\ref{I3})\\
%%%%%%%%%%%%%%%%%%%%%%%%%%%%%%%%%%%
\hline 
\parbox{0in}{\rule{0mm}{6.5ex}} $\displaystyle I_{4}(p)$ &
$\displaystyle I_{4}(p):=
\int_{k}\frac{1}{[(k+p/2)^{2}+m^{2}]^{2}}\frac{1}
{[(k-p/2)^{2}+m^{2}]^{2 }} $  & (\ref{I4})\\
%%%%%%%%%%%%%%%%%%%%%%%%%%%%%%%%%%%
\hline
 $C (p)$ & $\displaystyle C (p):= ( p^2+m^{2})^{-1}$ & \\
%%%%%%%%%%%%%%%%%%%%%%%%%%%%%%%%%%%
\hline \parbox{0in}{\rule{0mm}{6.5ex}} $i_{n} (p)$,\ $i_{n}$ &
$\displaystyle i_{n} (p):=\frac{I_{n} (p)}{A_{d}} \Bigg|_{m=1}$,\ \ 
$\displaystyle i_{n}:=\frac{I_{n}}{A_{d}}\Bigg|_{m=1}$ &
(\ref{DEFin}), (\ref{DEFi0})\\ 
%%%%%%%%%%%%%%%%%%%%%%%%%%%%%%%%%%%
\hline 
 $c (p)$ & $\displaystyle c (p):= ( 1+p^{2})^{-1}$ & \\
%%%%%%%%%%%%%%%%%%%%%%%%%%%%%%%%%%%
\hline 
$B (\ldots)$ & second cumulant of bare disorder &\\
%%%%%%%%%%%%%%%%%%%%%%%%%%%%%%%%%%%
\hline
$\tilde B (\dotsb)$ & second cumulant of  renormalized disorder (not
rescaled) & \\ 
%%%%%%%%%%%%%%%%%%%%%%%%%%%%%%%%%%%
\hline $B'_{ab}$, $B''_{ab}$, $\tilde B'_{ab}$, etc. & $B'_{ab}:= B'
(\bar \chi_{ab}) $,\ $B''_{ab}:= B'' (\bar \chi_{ab}) $,\ $\tilde
B'_{ab}:=\tilde B' (v_{ab}^{2})$,\ etc. &
\\
%%%%%%%%%%%%%%%%%%%%%%%%%%%%%%%%%%%
\hline 
$b (z)$ &$ b (z) := 4 A_{d}m^{4\zeta -\epsilon}\tilde B (z
m^{-2\zeta})$ &(\ref{changeofvar}) \\
%%%%%%%%%%%%%%%%%%%%%%%%%%%%%%%%%%%
\hline
$\chi_{ab} (x)$, $\lambda_{ab} (x)$ &
auxiliary fields &\\
%%%%%%%%%%%%%%%%%%%%%%%%%%%%%%%%%%%
\hline
 $\tilde \chi_{ab} (x)$ &  $\tilde \chi_{ab} (x) := \chi_{ab} (x)+\chi_{ba}
(x)-\chi_{aa} (x)-\chi_{bb} (x) $ &\\
%%%%%%%%%%%%%%%%%%%%%%%%%%%%%%%%%%%
\hline $\tilde \chi^{ab}_{v}\, $, $\bar \chi^{ab}_{v}$ & $\displaystyle \tilde
\chi^{ab}_{v}=\tilde \chi_{ab} (x)|_{v (x)=v}$,\ \  $\tilde \chi_v^{ab}=
\bar \chi^{ab}_{v}+ O
(\frac{1}{N})$ &  \\
%%%%%%%%%%%%%%%%%%%%%%%%%%%%%%%%%%%
\hline \parbox{0in}{\rule{0mm}{5ex}} $\bar \chi_{v}^{ab}=\bar
\chi_{v}=\bar \chi_{ab}$ & $\bar \chi_{v}^{ab}:=v_{ab}^{2}+2 T I_{1}+
4 I_{2}\left[\tilde B'_{ab}-\frac{1}{2}\left(\tilde B'_{aa}+\tilde
B'_{bb} \right) \right] $ & 
\\
%%%%%%%%%%%%%%%%%%%%%%%%%%%%%%%%%%%
\hline \parbox{0in}{\rule{0mm}{6.5ex}} $H_{v} (p)$ &$\displaystyle H_{v}
(p):=\frac{B''(\bar \chi_{v})}{1-4 I_{2}(p)\, B''(\bar \chi_{v})} $ &
(\ref{Hp}) \\
%%%%%%%%%%%%%%%%%%%%%%%%%%%%%%%%%%%
\hline \parbox{0in}{\rule{0mm}{6.5ex}} $\tilde H_{x} (p)$
&$\displaystyle \tilde H_{x} (p):=\frac{\tilde B''(x)}{1+4 [ I_{2}-
I_{2}(p)]\,\tilde B''(x)} $ &
(\ref{asdf}) \\
%%%%%%%%%%%%%%%%%%%%%%%%%%%%%%%%%%%
\hline \parbox{0in}{\rule{0mm}{6.5ex}} $h_{x} (p)$ &$\displaystyle h_{x}
(p):=\frac{b''(x)}{1+ [ i_{2}- i_{2}(p)]\,b''(x)} $ &
(\ref{DEF:hx(p)}) \\
%%%%%%%%%%%%%%%%%%%%%%%%%%%%%%%%%%%
\hline
$T$ & temperature &\\
%%%%%%%%%%%%%%%%%%%%%%%%%%%%%%%%%%%
\hline 
 $T_{m}$ & $\displaystyle T_{m}:=
{4T A_{d}m^{\theta}}/{\epsilon }$ & \\ 
%%%%%%%%%%%%%%%%%%%%%%%%%%%%%%%%%%%
\hline
\end{tabular}}

%\tableofcontents
%\bibliographystyle{../macros/KAY}
%\bibliography{../citation/citation}

\begin{thebibliography}{10}

\bibitem{ChauveLeDoussalWiese2000a}
P.~Chauve, P.~Le Doussal  and K.J. Wiese,
\newblock {\em Renormalization of pinned elastic systems: How does it work
  beyond one loop?},
\newblock Phys. Rev. Lett. {\bf 86} (2001)   1785--1788,
\newblock cond-mat/{\bf 0006056}.

\bibitem{LeDoussalWiese2001v1}
P.~Le Doussal and K.J. Wiese,
\newblock {\em Functional renormalization group at large {$N$} for random
  manifolds},
\newblock cond-mat\slash {\bf 0109204} (2001).

\bibitem{LeDoussalWiese2001}
P.~Le Doussal and K.J. Wiese,
\newblock {\em Functional renormalization group at large {$N$} for random
  manifolds},
\newblock Phys. Rev. Lett. {\bf 89} (2002)   125702,
\newblock cond-mat/{\bf 0109204v1}.

\bibitem{LeDoussalWieseChauve2002}
P.~Le Doussal, K.J. Wiese  and P.~Chauve,
\newblock {\em 2-loop functional renormalization group analysis of the
  depinning transition},
\newblock Phys. Rev. B {\bf 66} (2002)   174201,
\newblock cond-mat/{\bf 0205108}.

\bibitem{LeDoussalWiese2002a}
P.~Le Doussal and K.J. Wiese,
\newblock {\em Functional renormalization group for anisotropic depinning and
  relation to branching processes},
\newblock Phys. Rev. E {\bf 67} (2003)   016121,
\newblock cond-mat/{\bf 0208204}.

\bibitem{LeDoussalWiese2003a}
P.~Le Doussal and K.J. Wiese,
\newblock {\em Higher correlations, universal distributions and finite size
  scaling in the field theory of depinning},
\newblock Phys. Rev. E {\bf 68} (2003)   046118,
\newblock cond-mat/{\bf 0301465}.

\bibitem{LeDoussalWieseChauve2003}
P.~Le Doussal, K.J. Wiese  and P.~Chauve,
\newblock {\em Functional renormalization group and the field theory of
  disordered elastic systems},
\newblock cont-mat\slash {\bf 0304614} (2003).

\bibitem{LeDoussalWiese2003b}
P.~Le Doussal and K.J. Wiese,
\newblock {\em Functional renormalization group at large ${N}$ for disordered
  elastic systems, and relation to replica symmetry breaking},
\newblock Phys. Rev. B {\bf 68} (2003)   17402,
\newblock cond-mat/0305634.

\bibitem{RossoKrauthLeDoussalVannimenusWiese2003}
A.~Rosso, W.~Krauth, P.~Le Doussal, J.~Vannimenus  and K.J. Wiese,
\newblock {\em Universal interface width distributions at the depinning
  threshold},
\newblock Phys. Rev. E {\bf 68} (2003)   036128,
\newblock cond-mat{\slash\bf 0301464}.

\bibitem{ChauveLeDoussal2001}
P.~Chauve and P.~Le Doussal,
\newblock {\em Exact multilocal renormalization group and applications to
  disordered problems},
\newblock Phys. Rev. E {\bf 64} (2001)   051102/1--27,
\newblock cond-mat/{\bf 9602023}.

\bibitem{MezardParisi1991}
M.~M\'ezard and G.~Parisi,
\newblock {\em Replica field theory for random manifolds},
\newblock J. Phys. I (France) {\bf 1} (1991)   809--837.

\bibitem{MezardParisi1992}
M.~Mezard and G.~Parisi,
\newblock {\em Manifolds in random media: two extreme cases},
\newblock J. Phys. I (France) {\bf 2} (1992)   2231--42.

\bibitem{BrunetDerrida2000a}
E.~Brunet and B.~Derrida,
\newblock {\em Probability distribution of the free energy of a directed
  polymer in a random medium},
\newblock Phys. Rev. E {\bf 61} (2000)   6789--801.

\bibitem{BrunetDerrida2000}
E.~Brunet and B.~Derrida,
\newblock {\em Ground state energy of a non-integer number of particles with
  delta attractive interactions},
\newblock Physica A {\bf 279} (2000)   398--407.

\bibitem{DSFisher1985}
D.S. Fisher,
\newblock {\em Sliding charge-density waves as a dynamical critical phenomena},
\newblock Phys. Rev. {\bf B 31} (1985)   1396--1427.

\bibitem{Fisher1985b}
DS. Fisher,
\newblock {\em Random fields, random anisotropies, nonlinear sigma models and
  dimensional reduction},
\newblock Phys. Rev. B {\bf 31} (1985)   7233--51.

\bibitem{DSFisher1986}
D.S. Fisher,
\newblock {\em Interface fluctuations in disordered systems: {$5-\epsilon$}
  expansion},
\newblock Phys. Rev. Lett. {\bf 56} (1986)   1964--97.

\bibitem{Nattermann1987}
T.~Nattermann,
\newblock {\em Interface roughening in systems with quenched random
  impurities},
\newblock Europhys. Lett. {\bf 4} (1987)   1241--6.

\bibitem{NarayanDSFisher1990}
O.~Narayan and D.S. Fisher,
\newblock {\em Logarithmic effects on the critical behavior of superfluids in
  random media},
\newblock Phys. Rev. B {\bf 42} (1990)   7869--75.

\bibitem{NattermanStepanowTangLeschhorn1992}
T.~Nattermann, S.~Stepanow, L.H. Tang  and H.~Leschhorn,
\newblock {\em Dynamics of interface depinning in a disordered medium},
\newblock J. Phys. II (France) {\bf 2} (1992)   1483--1488.

\bibitem{NarayanDSFisher1992a}
O.~Narayan and D.S. Fisher,
\newblock {\em Dynamics of sliding charge-density waves in 4- epsilon
  dimensions},
\newblock Phys. Rev. Lett. {\bf 68} (1992)   3615--18.

\bibitem{NarayanDSFisher1992b}
O.~Narayan and D.S. Fisher,
\newblock {\em Critical behavior of sliding charge-density waves in 4- epsilon
  dimensions},
\newblock Phys. Rev. B {\bf 46} (1992)   11520--49.

\bibitem{NarayanDSFisher1993a}
O.~Narayan and D.S. Fisher,
\newblock {\em Threshold critical dynamics of driven interfaces in random
  media},
\newblock Phys. Rev. B {\bf 48} (1993)   7030--42.

\bibitem{NarayanDSFisher1993b}
O.~Narayan and D.S. Fisher,
\newblock {\em Nonlinear fluid flow in random media: critical phenomena near
  threshold},
\newblock Phys. Rev. B {\bf 49} (1993)   9469--502.

\bibitem{BlatterFeigelmanGeshkenbeinLarkinVinokur1994}
G.~Blatter, M.V. {Feigel'man}, V.B. Geshkenbein, A.I. Larkin  and V.M. Vinokur,
\newblock {\em Vortices in high-temperature superconductors},
\newblock Rev. Mod. Phys. {\bf 66} (1994)   1125.

\bibitem{ErtasKardar1994}
D.~Ertas and M.~Kardar,
\newblock {\em Anisotropic scaling in depinning of a flux line},
\newblock Phys. Rev. Lett. {\bf 73} (1994)   1703--6.

\bibitem{ErtasKardar1996}
D.~Ertas and M.~Kardar,
\newblock {\em Anisotropic scaling in threshold critical dynamics of driven
  directed lines},
\newblock Phys. Rev. {\bf B 53} (1996)   3520--42.

\bibitem{BalentsBouchaudMezard1996}
L.~Balents, J.P. Bouchaud  and M.~M\'ezard,
\newblock {\em The large scale energy landscape of randomly pinned objects},
\newblock J. Phys. I (France) {\bf 6} (1996)   1007--20.

\bibitem{Kardar1997}
M.~Kardar,
\newblock {\em Nonequilibrium dynamics of interfaces and lines},
\newblock Phys. Rep. {\bf 301} (1998)   85--112.

\bibitem{LeschhornNattermannStepanowTang1997}
H.~Leschhorn, T.~Nattermann, S.~Stepanow  and L.-H. Tang,
\newblock {\em Driven interface depinning in a disordered medium},
\newblock Annalen der Physik {\bf 6} (1997)   1--34.

\bibitem{BucheliWagnerGeshkenbeinLarkinBlatter1998}
H.~Bucheli, O.S. Wagner, V.B. Geshkenbein, A.I. Larkin  and G.~Blatter,
\newblock {\em {$(4+N)$}-dimensional elastic manifolds in random media: a
  renormalization-group analysis},
\newblock Phys. Rev. B {\bf 57} (1998)   7642--52.

\bibitem{DSFisher1998}
D.S. Fisher,
\newblock {\em Collective transport in random media: from superconductors to
  earthquakes},
\newblock Phys. Rep. {\bf 301} (1998)   113--150.

\bibitem{DincerDiplom}
Yusuf Dincer,
\newblock {\em Zur Universalit\"at der Struktur elastischer Mannigfaltigkeiten
  in Unordnung},
\newblock Master's thesis, Universit\"at K\"oln, 8 1999.

\bibitem{Scheidl2loopPrivate}
S.~Scheidl,
\newblock Private communication about 2-loop calculations for the random
  manifold problem. 2000-2004.

\bibitem{ScheidlDincer2000}
S.~Scheidl and Y.~Dincer,
\newblock {\em Interface fluctuations in disordered systems: Universality and
  non-gaussian statistics},
\newblock cond-mat\slash  {\bf 0006048} (2000).

\bibitem{NattermannScheidl2000}
T.~Nattermann and S.~Scheidl,
\newblock {\em Vortex-glass phases in type-ii superconductors},
\newblock Advances in Physics {\bf 49} (2000)   607--704.

\bibitem{GorokhovFisherBlatter2002}
DA. Gorokhov, DS. Fisher  and G.~Blatter,
\newblock {\em Quantum collective creep: a quasiclassical {Langevin} equation
  approach},
\newblock Phys. Rev. B {\bf 66} (2002)   214203.

\bibitem{SchwarzFisher2002}
J.~M. Schwarz and Daniel~S. Fisher,
\newblock {\em Depinning with dynamic stress overshoots: A hybrid of critical
  and pseudohysteretic behavior},
\newblock cond-mat\slash {\bf 0204623} (2002).

\bibitem{Gruner1988}
G.~Gruner,
\newblock {\em The dynamics of charge-density waves},
\newblock Rev. of Mod. Phys. {\bf 60} (1988)   1129--81.

\bibitem{LemerleFerreChappertMathetGiamarchiLeDoussal1998}
S.~Lemerle, J.~{Ferr\'e}, C.~Chappert, V.~Mathet, T.~Giamarchi  and P.~{Le
  Doussal},
\newblock {\em Domain wall creep in an {Ising} ultrathin magnetic film},
\newblock Phys. Rev. Lett. {\bf 80} (1998)   849.

\bibitem{PrevostRolleyGuthmann1999}
A.~Prevost, E.~Rolley  and C.~Guthmann,
\newblock {\em Thermally activated motion of the contact line of a liquid
  {$^4$He} meniscus on a cesium substrate},
\newblock Phys. Rev. Lett. {\bf 83} (1999)   348--51.

\bibitem{PrevostRolleyGuthmann2002}
A.~Prevost, E.~Rolley  and C.~Guthmann,
\newblock {\em Dynamics of a helium-4 meniscus on a strongly disordered cesium
  substrate},
\newblock Phys. Rev. B {\bf 65} (2002)   064517/1--8.

\bibitem{MoulinetGuthmannRolley2002}
S.~Moulinet, C.~Guthmann  and E.~Rolley,
\newblock {\em Roughness and dynamics of a contact line of a viscous fluid on a
  disordered substrate},
\newblock Eur. Phys. J. A {\bf 8} (2002)   437--43.

\bibitem{GiamarchiLeDoussalBookYoung}
T.~Giamarchi and P.~{Le~Doussal},
\newblock {\em Statics and dynamics of disordered elastic systems},
\newblock in A.P. Young, editor, {\em Spin glasses and random fields}, World
  Scientific, Singapore, 1997,
\newblock cond-mat/{\bf 9705096}.

\bibitem{KPZ}
M.~Kardar, G.~Parisi  and Y.-C. Zhang,
\newblock {\em Dynamic scaling of growing interfaces},
\newblock Phys. Rev. Lett. {\bf 56} (1986)   889--892.

\bibitem{LassigKinzelbach1997}
M.~Lassig and H.~Kinzelbach,
\newblock {\em Upper critical dimension of the {Kardar-Parisi-Zhang} equation},
\newblock Phys. Rev. Lett. {\bf 78} (1997)   903--6.

\bibitem{MarinariPagnaniParisi2000}
E.~Marinari, A.~Pagnani  and G.~Parisi,
\newblock {\em Critical exponents of the {KPZ} equation via multi-surface
  coding numerical simulations},
\newblock J. Phys. A {\bf 33} (2000)   8181--92.

\bibitem{Laessig1995}
M.~L{\"a}ssig,
\newblock {\em On the renormalization of the {Kardar-Parisi-Zhang} equation},
\newblock Nucl. Phys. {\bf B 448} (1995)   559--574.

\bibitem{Wiese1998a}
K.J. Wiese,
\newblock {\em On the perturbation expansion of the {KPZ}-equation},
\newblock J. Stat. Phys. {\bf 93} (1998)   143--154,
\newblock cond-mat/{\bf 9802068}.

\bibitem{Wiese2003a}
K.J. Wiese,
\newblock {\em The functional renormalization group treatment of disordered
  systems: a review},
\newblock Ann. Henri Poincar\'e {\bf 4} (2003)   473--496,
\newblock cond-mat/{\bf 0302322}.

\bibitem{BalentsLeDoussal2002}
L.~Balents and P.~Le Doussal,
\newblock {\em Field theory of statics and dynamics of glasses: rare events and
  barrier distributions},
\newblock cond-mat\slash  {\bf 0205358} (2002).

\bibitem{BalentsLeDoussal2003}
L.~Balents and P.~Le Doussal,
\newblock {\em Broad relaxation spectrum and the field theory of glassy
  dynamics for pinned elastic systems},
\newblock cond-mat\slash  {\bf 0312338} (2003).

\bibitem{LeDoussalWiesePREPf}
P.~Le Doussal and K.J. Wiese,
\newblock {\em 2-loop functional renormalization group treatment of random
  field models},
\newblock in preparation.

\bibitem{LeDoussalWiesePREPb}
K.J. Wiese and P.~Le Doussal,
\newblock {\em 3-loop {FRG} study of pinned manifolds},
\newblock in preparation.

\bibitem{EfetovLarkin1977}
K.B. Efetov and A.I. Larkin,
\newblock Sov. Phys. JETP {\bf 45} (1977)   1236.

\bibitem{MezardParisiVirasoro}
M.~M\'ezard, G.~Parisi  and M.A. Virasoro,
\newblock {\em Spin Glas Theory and Beyond},
\newblock World Scientific, Singapore, 1987.

\bibitem{Middleton2001}
A.A. Middleton,
\newblock {\em Energetics and geometry of excitations in random systems},
\newblock Phys. Rev. B {\bf 63} (2001)   060202.

\bibitem{Goldschmidt1993}
YY. Goldschmidt,
\newblock {\em The $1/d$ expansion for the quantum mechanical $n$-body problem.
  application for directed polymers in a random medium},
\newblock Nucl. Phys. B {\bf 393} (1993)   507--22.

\bibitem{DeDominicisKondorTemesvari1994}
C.~De Dominicis, I.~Kondor  and T.~Temesvari,
\newblock {\em Dyson's equations for the {Ising} spin-glass},
\newblock J. Phys. I (France) {\bf 4} (1994)   1287--308.

\bibitem{CarlucciDeDominicisTemesvari1996}
D.M. Carlucci, C.~De Dominicis  and T.~Temesvari,
\newblock {\em Stability of the {Mezard-Parisi} solution for random manifolds},
\newblock J. Phys. I (France) {\bf 6} (1996)   1031--41.

\bibitem{DeDominicisEtAlBookYoung}
C.~De Dominicis,
\newblock {\em Beyond the {Sherrington-Kirkpatrick Model}},
\newblock in A.P. Young, editor, {\em Spin glasses and random fields}, World
  Scientific, Singapore, 1997.

\bibitem{BrezinDeDominicis1998}
E.~Brezin and C.~De Dominicis,
\newblock {\em New phenomena in the random field {Ising} model},
\newblock Europhys. Lett. {\bf 44} (1998)   13--19.

\bibitem{BrezinDeDominicis2001}
E.~Brezin and C.~De Dominicis,
\newblock {\em Interactions of several replicas in the random field ising
  model},
\newblock Eur. Phys. J. B {\bf 19} (2001)   467--71.

\bibitem{BrezinDeDominicis2002a}
E.~Br\'ezin and C.~De Dominicis,
\newblock {\em Twist free energy in a spin glass},
\newblock cond-mat\slash  {\bf 0201066} (2002).

\bibitem{BrezinDeDominicis2002b}
E.~Br\'ezin and C.~De Dominicis,
\newblock {\em Twist free energy},
\newblock cond-mat\slash  {\bf 0201069} (2002).

\bibitem{DeDominicisBrezin2004}
C.~De Dominicis and E.~B\'rezin,
\newblock {\em On a dynamical-like replica-symmetry-breaking scheme for the
  spin glass},
\newblock cond-mat\slash {\bf 0402629} (2004).

\bibitem{Zimmermann1969}
W.~Zimmermann,
\newblock {\em Convergence of {Bogoliubov's} method of renormalization in
  monmentum space},
\newblock Commun. Math. Phys. {\bf 15} (1969)   208--234.

\bibitem{Hepp1966}
K.~Hepp,
\newblock {\em Proof of the {Bogoliubov-Parasiuk} theorem on renormalization},
\newblock Comm. Math. Phys. {\bf 2} (1966)   301--326.

\bibitem{BogoliubovParasiuk1957}
N.N. Bogoliubov and O.S. Parasiuk,
\newblock {\em {\"Uber die Multiplikation der Kausalfunktionen in der
  Quantentheorie der Felder}},
\newblock Acta Math. {\bf 97} (1957)   227.

\bibitem{BergereLam1975}
M.C. Bergere and Y.-M.P. Lam,
\newblock {\em {Bogoliubov-Parasiuk} theorem in the $\alpha$-parametric
  representation},
\newblock J. Math. Phys. {\bf 17} (1976)   1546--1557.

\end{thebibliography}

\end{document}